\documentclass[preprint]{ptephy_v1}

\preprintnumber{XXXX-XXXX} 

\usepackage{graphics} 
\usepackage[dvipdfmx]{}
\usepackage{url} 
\usepackage{natbib}
\usepackage{rotating}
\usepackage{longtable}
\usepackage{lscape}
\usepackage{color}
\usepackage[hang,small,bf]{caption}
\usepackage[subrefformat=empty]{subcaption}
\def\vector#1{\mbox{\boldmath $#1$}}

\begin{document}

\title{J-GEM optical and near-infrared follow-up of gravitational wave events during LIGO's and Virgo's third observing run}

\shorttitle{J-GEM optical and near-infrared follow-up in O3}


\author[1,2]{Mahito Sasada}
\affil[1]{Hiroshima Astrophysical Science Center, Hiroshima University, 1-3-1 Kagamiyama, Higashi-Hiroshima, Hiroshima 739-8526, Japan\email{sasadam@hiroshima-u.ac.jp}}
\affil[2]{Mizusawa VLBI Observatory, National Astronomical Observatory of Japan, 2-12 Hoshigaoka, Mizusawa, Oshu, Iwate 023-0861, Japan}
\author[3]{Yousuke Utsumi}
\affil[3]{Kavli Institute for Particle Astrophysics and Cosmology (KIPAC), SLAC National Accelerator Laboratory, Stanford University, 2575 Sand Hill Road, Menlo Park, CA 94025, USA}
\author[4]{Ryosuke Itoh}
\affil[4]{Bisei Astronomical Observatory, 1723-70 Ookura, Bisei-cho, Ibara, Okayama 714-1411, Japan}
\author[5,6]{Nozomu Tominaga}
\affil[5]{Department of Physics, Faculty of Science and Engineering, Konan University, 8-9-1 Okamoto, Kobe, Hyogo 658-8501, Japan}
\affil[6]{Kavli Institute for the Physics and Mathematics of the Universe (WPI), The University of Tokyo, 5-1-5 Kashiwanoha, Kashiwa, Chiba, 277-8583, Japan}
\author[7]{Masaomi Tanaka}
\affil[7]{Astronomical Institute, Tohoku University, Sendai 980-8578, Japan}
\author[8]{Tomoki Morokuma}
\affil[8]{Institute of Astronomy, Graduate School of Science, The University of Tokyo, 2-21-1 Osawa, Mitaka, Tokyo 181-0015, Japan}
\author[1,9]{Kenshi Yanagisawa}
\affil[9]{National Astronomical Observatory of Japan, 2-21-1 Osawa, Mitaka, Tokyo 181-8588, Japan}
\author[1]{Koji S. Kawabata}
\author[5]{Takayuki Ohgami}
\author[10]{Michitoshi Yoshida}
\affil[10]{Subaru Telescope, National Astronomical Observatory of Japan, 650 North A’ohoku Place, Hilo, HI 96720,USA}
\author[11]{Fumio Abe}
\affil[11]{Institute for Space-Earth Environmental Research, Nagoya University, Chikusa-ku, Aichi 464-8601, Japan}
\author[12]{Ryo Adachi}
\affil[12]{Department of Physics, Tokyo Institute of Technology, 2-12-1 Ookayama, Meguro-ku, Tokyo 152-8551, Japan}
\author[1]{Hiroshi Akitaya}
\author[13]{Yang Chong}
\affil[13]{Department of Physical Science, Hiroshima University, Kagamiyama 1-3-1, Higashi-Hiroshima 739-8526, Japan}
\author[7,14]{Kazuki Daikuhara}
\affil[14]{Department of Physics, Faculty of Science, Toho University, 2-2-1 Miyama, Funabashi, Chiba 274-8510, Japan}
\author[5]{Ryo Hamasaki}
\author[15]{Satoshi Honda}
\affil[15]{Nishi-harima Astronomical Observatory, Center for Astronomy, University of Hyogo, 407-2 Nishi-gaichi, Sayo, Hyogo 679-5313, Japan}
\author[12]{Ryohei Hosokawa}
\author[12]{Kota Iida}
\author[13]{Fumiya Imazato}
\author[16]{Chihiro Ishioka}
\affil[16]{Faculty of education, Saitama University, Simo-okubo 255, Sakura-ku, Saitama 338-8570, Japan}
\author[5]{Takumi Iwasaki}
\author[17]{Mingjie Jian}
\affil[17]{Department of Astronomy, School of Science, The University of Tokyo, 7-3-1 Hongo, Bunkyo-ku, Tokyo 113-0033, Japan}
\author[11]{Yuhei Kamei}
\author[14]{Takahiro Kanai}
\author[18]{Hidehiro Kaneda}
\affil[18]{Graduate School of Science, Nagoya University, Chikusa-ku, Nagoya, Aichi 464-8602, Japan}
\author[5,13]{Ayane Kaneko}
\author[15]{Noriyuki Katoh}
\author[12]{Nobuyuki Kawai}
\author[14,19]{Keiichiro Kubota}
\affil[19]{Department of Astronomy, Graduate School of Science, Kyoto University, Kitashirakawa-Oiwake-cho, Sakyo-ku, Kyoto, Kyoto 606-8502, Japan}
\author[16]{Yuma Kubota}
\author[12]{Hideo Mamiya}
\author[20]{Kazuya Matsubayashi}
\affil[20]{Okayama Observatory, Kyoto University, 3037-5 Honjo, Kamogata-cho, Asakuchi, Okayama 719-0232, Japan}
\author[18,21]{Kumiko Morihana}
\affil[21]{Institute of Liberal Arts and Sciences, Nagoya University, Furo-cho, Chikusa-ku,Nagoya, 464-8601, Japan}
\author[12]{Katsuhiro L. Murata}
\author[22]{Takahiro Nagayama}
\affil[22]{Graduate School of Science and Engineering, Kagoshima University, 1-21-40 Korimoto Kagoshima-city Kagoshima, 890-0065, Japan}
\author[12]{Noriatsu Nakamura}
\author[1]{Tatsuya Nakaoka}
\author[8,23]{Yuu Niino}
\affil[23]{Research Center for the Early Universe, the University of Tokyo, 7-3-1 Hongo, Bunkyo, Tokyo 113-0033, Japan} 
\author[13]{Yuki Nishinaka}
\author[12]{Masafumi Niwano}
\author[19]{Daisaku Nogami}
\author[16,24]{Yumiko Oasa}
\affil[24]{Graduate School of Science and Engineering, Saitama University, Simo-okubo 255, Sakura-ku, Saitama 338-8570, Japan}
\author[12]{Miki Oeda}
\author[12]{Futa Ogawa}
\author[8]{Ryou Ohsawa}
\author[19]{Kouji Ohta}
\author[16]{Kohei Oide}
\author[15]{Hiroki Onozato}
\author[8]{Shigeyuki Sako}
\author[15]{Tomoki Saito}
\author[14]{Yuichiro Sekiguchi}
\author[23]{Toshikazu Shigeyama}
\author[16]{Takumi Shigeyoshi}
\author[23,25]{Minori Shikauchi}
\affil[25]{Department of Physics, School of Science, The University of Tokyo, 7-3-1 Hongo, Bunkyo, Tokyo 113-0033, Japan}
\author[12]{Kazuki Shiraishi}
\author[26]{Daisuke Suzuki}
\affil[26]{Department of Earth and Space Science, Graduate School of Science, Osaka University, 1-1 Machikaneyama, Toyonaka, Osaka 560-0043, Japan}
\author[13]{Kengo Takagi}
\author[15]{Jun Takahashi}
\author[24,27]{Takuya Takarada}
\affil[27]{Astrobiology Center, NINS, 2-21-1 Osawa, Mitaka, Tokyo 181-8588, Japan}
\author[15]{Masaki Takayama}
\author[24]{Himeka Takeuchi}
\author[16]{Yasuki Tamura}
\author[14]{Ryoya Tanaka}
\author[12]{Sayaka Toma}
\author[15]{Miyako Tozuka}
\author[13]{Nagomi Uchida}
\author[16]{Yoshinori Uzawa}
\author[20]{Masayuki Yamanaka}
\author[14]{Moeno Yasuda}
\author[12]{Yoichi Yatsu}



\begin{abstract}
The Laser Interferometer Gravitational-wave Observatory Scientific Collaboration and Virgo Collaboration (LVC) sent out 56 gravitational-wave (GW) notices during the third observing run (O3). Japanese collaboration for Gravitational wave ElectroMagnetic follow-up (J-GEM) performed optical and near-infrared observations to identify and observe an electromagnetic (EM) counterpart. We constructed web-based system which enabled us to obtain and share information of candidate host galaxies for the counterpart, and status of our observations. Candidate host galaxies were selected from the GLADE catalog with a weight based on the three-dimensional GW localization map provided by LVC.
We conducted galaxy-targeted and wide-field blind surveys, real-time data analysis, and visual inspection of observed galaxies. We performed galaxy-targeted follow-ups to 23 GW events during O3, and the maximum probability covered by our observations reached to 9.8\%. Among them, we successfully started observations for 10 GW events within 0.5 days after the detection. This result demonstrates that our follow-up observation has a potential to constrain EM radiation models for a merger of binary neutron stars at a distance of up to $\sim$100~Mpc with a probability area of $\leq$ 500~deg$^2$.
\end{abstract}

\subjectindex{F12, F33}

\maketitle

\section{Introduction} \label{sec:intro}
In 2015, gravitational waves (GWs) were detected for the first time by the Laser Interferometer Gravitational-wave Observatory (LIGO) \citep{2015CQGra..32k5012A}, which marked the dawn of GW astronomy. The most plausible targets for the current GW detectors are signals emitted from compact binary coalescences (CBCs). CBCs include three types: Mergers of binaries composed of two black holes (BHs), two neutron stars (NSs), and one NS and one BH, called BBH, BNS, and NSBH, respectively. 
The first GW event, GW150914, detected by Advanced LIGO was a coalescing BBH \citep{2016PhRvL.116x1102A}. In 2017, the first GW event from a BNS, GW170817, was also detected by both the Advanced LIGO and Advanced Virgo \citep{2015CQGra..32b4001A} detectors \citep{2017PhRvL.119p1101A}.

The direction of a GW signal is predominantly determined by the differences of the arrival times of the GW to more than two detectors. The uncertainty of the localization depends on the number of detectors that observe the GW signal, the source direction, and the signal-to-noise ratio. The median 90\% credible region is 120--180 square degrees \cite{2018LRR....21....3A}, and is regularly exceed $\sim$1000 square degrees.

Depending on the nature of the coalescing source, a GW source may accompany electromagnetic (EM) emission \cite{1989Natur.340..126E,1998ApJ...507L..59L,2010MNRAS.406.2650M}. The identification and observation of the EM counterpart to a GW signal are important not only for understanding the physics of the merging event but also for maximally utilizing information in the GW signal.
However, identifying an EM counterpart to a GW source is challenging, particularly in optical bands because of limited field-of-views (FoVs) of instruments.
A wide-field blind survey using a wide FoV instrument can search any EM counterpart associated with the GW source, even without any association with host-like galaxy. Another approach is to survey cataloged galaxies within the three-dimensional (3D) localization map under the assumption that the EM counterpart is located close to or on a galaxy \citep{2016ApJ...820..136G}. This approach is called a galaxy-targeted survey. 

In fact, EM follow-up surveys to GW150914 were performed across the entire EM spectrum, from radio to gamma-ray bands \citep{2016ApJ...826L..13A}. However, no evident EM counterpart was discovered \citep{2016MNRAS.462.4094S,2016ApJ...824L..24K,2016ApJ...823L..33S,2016PASJ...68L...9M}. 
The non-detection in GW150914 may be a natural consequence of BBH merger. Note that a possible gamma-ray counterpart to GW150914 was tentatively claimed by \cite{2016ApJ...826L...6C}, which triggered the discussion on the possible presence of relativistic jets in BBH mergers as in gamma-ray bursts (e.g. \citep{2016PTEP.2016e1E01Y}, \citep{2017NewA...51....7J}, though see the counterargument presented by \cite{2016ApJ...827L..38G}). 

Identification of EM counterpart to a GW event was achieved for the first time for GW170817, where a gamma-ray signal was detected 1.74$\pm$0.05 sec after the GW detection (GRB~170817A, \cite{2017ApJ...848L..13A,2017ApJ...848L..15S}). Subsequently, the optical counterpart (AT~2017gfo) was identified by galaxy-targeted surveys, at 10.86 hours after the GW detection \citep{2017ApJ...848L..12A,2017Sci...358.1556C}. The optical emission of AT~2017gfo rapidly declined, while the near-infrared (NIR) emission lasted relatively longer (e.g. \cite{2017ApJ...848L..16S,2017ApJ...848L..17C,2017ApJ...848L..24V,2017ApJ...850L...1L,2017PASJ...69..101U}.) The optical spectra of AT~2017gfo rapidly evolved from blue to red \citep{2017Sci...358.1574S,2017Natur.551...67P}. These properties are clearly different from those of supernovae. Also, EM emission was detected in X-ray (e.g. \citep{2017ApJ...848L..25H}, \citep{2017Sci...358.1565E}) and radio wavelengths (e.g. \citep{2017ApJ...848L..21A}) at 9 and 16 days after the GW detection, respectively. The X-ray and radio emission were detected for hundreds of days \citep{2019ApJ...886L..17H}. At the same time, the Very Long Baseline Interferometric observation detected a knot with a superluminal motion \citep{2018Natur.561..355M}.

When a BNS coalesces, a portion of matter of the merger material is ejected into interstellar space \citep{1986ApJ...308L..43P}. In this material, $r$-process elements including lanthanides are likely to be synthesized. This matter gives rise to thermal emission mainly in the optical and IR bands, which is powered by the radioacitve decay of the $r$-process nuclei \citep{1989Natur.340..126E}, called a "kilonova". The observed temporal evolution of the optical and NIR emission of AT~2017gfo was consistent with simple kilonova models \citep{2017PhRvD..96l3012S,2017PASJ...69..102T,2017ApJ...851L..21V}. However, the blue component seen in the early phase may not be fully reproduced by the simple kilonova model. To explain the blue emission, two models were proposed; (1) a kilonova model with a higher electron fraction \citep{2017PASJ...69..102T}, or (2) a cocoon generated by interactions between the NS ejecta and the jet that powered the emission from radio to gamma-ray wavelengths \citep{2017ApJ...848L..20M,2018ApJ...855..103P,2018MNRAS.478L..18T,2018MNRAS.479..588G,2018PhRvL.120x1103L}. Both models can explain the light curve observed after 0.5 days. But these models predict different brightness at $\leq$ 0.5 days \citep{2018ApJ...855L..23A}. Thus, early-phase observations are required to determine the origin of the optical blue emission.

The third LIGO/Virgo observing run (O3) started on 1st April, 2019. During O3, 56 GW alerts were announced \footnote{\url{https://gracedb.ligo.org/superevents/public/O3/}}. A GW signal that is consistent with the CBCs can be estimated plausible masses of its components. Based on the masses, the detected GW signals in O3 were classified into 37 BBHs, six BNSs, five NSBHs, and four MassGap objects of which masses consist of between 3$\;M_{\odot}$ and 5$\;M_{\odot}$. Details of four GW events were already published by the LIGO Scientific Collaboration and the Virgo Collaboration (LVC): BBH event GW190412 was likely originated from the coalescence of BBH with asymmetric masses of $\sim8\;M_{\odot}$ and $\sim30\;M_{\odot}$ \citep{2020PhRvD.102d3015A}, and GW190521 was originated from the BBH merger with a total mass of 150 $M_{\odot}$ \cite{2020PhRvL.125j1102A}. In this event, a candidate of an optical counterpart was claimed \cite{2020PhRvL.124y1102G}. A BNS event GW190425, which was the first event with a NS in O3, was likely originated from a coalescence of BNS with a total mass of $3.4\;M_{\odot}$ \citep{2020ApJ...892L...3A}. Finally, a NSBH event GW190814 was likely originated from the coalescence of a 23 $M_{\odot}$ BH with a 2.6 $M_{\odot}$ compact object \cite{2020ApJ...896L..44A}.

In the O3, EM follow-up observations have been performed by various telescopes and satellites from a gamma-ray to radio wavelengths. In the optical and NIR wavelengths, a lot of follow-up campaigns have been performed for detected GW signals: GROWTH \cite{2019ApJ...885L..19C,2019ApJ...881L...7G,2019ApJ...881L..16A,2020ApJ...890..131A,2020arXiv200305516A}, GRANDMA \cite{2020MNRAS.492.3904A,2020MNRAS.497.5518A}, GOTO \cite{2020MNRAS.497..726G}, Magellan \cite{2019ApJ...884L..55G}, ZTF \cite{2020NatAs.tmp..179A}, MMT and SOAR \cite{2019ApJ...880L...4H}.

The Japanese collaboration for Gravitational-wave ElectroMagnetic follow-up (J-GEM) has conducted optical and NIR follow-up campaigns of GW events since 2015 \citep{2016PASJ...68L...9M,2017PASJ...69....9Y,2018PASJ...70....1U,2017PASJ...69..101U,2017PASJ...69..102T,2018PASJ...70...28T}. In this paper, we introduce J-GEM facilities and infrastructures which realize rapid follow-up observations, and describe details of our galaxy-targeted surveys to GW events in O3 run. Results of wide-field blind surveys by Hyper Suprime-Cam (HSC) on the 8.2-m Subaru telescope and Tomo-e Gozen on the 1.05-m Kiso Schmidt telescope will be reported by \cite{Ohgami2020,Niino2020}, respectively. We describe our observational methods in Section~\ref{sec:method}, and the results of the follow-up observations  in Section~\ref{sec:O3}. In Section~\ref{sec:discussion}, we discuss the results and future prospects of our follow-up campaigns. Finally, we give our conclusions in Section~\ref{sec:conclusion}.

\section{Observation Methods} \label{sec:method}
J-GEM consists of a consortium of facilities that cover optical, NIR, and radio wavelengths. To enable rapid follow-up observations for GW sources detected at arbitrary time from various directions, we form a network of multi-latitudinal and longitudinal observatories (see figure~20 in \cite{Tanaka2020}). Our network enables an effective search for a transient in a large GW localization area by involving multiple telescopes and cameras. The telescopes used for EM follow-up during O3 are listed in Table~\ref{tab:telescope}. With these multi-located telescopes, we are able to perform the stable and quick follow-up avoiding the effect of bad weather. Optical data were obtained with nine instruments on eight telescopes, and NIR data were obtained with four instruments on four telescopes.

We conducted both galaxy-targeted and wide-field surveys in response to GW event alerts. These two approaches have different advantages. The wide-field survey can provide a search of the counterpart regardless of having or not having the host galaxy if the FoV of the camera is sufficiently large. The galaxy-targeted survey can also be effective if the counterpart is associated with the host galaxy. To maxmize the efficiency, the cataloged galaxies within 3D localization map can be ranked according to the position in the localization area as well as a luminosity of the galaxy. By coordinating galaxy-targeted observations with multiple telescopes distributed over the world, we can further enhance the capability of our survey by reducing the risk of weather loss. A drawback of galaxy-targetted survey is an incompleteness of the cataloged galaxies, which is  severe at $\geq$200~Mpc (Fig.~\ref{fig:completeness}). On the other hand, the wide-field surveys do not rely on the information of cataloged galaxies.

\clearpage
\begin{landscape}
\begin{table}[!h]
\caption{J-GEM telescopes which are used in O3}
\label{tab:telescope}
\centering
\footnotesize
\begin{tabular}{cccccccc}
\hline
Telescope & Diam. [m] & Instrument & Obs. Mode & Limiting mag. & exp. time & Zero point accuracy & Note \\
(1) & (2) & (3) & (4) & (5) & (6) & (7) & (8) \\
\hline\hline
Subaru & 8.2 & Hyper Suprime-Cam (HSC)(a) & Imaging ($r2$, $z$,$Y$) & 25.3, 23.6, 22.9 & & 0.1 & (A) \\
 &  & FOCAS(b) & Imaging ($R$), spectroscopy (Optical) & ($\sim$23) & 10 & --- & (B), (C) \\
Seimei & 3.8 & KOOLS-IFU(c) & Spectroscopy (Optical) & ($\sim$18) & 600 & --- & (C) \\
Nayuta & 2.0 & NIC & Imaging ($J$, $H$, $K_{\rm S}$) & 20.09, 20.73, (19.0) & 600 & 0.14(0.02), 0.07(0.02), --- & (B) \\
Kanata & 1.5 & HONIR(d) & Imaging ($R_{\rm C}$, $H$) & 19.40, 20.00 & 1040 & 0.17(0.02), 0.14(0.01) & (B)  \\
IRSF & 1.4 & SIRIUS(e),(f) & Imaging ($J$, $H$, $K_{\rm S}$) & (20.9), (20.8), (20.0) & 900 & --- & (B) \\
Kiso Schimidt & 1.05 & Tomo-e Gozen(g) & Imaging (Blank) & 18.67 & 12 & 0.09(0.02) & (A), (B) \\
B\&C 61cm & 0.61 & Tripole5 & Imaging ($g,r,i$) & 19.14, 19.31, 18.29 & 1080 & 0.03(0.02), 0.11(0.08), 0.07(0.01) & (B) \\
MITSuME Akeno & 0.5 & ($g,R_{\rm C},I_{\rm C}$ imager) & Imaging ($g,R_{\rm C},I_{\rm C}$) & 16.77, 16.41, 16.17 & 480 & 0.07(0.01), 0.08(0.01), 0.10(0.01) & (B) \\
OAO 91cm & 0.91 & OAOWFC(h) & Imaging ($J$) & 18.74 & 3416 & 0.09(0.01) & (B) \\
MITSuME Okayama & 0.5 & ($g,R_{\rm C},I_{\rm C}$ imager)(i),(j) & Imaging ($g,R_{\rm C},I_{\rm C}$) & 14.20, 14.47, 15.81 & 540 & 0.27(0.05), 0.29(0.03), 0.12(0.01) & (B) \\
SaCRA & 0.55 & MuSaSHI & Imaging ($r, i, z$) & 17.58, 18.78, (17.1) & 1800 & 0.14(0.03), 0.05(0.02), --- & (B) \\
HinOTORI & 0.5 &  simultaneous imager & Imaging ($u, R_c, I_c$) & commissioning & --- & --- & (B) \\
\hline
\end{tabular}
\begin{flushleft}
(1) Telescope; (2) diameter of the telescope; (3) name of instrument; (4) observation mode and band(s); (5) typical 5-$\sigma$ limiting magnitude in the AB magnitude system. The limiting magnitudes of individual instruments were the median values at the EM follow-up observations in O3. An exception is that the values in parentheses was estimated from outside of EM follow-up observations. ; (6) typical exposure time in seconds; (7) accuracy of obtained zero point, which is the standard deviation of differences between the observed and catalog magnitudes. Values in parentheses are uncertainties of the accuracy, which were calculated by a bootstrap resampling method to the obtained and catalog magnitudes.; (8) Observation strategy: (A) wide-field survey, (B) galaxy-targeted survey, (C) integral field spectroscopy.
{\dag}References: (a)\cite{2018PASJ...70S...1M}; (b)\cite{2002PASJ...54..819K}; (c)\cite{2019PASJ...71..102M}; (d)\cite{2014SPIE.9147E..4OA}; (e)\cite{1999sf99.proc..397N}; (f)\cite{2003SPIE.4841..459N}; (g)\cite{2018SPIE10702E..0JS}; (h)\cite{2019PASJ...71..118Y}; (i)\cite{2005NCimC..28..755K}; (j)\cite{2010AIPC.1279..466Y}
\end{flushleft}
\end{table}
\end{landscape}

\clearpage

\begin{figure}[ht!]
\includegraphics[scale=0.5]{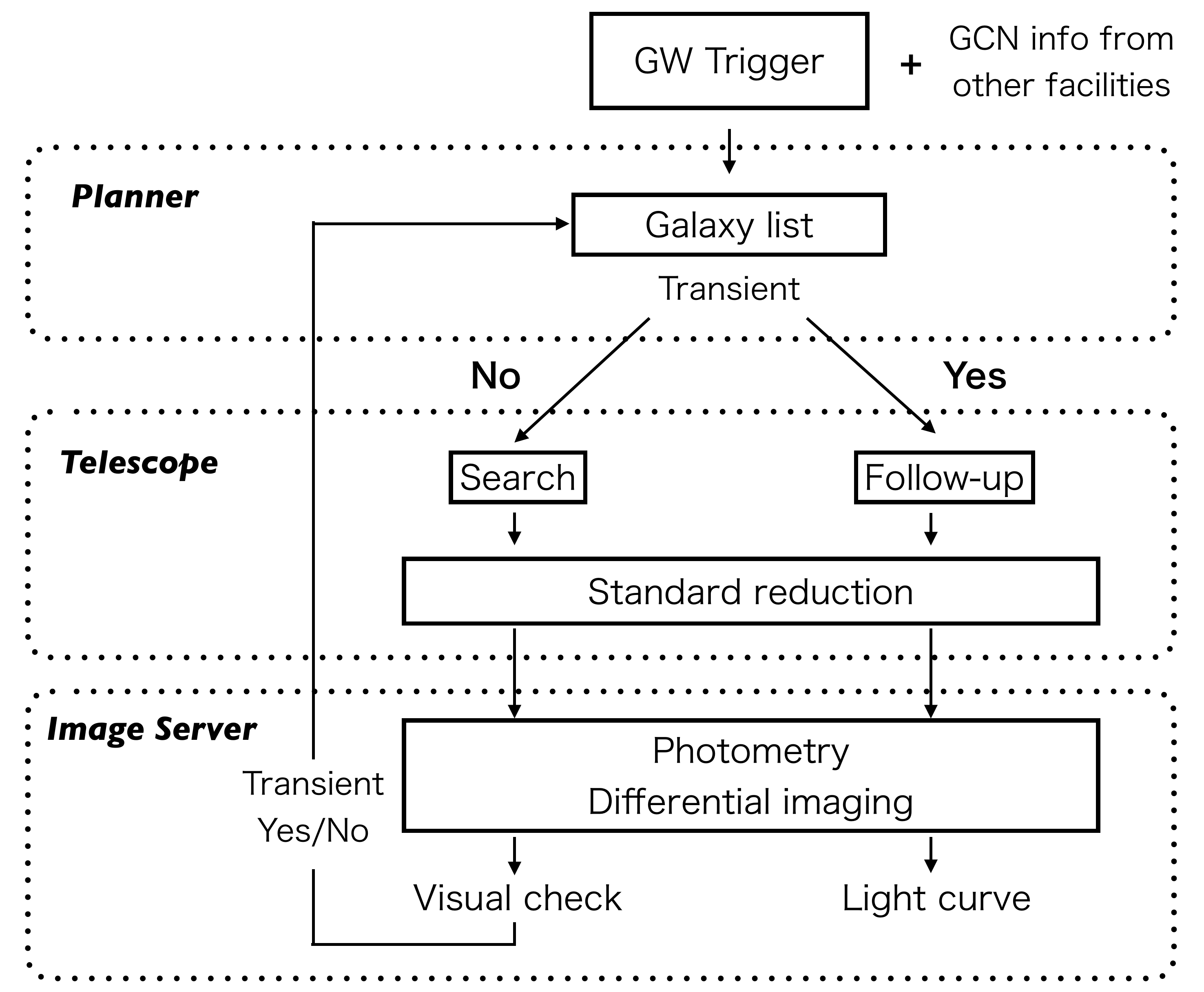} 
\caption{A flowchart of our EM follow-up scheme for GW events. \label{fig:flow}}
\end{figure}

\subsection{Galaxy-targeted observations}

For galaxy-targeted surveys, it is important to observe as many high-probability galaxies as possible in the 3D localization map of a GW source. Therefore, an effective observation scheme is required to increase the completeness of observations with respect to the candidate galaxies. To perform quick and efficient follow-up without duplication, we developed web-based systems (\emph{planner} and \emph{Image Server}), which enable us to obtain and share information of galaxies, and transients, as well as status of observations.

A flowchart of our follow-up scheme is shown in Fig.~\ref{fig:flow}. When a GW alert is announced, we receive triggers through notices in NASA's Gamma-ray Coordinates Network (GCN) using a python-based VOEvent client program (based on \texttt{pyGCN}\footnote{ \url{https://github.com/lpsinger/pygcn} }). A part of the GW alert, a 2D or 3D localization map is provided. Once the VOEvent client program generates a list of galaxies in the order of probability, this information is automatically saved in an internal database named \emph{planner}. Initiated by the GW trigger, we rank the galaxies in the catalog in terms of their possibility of hosting the GW source and their brightnesses. Information whether or not a transient exists in each galaxy is also registered in \emph{planner}. Through this system, we are able to start observations shortly after receiving a GCN notice. 

Each facility reduces the obtained data by the procedure of dark subtraction, flat fielding, and calculating the sky coordinates, soon after the observations are made. 
The reduced images are uploaded to a system, named \emph{Image Server}, which performs detection and photometry of sources in the uploaded image. The limiting magnitude is also evaluated. Differential imaging is performed to visually check whether a new transient exists or not by comparing the image with a reference image. In the case where a transient is identified, information on the presence (or absence) of transients is sent to the database. 
If a galaxy which has a transient is registered, a high priority is assigned to the galaxy. if there is no galaxies with transients, normal transient searches are performed for galaxies that have not been observed yet. In the case of follow-up for a discovered transient, the time variation of the target's brightness is derived from the database. \emph{planner} also collects information of the limiting magnitude of each observation. After the observation sequence, the system generates and distributes a GCN report. This system enables us to operate the rapid galaxy survey, and it allows us to assess our data quality. The data presented in this paper were all extracted from \emph{planner} and \emph{Image Server}. 

\subsubsection{Selection of galaxies} \label{subsec:tables}
For our galaxy-targeted survey, we employ the available galaxy catalog, Galaxy List for the Advanced Detector Era (GLADE) \citep{2018MNRAS.479.2374D}, to make a list of the candidate galaxies hosting the GW source. For the selection of candidate galaxies to search the host galaxy having an EM counterpart, weighting with the stellar mass of the galaxy estimated from the NIR brightness is a good way. However, in the GLADE catalog, the number of galaxies listed in the $K_{\rm S}$-band catalog is only about 60\% of that in the $B$-band one. Thus,  the $K_{\rm S}$-band candidate galaxies will result in a large number of galaxies being omitted from the targets. Although the selected $B$-band candidate galaxies will prefer blue star-forming galaxies, it can be acceptable, because it is currently not entirely clear what types of galaxies are associated with GW events. Furthermore, for short gamma-ray bursts, of which NS coalescence is though to be a possible origin, both early-type star-forming and late-type elliptical galaxies have been reported to be the host galaxies \cite{2015ApJ...815..102F}. This suggests that not a few GW events caused by NS coalescence happen in star-forming galaxies as their hosts. In this case, selecting galaxies in the $B$-band catalog with a large number of samples is more advantageous for searching the EM counterpart than in the near-infrared catalog, which is biased towards non-star-forming, inactive galaxies and has a small sample. Therefore, we decided to select candidate galaxies from the $B$-band catalog.

We rank the GLADE galaxies with the $B$-band luminosity weighted probability, $P^{(j)}$, using the following formula:
\begin{equation}
    P^{(j)} = \frac{ L_B^{(j)} p(\vector{r}^{(j)}) }{ \sum_i L_B^{(i)} \frac{p(\vector{r}^{(i)})}{C(\vector{r}^{(i)})}},
\end{equation}
where $\vector{r}^{(j)}$ is the position of the GLADE galaxy, $p(\vector{r}^{(j)})$ is the localization probability of the GW source at $\vector{r}^{(j)}$, $L^{(j)}_B$ is the $B$-band luminosity of the galaxy, and $C(\vector{r}^{(i)})$ is the completeness of the GLADE catalog at the distance of $\vector{r}^{(i)}$ (see appendix~\ref{sec:completeness}). For events that exhibit 3D localization maps provided by BAYESTAR or LALInference \citep{2016PhRvD..93b4013S,2013PhRvD..88f2001A}, the luminosity distance is also taken into account. If either the distance or $B$-band luminosity is not available in the GLADE catalog, we exclude the galaxy.
For the Burst event of which the signal is detected without a template and without prior knowledge of the waveform \citep{2020GCN.26734....1L}, we use the GWTC-1 catalog \cite{2019PhRvX...9c1040A} instead, and we only use the 2D localization map. Although we calculate the galaxy ranks in a similar way, this may not be relevant if the GW signal is emitted from a Galactic supernova \citep{2009CQGra..26f3001O}.

\subsubsection{Data reduction and candidate screening}
Observed images of galaxies are visually checked by comparing with a reference image, and an image subtracted with a reference frame. We configured a web-based application, \emph{Image Server}, which shows an uploaded image together with the reference and subtracted images. Additionally, the brightness of the candidate source and the limiting magnitude are measured by \emph{Image Server}.

The obtained data are reduced using standard procedures (dark-subtracting and flat-fielding.) Then the data are processed by the \textsf{Astrometry.net} application \citep{2010AJ....139.1782L} to perform astrometry. The World Coordinate System parameters are set in the header of each FITS image. Then, the reduced data are uploaded to \emph{Image Server}. At the server, the typical size of the point-spread-function (PSF) of point sources is measured using \texttt{SExtractor} \citep{1996A&AS..117..393B}. We perform aperture photometry with an aperture size of 2.5 times the estimated full-width-half-maximum size of the PSF. The photometric zero point, $ZP$, of each image in the AB magnitude system is calculated using photometric data of field stars cataloged in either Panoramic Survey Telescope and Rapid Response System (Pan-STARRS) for the $g$, $r$, and $i$ bands \citep{2016arXiv161205560C}, SkyMapper for the $g$, $r$, and $i$ bands \citep{2018PASA...35...10W}, the fourth United States Naval Observatory (USNO) CCD Astrograph Catalog (UCAC4) for the $g$, $r$, and $i$ bands \citep{2013AJ....145...44Z}, and the Two Micron All Sky Survey (2MASS) for the $J$, $H$ and $K_{\rm S}$ bands \citep{2006AJ....131.1163S}. The $R_{\rm C}$-, and $I_{\rm C}$-band fluxes are derived from the $g$-, $r$-, $i$-band fluxes, using transformation equations\footnote{\url{http://classic.sdss.org/dr4/algorithms/sdssUBVRITransform.html}} \citep{2005AJ....130..873J}.

The sky counts within a circle with the same size as used for the aperture photometry are used to estimate the sky background level. The average and standard deviation ($\sigma$) of the sky background levels are calculated from the summed sky counts at 2,000 random aperture positions. The limiting magnitude of the reduced data is estimated as $-2.5{\rm log}_{10}(n\sigma)+ZP$, where $n=5$, i.e., the 5-$\sigma$ detection limit is adopted.

Image subtraction between the new and reference images is performed using \emph{HOTPANTS}\footnote{ \url{https://github.com/acbecker/hotpants} } \citep{2015ascl.soft04004B}, which employs the algorithm presented by \cite{1998ApJ...503..325A,1999A&A...343...10A}. We visually check the new, reference, and differential images to see whether a new transient exists or not. In addition, the sources in the image are matched automatically with the PS1 catalog, and mark circles to manually check whether the sources are cataloged or not \citep{2016arXiv161205560C}. All measurements and results obtained with \emph{Image Server} are reported to \emph{planner} immediately.

\subsection{Wide-field follow-up observations}

We perform blind, wide-field follow-up observations using optical imaging instruments with wide FoVs: Tomo-e Gozen (20.7~deg$^{2}$) \citep{2018SPIE10702E..0JS} and HSC (1.8~deg$^{2}$) \citep{2018PASJ...70S...1M}. We have performed tiling observations with these instruments to cover as much of the GW localization area as possible because the FoVs of these instruments are still smaller than the 90\% localization areas of GW events in O3. The sky positions of the telescope pointings for the tiling observations are determined to effectively cover the high-probability regions of the 2D skymap data available in GraceDB by considering the visibility from each telescope (see appendix 1 in \cite{2018PASJ...70....1U} for the case of HSC). The obtained optical data are reduced using standard procedures. Then, we perform image subtraction between the reduced images and the reference image, and we extract variable sources from the difference images.  
Finally, we identify candidate EM counterparts from the list of screened sources after a visual inspection. The details of wide-field follow-up observations in O3 will be presented in forthcoming papers \citep{Ohgami2020,Niino2020}.

In addition to the above blind searches, images obtained with Tomo-e Gozen are cut out around the galaxies listed in \emph{plannar} with a size of 10$\times$10 arcmin$^2$. Then, the cutout images are uploaded to \emph{Image Server} and visually checked as in the galaxy-targeted approach.

\section{Results of our EM Follow-up Campaigns} \label{sec:O3}

During O3, we performed galaxy-targeted follow-up for 23 GW events (Table~\ref{tab2}). Depending on the properties of the signals and estimated masses, each event is categorized by LVC into six classes: BBH, BNS, NSBH, MassGap, Burst, and Terrestrial. A Terrestrial event is likely to be an instrumental or environmental origin. In addition, LVC estimates the probabilities, HasNS, in which the GW source includes at least one NS component, and HasRemnant, in which the GW source has a non-zero amount of neutron star matter. Each facilities took into account these parameters to decide observations. Eventually, we observed 11 BBH, five BNS, four NSBH, one MassGap, one Burst, and one Terrestrial (having $\sim$40\% probability of BNS) during O3. 

\begin{landscape}
\begin{table}[!h]
\caption{Followed-up GW events during O3}
\label{tab2}
\centering
\hspace{-18mm}
\begin{tabular}{ccccccc}
\hline
Superevent ID & Event time in UT & Classification & HasNS & HasRemnant & 90\% prob. area & Distance \\
 & (MJD) & (\%) &  & & (deg$^2$) & (Mpc) \\
 \hline\hline
S190408an & 2019 Apr 08 18:18:02 (58581.76253) & BBH ($\geq$99) & 0.0 & 0.12 & 387 & 1473$\pm$358 \\
S190412m (GW190412) & 2019 Apr 12 05:31:21 (58585.23011) & BBH (100) & 0.0 & 0.12 & 156 & 812$\pm$194 \\
S190421ar & 2019 Apr 21 21:39:33 (58594.90247) & BBH (97) & 0.0 & 0.0 & 1444 & 1628$\pm$535 \\
S190425z (GW190425) & 2019 Apr 25 08:18:42 (58598.34632) & BNS ($\geq$99) & 1.0 & 1.0 & 7461 & 156$\pm$41 \\
S190426c & 2019 Apr 26 15:21:55 (58599.64022) & BNS (49), MassGap (24) & 1.0 & 1.0 & 1131 & 377$\pm$100 \\
S190510g & 2019 May 10 02:59:39 (58613.12476) & Terrestrial (58), BNS (42) & 1.0 & 1.0 & 1166 & 227$\pm$92 \\
S190521r (GW190521) & 2019 May 21 07:43:59 (58624.32222) & BBH ($\geq$99) & 0.0 & 0.0 & 488 & 1136$\pm$279 \\
S190720a & 2019 Jul 20 00:08:37 (58684.00598) & BBH (99) & 0.0 & 0.0 & 443 & 869$\pm$283 \\
S190728q & 2019 Jul 28 06:45:48 (58692.28180) & BBH (95) & 0.0 & 0.0 & 104 & 874$\pm$171 \\
S190814bv (GW190814) & 2019 Aug 14 21:11:16 (58709.88282) & NSBH ($\geq$99) & 1.0 & 0.0 & 23 & 267$\pm$52 \\
S190901ap & 2019 Sep 01 23:31:39 (58727.98031) & BNS (86), Terrestrial (14) & 1.0 & 1.0 & 14753 & 241$\pm$79 \\
S190923y & 2019 Sep 23 12:56:37 (58749.53931) & NSBH (68), Terrestrial (32) & 1.0 & 0.0 & 2107 & 438$\pm$133 \\
S190930s & 2019 Sep 30 13:36:18 (58756.56688) & MassGap (95) & 0.0 & 0.0 & 1748 & 709$\pm$191 \\
S190930t & 2019 Sep 30 14:34:45 (58756.60746) & NSBH (74), Terrestrial (26) & 1.0 & 0.0 & 24220 & 108$\pm$38 \\
S191105e & 2019 Nov 05 14:35:59 (58792.60832) & BBH (95) & 0.0 & 0.0 & 643 & 1183$\pm$281 \\
S191205ah & 2019 Dec 05, 21:52:45 (58822.91164) & NSBH (93) & 1.0 & 0.0 & 6378 & 385$\pm$164 \\
S191213g & 2019 Dec 13 04:34:45 (58830.19080) & BNS (77), Terrestrial (23) & 1.0 & 1.0 & 4480 & 201$\pm$81 \\
S191216ap & 2019 Dec 16 21:34:15 (58833.89879) & BBH ($\geq$99) & 0.19 & 0.0 & 253 & 376$\pm$70 \\
S200114f & 2020 Jan 14 02:08:55 (58862.08953) & Burst & --- & --- & 403 & --- \\
S200213t & 2020 Feb 13 04:11:17 (58892.17451) & BNS (63), Terrestrial (37) & 1.0 & 1.0 & 2326 & 201$\pm$80 \\
S200219ac & 2020 Feb 19 09:44:52 (58898.40616) & BBH (96) & 0.0 & 0.0 & 781 & 3533$\pm$1031 \\
S200224ca & 2020 Feb 24 22:23:11 (58903.93277) & BBH ($\geq$99) & 0.0 & 0.0 & 72 & 1575$\pm$322 \\
S200225q & 2020 Feb 25 06:04:58 (58904.25345) & BBH (96) & 0.0 & 0.0 & 22 & 995$\pm$188 \\\hline
\end{tabular}
\end{table}
\end{landscape}

Data of our follow-up campaigns were recorded in \emph{planner} and \emph{Image Server}, and the summary is shown in Table~\ref{tab3}. The maximum number of the observed galaxies for each GW event is 170, which is for GW190425. The total coverage of each GW event is estimated by summing the probabilities of the observed galaxies, where the probability of each galaxy, $P^{(j)}$, is calculated using Eq. (1). The total sky coverage of our galaxy-targeted follow-up campaign for each GW event during O3 reaches a maximum of 9.8\% and a median value of 0.012\%.

\begin{table}[!h]
\caption{J-GEM galaxy-targeted follow-up during O3}
\label{tab3}
\centering
\small
\begin{tabular}{ccccccc}
\hline
Superevent & Classification & N$^{1}$ & Start time$^{2}$ & Coverage$^{3}$ & Telescope & GCN\\
 &  &  & (days) & (\%) &  & \\\hline\hline 
S190408an & BBH & 1 & 0.042 & 0.0042 & HONIR, MITSuME-Akeno & \cite{2019GCN.24064....1M}\\
 & & & & & MITSuME-Okayama & \\
S190412m & BBH & 157 & 0.211 & 0.62 & NIC, HONIR & \cite{2019GCN.24113....1T}\\
 & & & & & Tomo-e Gozen, Tripole5 & \cite{2019GCN.24350....1K}\\
 & & & & & MITSuME-Akeno, OAOWFC & \\
 & & & & & MuSaSHI & \\
S190421ar & BBH & 3 & 0.796 & 0.00078 & Tomo-e Gozen, OAOWFC & \\
S190425z & BNS & 170 & 0.093 & 1.9 & FOCAS, SIRIUS& \cite{2019GCN.24192....1S}\\
 & & & & & Tripole5, MITSuME-Akeno & \cite{2019GCN.24219....1M}\\
 & & & & & & \cite{2019GCN.24230....1M}\\
 & & & & & & \cite{2019GCN.24328....1M}\\
S190426c & BNS/MassGap & 64 & 0.077 & 2.0 & NIC, HONIR & \cite{2019GCN.24299....1N}\\
 & & & & & Tomo-e Gozen, OAOWFC & \\
S190510g & Ter/BNS & 15 & 0.308 & 1.3 & (HSC)$^{4}$, KOOLS-IFU & \cite{2019GCN.24464....1K}\\
 & & & & & NIC, HONIR & \\
 & & & & & Tomo-e Gozen, Tripole5 & \\
 & & & & & MITSuME-Akeno, OAOWFC & \\
 & & & & & MITSuME-Okayama, MuSaSHI & \\
S190521r & BBH & 1 & 0.203 & 0.0 & HONIR & \cite{2019GCN.24661....1S}\\
S190720a & BBH & 1 & 2.776 & 0.0017 & MITSuME-Akeno & \\
S190728q & BBH & 7 & 0.226 & 0.0070 & HONIR, MITSuME-Akeno & \cite{2019GCN.25226....1Y}\\
 & & & & & MITSuME-Okayama & \\
S190814bv & NSBH & 24 & 1.767 & 9.8 & HONIR, Tripole5 & \cite{2019GCN.25389....1S}\\
 & & & & & MITSuME-Akeno, OAOWFC & \cite{2019GCN.25377....1N}\\
 & & & & & MuSaSHI & \\
S190901ap & BNS/Ter & 2 & 0.456 & 0.010 & OAOWFC, MuSaSHI & \\
S190923y & NSBH/Ter & 1 & 1.151 & 0.014 & OAOWFC & \\
S190930s & MassGap & 4 & 0.041 & 0.0058 & MITSuME-Akeno & \\
S190930t & NSBH/Ter & 18 & 4.841 & 0.27 & Tripole5, MITSuME-Akeno & \cite{2019GCN.25907....1M}\\
 & & & & & & \cite{2019GCN.25920....1M}\\
 & & & & & & \cite{2019GCN.25941....1K}\\
S191105e & BBH & 3 & 1.179 & 0.00036 & MITSuME-Akeno, MITSuME-Okayama & \\
S191205ah & NSBH & 1 & 1.886 & 0.0 & MITSuME-Akeno & \cite{2019GCN.26381....1M}\\
S191213g & BNS/Ter & 45 & 1.587 & 1.5 & NIC, HONIR & \cite{2019GCN.26477....1T}\\
 & & & & & MITSuME-Akeno, OAOWFC & \\
 & & & & & MITSuME-Okayama & \\
S191216ap & BBH & 1 & 1.449 & 0.00065 & (HSC)$^{4}$, NIC & \cite{2019GCN.26496....1Y}\\
 & & & & & MITSuME-Akeno, OAOWFC & \cite{2019GCN.26509....1O}\\
 & & & & & BAO101$^{5}$ & \\
S200114f & Burst & 42 & 0.659 & --- & NIC, HONIR & \cite{2020GCN.26803....1T}\\
 & & & & & OAOWFC & \\
S200213t & BNS/Ter & 74 & 0.238 & 5.0 & NIC, HONIR & \cite{2020GCN.27066....1K}\\
 & & & & & Tomo-e Gozen, OAOWFC & \\
S200219ac & BBH & 23 & 0.130 & 0.00015 & Tomo-e Gozen & \\
S200224ca & BBH & 108 & 0.558 & 0.069 & (HSC)$^{4}$, Tomo-e Gozen & \cite{2020GCN.27205....1O}\\
S200225q & BBH & 75 & 2.398 & 0.38 & Tomo-e Gozen & \\\hline
\end{tabular}
\begin{flushleft}
$^{1}$ Number of observed galaxies.
$^{2}$ Start time of observations after detection of GW signals.
$^{3}$ Total covered probability of observed galaxies by our follow-up campaigns.
$^{4}$ Data obtained by HSC is not included in the calculation of the coverage.
$^{5}$ BAO101cm: 101~cm telescope at Bisei Astronomical Observatory and CCD optical camera ($I_{\rm C}$ band)
\end{flushleft}
\end{table}

\begin{figure}[ht!]
\centering
\includegraphics[scale=0.8]{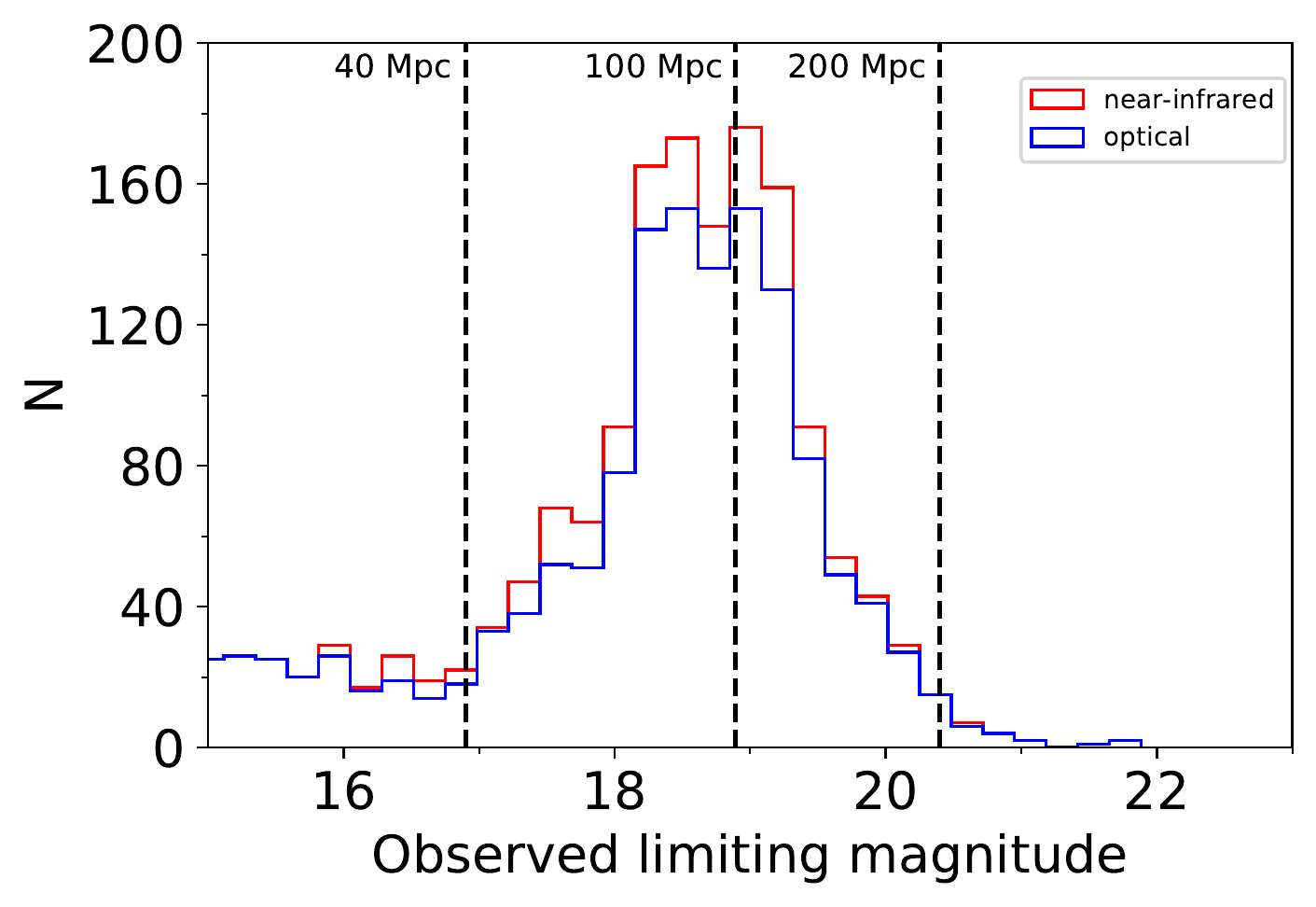}
\caption{Distributions of the obtained limiting magnitudes in the optical ($g$, $r$, $R_{\rm C}$, $i$, $I_{\rm C}$, and Blank bands; blue) and NIR ($J$, $H$, and $K_{\rm S}$ bands; red) in the galaxy-targeted follow-up campaigns during O3. Dashed lines show the peak magnitudes of GW170817 in the $g$ band scaled by distances of 40 Mpc, 100 Mpc, and 200 Mpc, respectively. In the distribution, the limiting magnitudes obtained by FOCAS are not included. \label{fig:lim}}
\end{figure}

The limiting magnitudes of our observations depends on the telescopes, instruments, and the sky condition. 
The deepest limiting magnitudes of our follow-up observations in the optical and NIR bands are 21.2 and 20.6, and median limiting magnitudes are 18.7 and 18.6, respectively. Here, the limiting magnitude of the FOCAS observation, 23.5~mag in $R$-band, is not included in the calculation of median values. The peak magnitudes of GW170817 scaled by distances of 40~Mpc ($g=$16.8), 100~Mpc ($g=$18.8), and 200~Mpc ($g=$20.3) are shown together with the distribution of the obtained limiting magnitudes in Fig.~\ref{fig:lim}. Thus, with our observations, we could potentially detect emission from a GW170817-like BNS source at a distance of $\sim$100~Mpc.

\begin{figure}[ht!]
\centering
\includegraphics[scale=0.6]{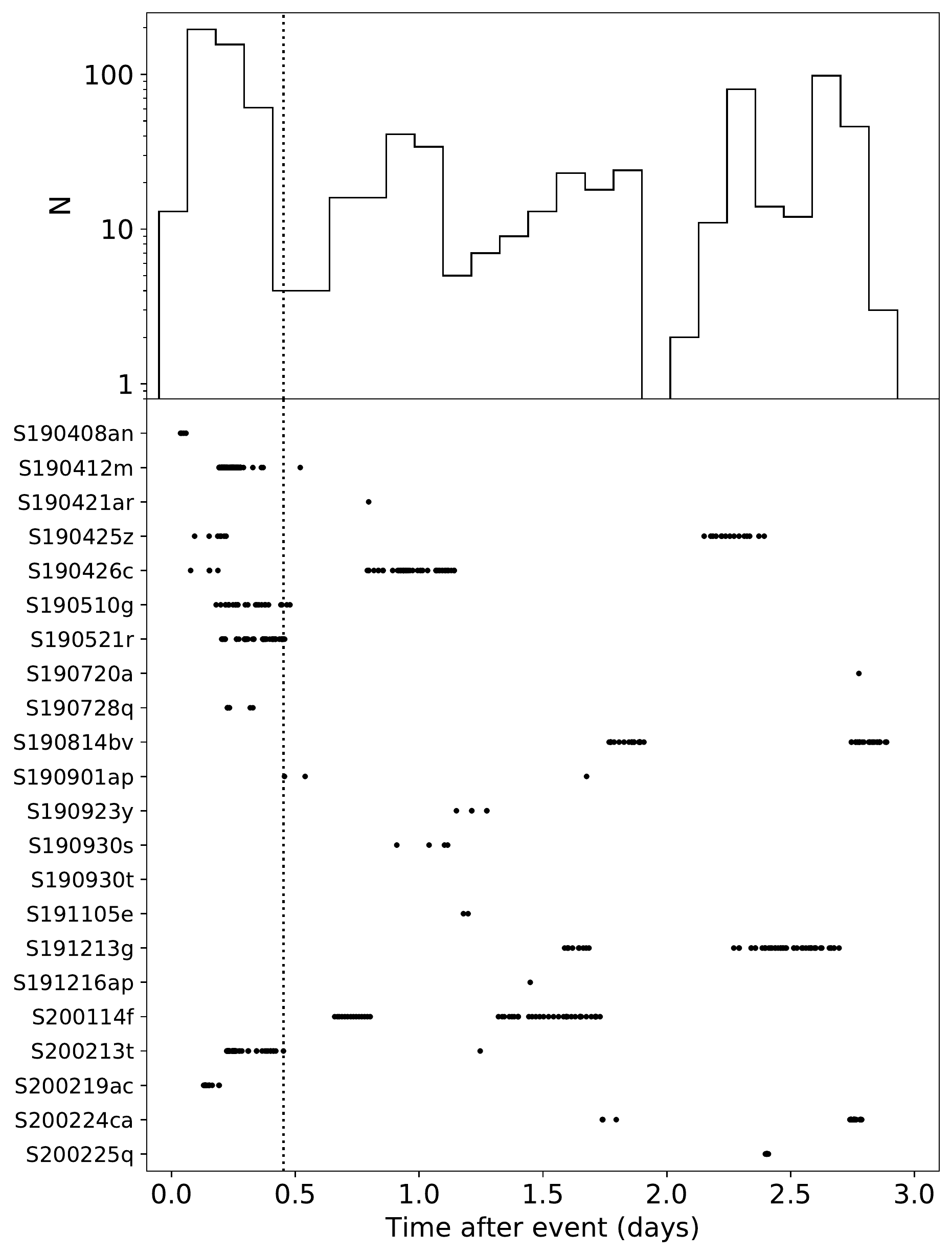}
\caption{Elapsed times of the J-GEM observations being obtained after the GW detection. A histogram of the times of observations is shown in the top panel. The filled circles show the observation time from each campaign in the bottom panel. The dotted line shows the elapsed time of 10.86 hours in which the optical counterpart of GW170817 was first identified as AT~2017gfo \citep{2017ApJ...848L..12A}. \label{fig:hist}}
\end{figure}

Since early detection of the EM counterpart is desired to provide better understanding of the emission mechanisms at this epoch (see Section~4.2), it is worth analyzing the starting time after the GW detection. The starting times of our observations after GW detections are distributed with a range from 0.041~days to 4.841~days, with a median value of 0.558~days. The starting times of the events are shown in Fig.~\ref{fig:hist}. We succeeded in starting follow-up observations within 0.5~days after detection for 10 GW events.

\clearpage

\section{Discussion} \label{sec:discussion}

\subsection{Detectability for BBH events}
We performed galaxy-targeted follow-up for 11 BBH events during O3. One hundred and fifty seven galaxies were observed, and the coverage reached to 0.62~\%. In our follow-up campaigns, we did not find any optical counterpart associated with any BBH source.

The BBH events we followed up are located at distances greater than 400~Mpc, except for S191216ap (Table~\ref{tab2}.) If BBH mergers are possibly associated with GRB-like jets \citep{2016ApJ...826L...6C}, afterglow-like optical emission might be produced, which is expected to exhibit brightnesses of 23--24~mag roughtly one day after the merger at a distance of 400~Mpc \citep{2016PTEP.2016e1E01Y}. The mean limiting magnitude of our observations was 18.58~mag in the optical band. This means that most of our observations were not sensitive enough to detect such emission.

\subsection{Detectability of events containing a NS component}
We performed EM follow-up observations for 10 events with a probably having at least one NS in the coalescing system. For these events, we did not find any EM counterpart. The GW sources are located at distances from 108~Mpc to 438~Mpc. Distances of 8 out of 10 events are greater than 200~Mpc. Although EM follow-up at all wavelengths were also conducted for these GW events by other facilities around the world, no evident EM counterpart was discovered (e.g. see the results for GW190408an, GW190425, S190426c, S190510g, and S190814bv reported by
\cite{2019ApJ...885L..19C,2019ApJ...881L..26L,2019ApJ...881L...7G,2019ApJ...880L...4H,2019ApJ...887L..13D,2019ApJ...884L..55G,2020ApJ...890..131A}).

\begin{figure}[ht!]
\centering
\includegraphics[scale=0.65]{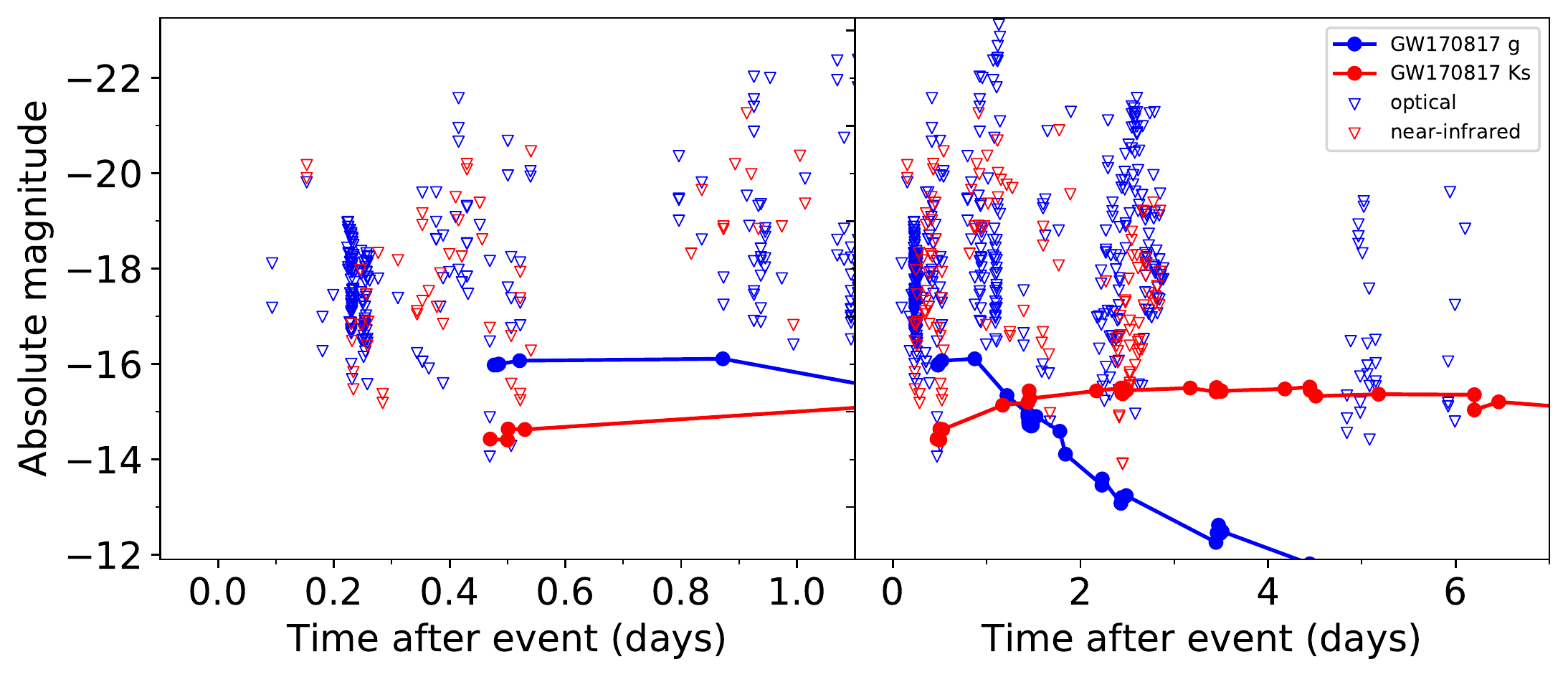}
\caption{Time series of the optical ($g$, $r$, $R_{\rm C}$, $i$, $I_{\rm C}$, and BLANC bands) and NIR ($J$, $H$, and $K_{\rm S}$ bands) limiting magnitudes (open blue and red triangles, respectively) of our follow-up observations of GW events inferred to have at least one NS together with light curves of GW170817 in the $g$ and $K_{\rm S}$ bands in the absolute magnitude system (filled blue and red circles, respectively.) The limiting magnitudes are calculated from distances of targeted galaxies listed in the GLADE catalog. Whole light curves, and those of first day after the GW detection, are respectively shown in the right and left panels. \label{fig:lc_bns}}
\end{figure}

Some of the limiting magnitudes obtained in our follow-up are deeper than the light curve of AT~2017gfo at the luminosity distance estimated for the corresponding GW events (Fig.~\ref{fig:lc_bns}). For example, in the cases of S191213g and S200213t, whose luminosity distances are estimated to be $\sim$200~Mpc, some of the absolute limiting magnitudes, scaled by the distances of the observed galaxies, are deeper than the absolute magnitude of AT~2017gfo (see appendix~\ref{sec:skymap}). Assuming the observed magnitudes of AT~2017gfo-like objects at distances of 40, and 100~Mpc, the expected optical brightness of any EM counterpart could have been detected by our observations (see Fig.~\ref{fig:lim}). Especially, the evolution of the optical brightness of an AT~2017gfo-like EM counterpart at a distance of less than 100~Mpc could have been observed by our observations.

In the follow-up campaigns, the maximum number of observed galaxies per event was 170, and the coverage was up to 9.8~\%. One of the reasons for the low coverage is that the 90\% probability areas of the GWs are larger than 1,000 deg$^2$, except for the case of S190814bv (23 deg$^2$.) Note that the main probability area for S190814bv is located in the southern hemisphere. Additionally, the completeness of the GLADE catalog for S190814bv is lower than 50\% because it is relatively distant (267~Mpc.) As a result, coverage of this event was also considerably reduced.

To detect an EM counterpart, not only deeper limiting magnitudes but also larger numbers of galaxies (or wider coverage) is important. Here, we evaluate the feasibility of detecting an EM counterpart by comparing the observed and expected numbers of galaxies within a GW localization region. The expected number of galaxies within the 3D localization map of a GW source is estimated from a multiplication between the volume, $\Delta V$, and the number density of galaxies. The volume is calculated as $\Delta V=(4/3){\pi}d^{3}(\Delta \Omega /4\pi)$, where $d$ is a distance of the GW source and $\Delta \Omega$ is an area of GW localization region. The number density of galaxies is derived from the Schechter function as $\int_{x_{1/2}}^{\infty} {\phi}^{*}\,x^{a}\,e^{-x} \,dx=2.35\times10^{-3}$~Mpc$^{-3}$, where we assume $x=L_{B}/L_{B}^{*}$, $L_{B}$ is the $B$-band luminosity and $L_{B}^{*}$ is the characteristic luminosity\citep{2016ApJ...820..136G}.

\begin{figure}[ht!]
\centering
\includegraphics[scale=0.7]{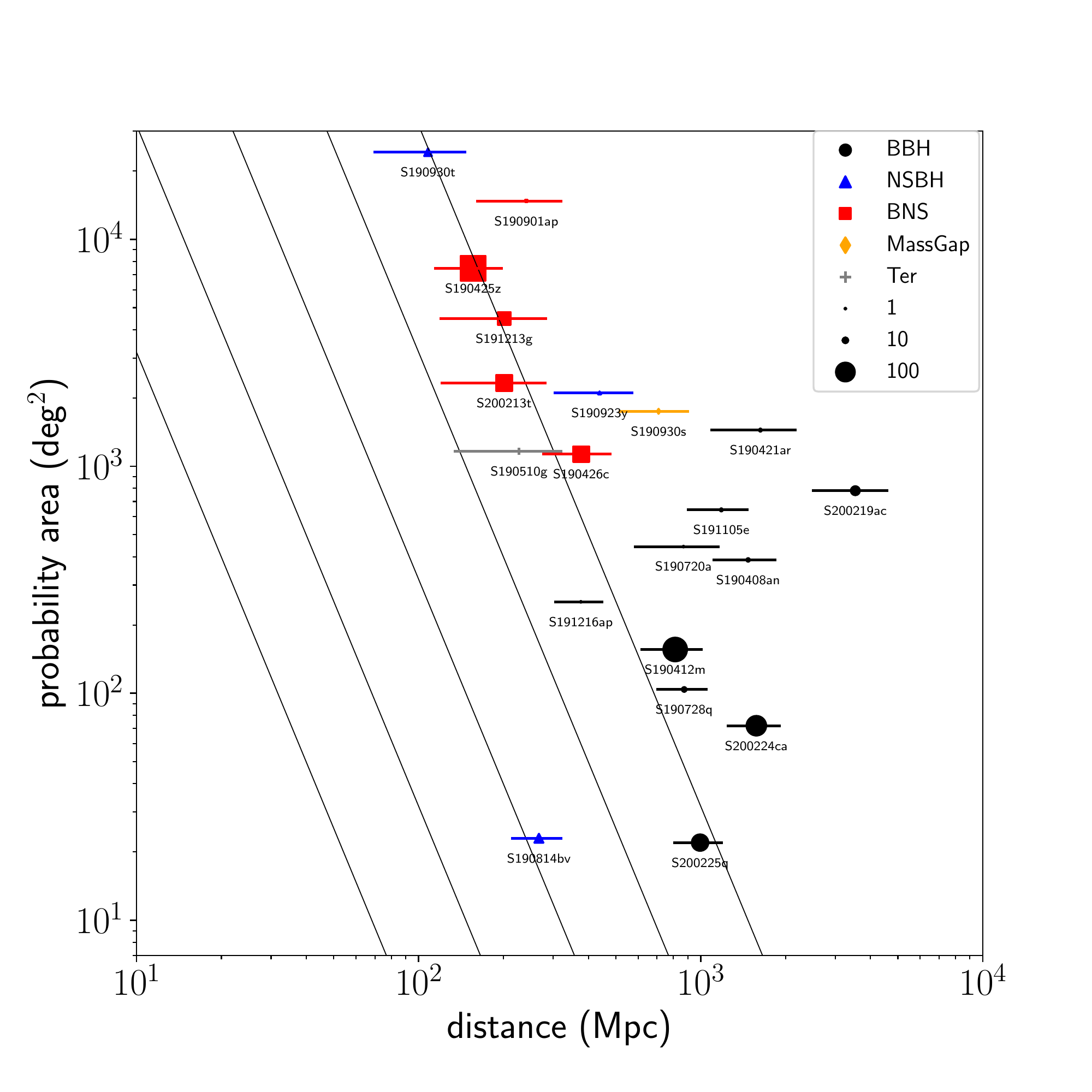}
\caption{Distribution of numbers of observed galaxies for GW events followed up by J-GEM in the distance-probability area plane. Points are shown for BBH (filled circle), NSBH (filled triangle), BNS (filled square), MassGap object (filled diamond), and Terrestrial event (cross). The point sizes represent the numbers of observed galaxies within the 3D localization maps of each GW event. Solid lines show the expected numbers of galaxies within the 3D localization map, from left to right, of 1, 10, 100, 1,000, and 10,000 \label{fig:ngal}}
\end{figure}

Based on the above assumptions, the numbers of galaxies within the 3D localization maps for most GW events are estimated to be over 1,000 as shown in Fig.~\ref{fig:ngal}. Thus, with our follow-up campaigns, observing all the galaxies within the huge 3D localization maps of such GW events is not feasible. In our follow-up campaigns, the maximum number of observed galaxies was 170. If the localization accuracy is about 500~deg$^2$ and a distance to the event is $\sim$100~Mpc, a typical number of galaxies is about 100. Therefore, our galaxy-targeted follow-up system is effective for detecting an EM counterpart for a localization area of 500~deg$^{2}$ ($\sim$20 times larger than that of GW170817) and a distance of 100~Mpc.

To understand the mechanism of the EM emission originating from a merger including at least one NS, EM observations soon after the GW detection is important. During the O3 run, our follow-up observations started within one day after GW detection for 13 out of 23 events. In particular, observations for 10 events started within half a days.

\begin{figure}[ht!]
\centering
\includegraphics[scale=0.8]{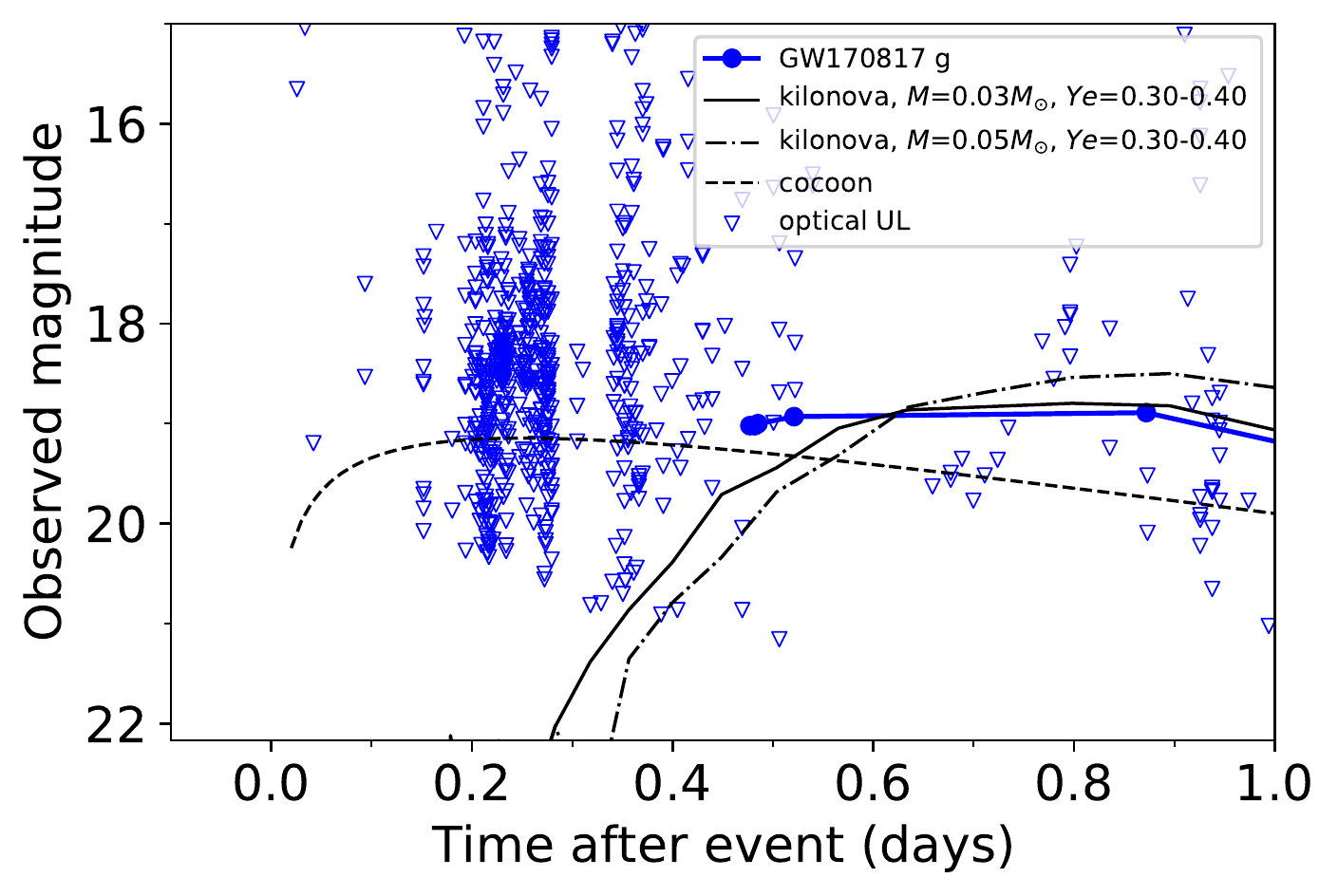}
\caption{Time series of our limiting magnitudes in the optical ($g$, $r$, $R_{\rm C}$, $i$, $I_{\rm C}$, and BLANC) bands (open triangle) taken for all galaxy-targeted observations in O3 run together with the light curve of GW170817 scaled to a distance to 100 Mpc (filled circle), and model curves based on the merger of a binary NS. Curves were calculated based on two kilonova models, where $M_{\rm ej}=$ 0.03 or 0.05 $M_{\odot}$ and $Y_{\rm e}=$0.30--0.40 (solid and dot-dashed lines) and a cocoon model (dashed line.) \label{fig:obs_lc}}
\end{figure}

We compare the results of our observations with possible light curve models of EM emission from NS mergers. The lines in Fig.~\ref{fig:obs_lc} show the expected $g$-band magnitudes at 100 Mpc for a radioactive model (solid) and a cocoon model (dashed). The radioactive model assumes that the heating source is purely the radioactive decay of $r$-process elements in the lanthanide-free ejecta with a mass of $M_{\rm ej}=$0.05 or 0.03 $M_{\odot}$ and electron fraction of $Y_{\rm e}=$0.30--0.40 \citep{2020arXiv200805495B}. The cocoon model incorporates heating via interactions between the relativistic jet and the ejecta \citep{piro18}, where the parameters are chosen to reproduce the early observations of GW170817.

The behavior of each model during the early phase, i.e. $\le$ 1 day after the merger, are different (see also \cite{2018ApJ...855L..23A}). The optical brightness of the radioactive model tends to show a rise to a peak, and subsequent decay due to photon diffusion and decreasing temperature. Conversely, the optical brightness of the cocoon model tends to show more monotonic behavior due to the temporal decrease in both the bolometric luminosity and temperature.

The limiting magnitudes of our observations are shown with triangles in Fig.~\ref{fig:obs_lc}. Some of our observations within 0.5~days are deep enough to detect the emission predicted by these models at 100 Mpc. This means that the early follow-up observations made with the J-GEM facilities can distinguish the emission mechanisms of the NS merger up to about 100 Mpc.

\section{Conclusions} \label{sec:conclusion}
J-GEM is a consortium of astronomical observation facilities that covers at optical, NIR, and radio wavelengths. 
We developed an efficient data management and candidate screening system that perform the following: (1) ranking cataloged galaxies in the 3D localization map, (2) coordinating and collecting information of observations with different facilities, (3) measuring the characteristics of obtained images, and (4) producing cut-out images of new images as well as reference and subtracted images. The system enables efficient search of a possible transient in the GW localization volume and subsequent follow-up observations of potential candidates.

We performed follow-up for 23 GW events (11 BBHs, five BNSs, four NSBHs, one MassGap, one Burst, and one Terrestrials) during O3. In our follow-up campaigns, the shortest starting time between the detection of the GW signal and the onset of observations was as short as 0.041~days, and observations of 10 GW events started within 0.5~days after GW detection. Some of the limiting magnitudes obtained in the follow-up to BNS and NSBH were deeper than the scaled luminosity of GW170817. We show that our follow-up system exhibits the potential to detect a GW170817-like EM counterpart in the GW event with a localization accuracy of $\sim$500~deg$^{2}$ and at a distance of $\sim$100~Mpc.

LVC will schedule the next observing run that will operate together with KAGRA \cite{2020JPhCS1342a2014A}. The expected localization accuracy will be narrower to 110---180 deg$^2$ at 115~Mpc in the case of BNS thanks to the increased number of GW detectors \citep{2018LRR....21....3A}. Out studies demonstrate that J-GEM facilities will contribute to follow up GW events detected by GW detectors during the next run.

\section*{Acknowledgments}

This work was supported by MEXT KAKENHI Grant Numbers JP17H06363, and JP24103003, JSPS KAKENHI Grant Numbers JP15H02069, JP16H02158, JP17H06362, JP18H03720, JP19H00694, JP19K14761, JP20H00158, and JP26800103, the U.S. Department of Energy under contract number DE-AC02-76-SF00515, Optical and Near-Infrared Astronomy Inter-University Cooperation Program, and the joint research program of the Institute for Cosmic Ray Research (ICRR).

Based in part on data collected at Subaru Telescope, which is operated by the National Astronomical Observatory of Japan.

The Pan-STARRS1 Surveys (PS1) and the PS1 public science archive have been made possible through contributions by the Institute for Astronomy, the University of Hawaii, the Pan-STARRS Project Office, the Max-Planck Society and its participating institutes, the Max Planck Institute for Astronomy, Heidelberg and the Max Planck Institute for Extraterrestrial Physics, Garching, The Johns Hopkins University, Durham University, the University of Edinburgh, the Queen's University Belfast, the Harvard-Smithsonian Center for Astrophysics, the Las Cumbres Observatory Global Telescope Network Incorporated, the National Central University of Taiwan, the Space Telescope Science Institute, the National Aeronautics and Space Administration under Grant No. NNX08AR22G issued through the Planetary Science Division of the NASA Science Mission Directorate, the National Science Foundation Grant No. AST-1238877, the University of Maryland, Eotvos Lorand University (ELTE), the Los Alamos National Laboratory, and the Gordon and Betty Moore Foundation.

This publication makes use of data products from the Two Micron All Sky Survey, which is a joint project of the University of Massachusetts and the Infrared Processing and Analysis Center/California Institute of Technology, funded by the National Aeronautics and Space Administration and the National Science Foundation.

\bibliographystyle{ptephy}
\bibliography{sample63}{}

\appendix

\section{Completeness of the GLADE catalog} \label{sec:completeness}

In the follow-up of a GW event, a galaxy catalog is used to select candidate host galaxies. The GW source should be located on a galaxy listed in a catalog, if the catalog is complete. However, the actual catalog is incomplete, where the fraction of galaxies in the catalog to the total number of galaxies is expressed as the completeness. Probability of galaxies located in a 3D localization map of a GW event is calculated using the completeness of the catalog. The total coverage of a galaxy-targeted survey performed by J-GEM can be estimated by summing the probabilities of the observed galaxies.

\begin{figure}[ht!]
\centering
\includegraphics[scale=0.5]{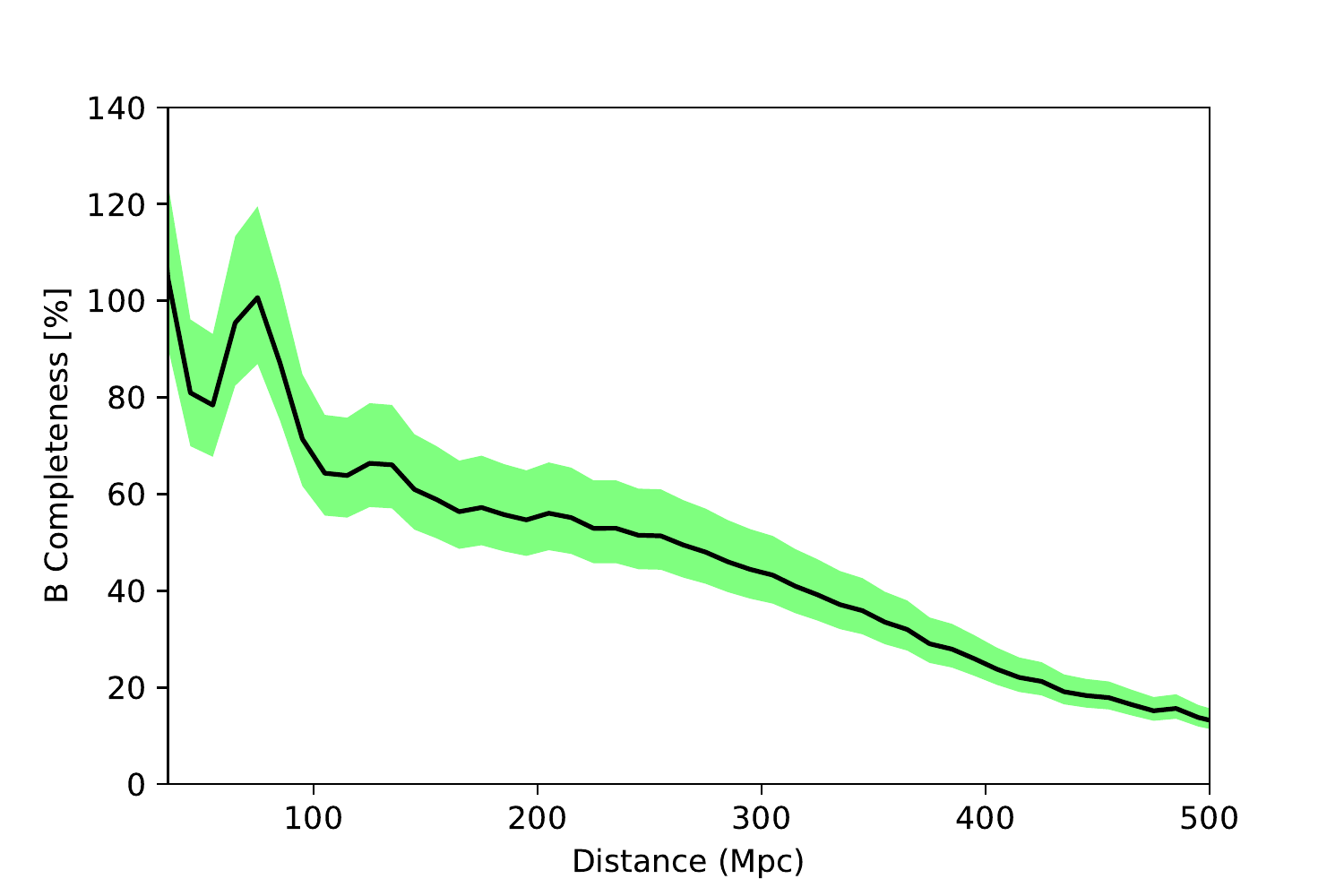}
\caption{$B$-band luminosity-weighted completeness of the GLADE catalog as a function of the luminosity distance. The distances are shown from 35~Mpc because the completeness values exceed 100 percent at smaller distances due to the local overdensity of galaxies around the Milky Way. Note that similar results were found by \cite{2018MNRAS.479.2374D}.
\label{fig:completeness}}
\end{figure}

The completeness of the GLADE catalog used in O3 depends on the known distances to the galaxies in the catalog. We calculated the $B$-band luminosity-weighted completeness at each distance ($d_{L,i}$). The $B$-band luminosity of galaxies at each distance were summed to calculate the total $B$-band luminosity $L_{B}(V_{i})$. Then, the averaged $B$-band luminosity $L_{B}(V_{i})^{\ast}$ was calculated as $j_{B}\times\Delta V_{i}$, where the total $B$-band luminosity density $j_{B}$ is 
$(1.9{\pm}0.3)\times10^{8}\;h\;L_{B,\odot}\;Mpc^{-3}$, adopting the method of \cite{2016ApJ...820..136G}, and the volume is 
$\Delta V_{i}=V_{i+1}-V_{i}=(4/3)\pi(d_{L,i+1}^{3}-d_{L,i}^{3})$. Hence, the completeness was estimated as
$C(d_{L,i})=L_{B}({\Delta}V_{i}) / L_{B}^{\ast}({\Delta}V_{i})$. 
Using these calculations, we determined the distribution of completeness as a function of galaxy distance in the GLADE catalog, as shown in Fig.~\ref{fig:completeness}.

\section{Skymap and time series of limiting magnitudes for each event} \label{sec:skymap}

\begin{figure}[htbp]
  \begin{minipage}[b]{0.24\linewidth}
    \centering
    \includegraphics[keepaspectratio, scale=0.24]{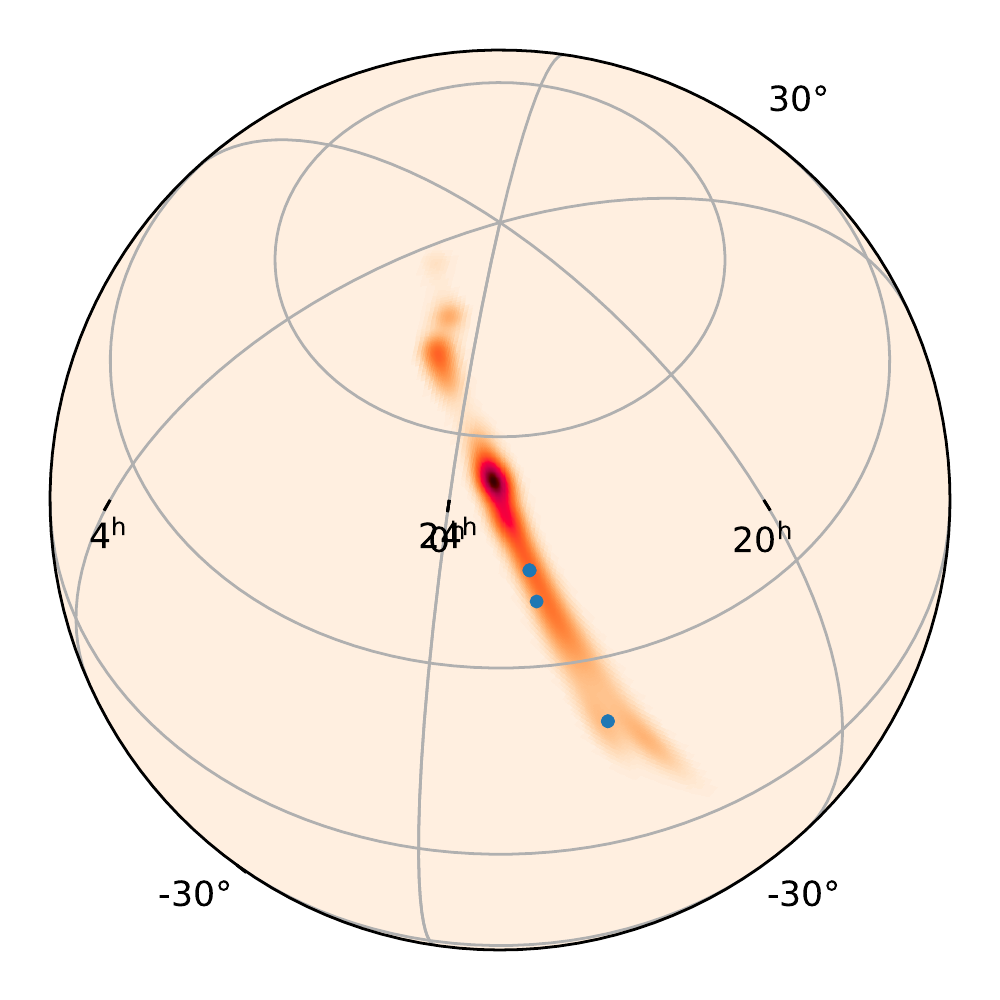}
    \subcaption{S190408an Skymap}
  \end{minipage}
  \begin{minipage}[b]{0.24\linewidth}
    \centering
    \includegraphics[keepaspectratio, scale=0.24]{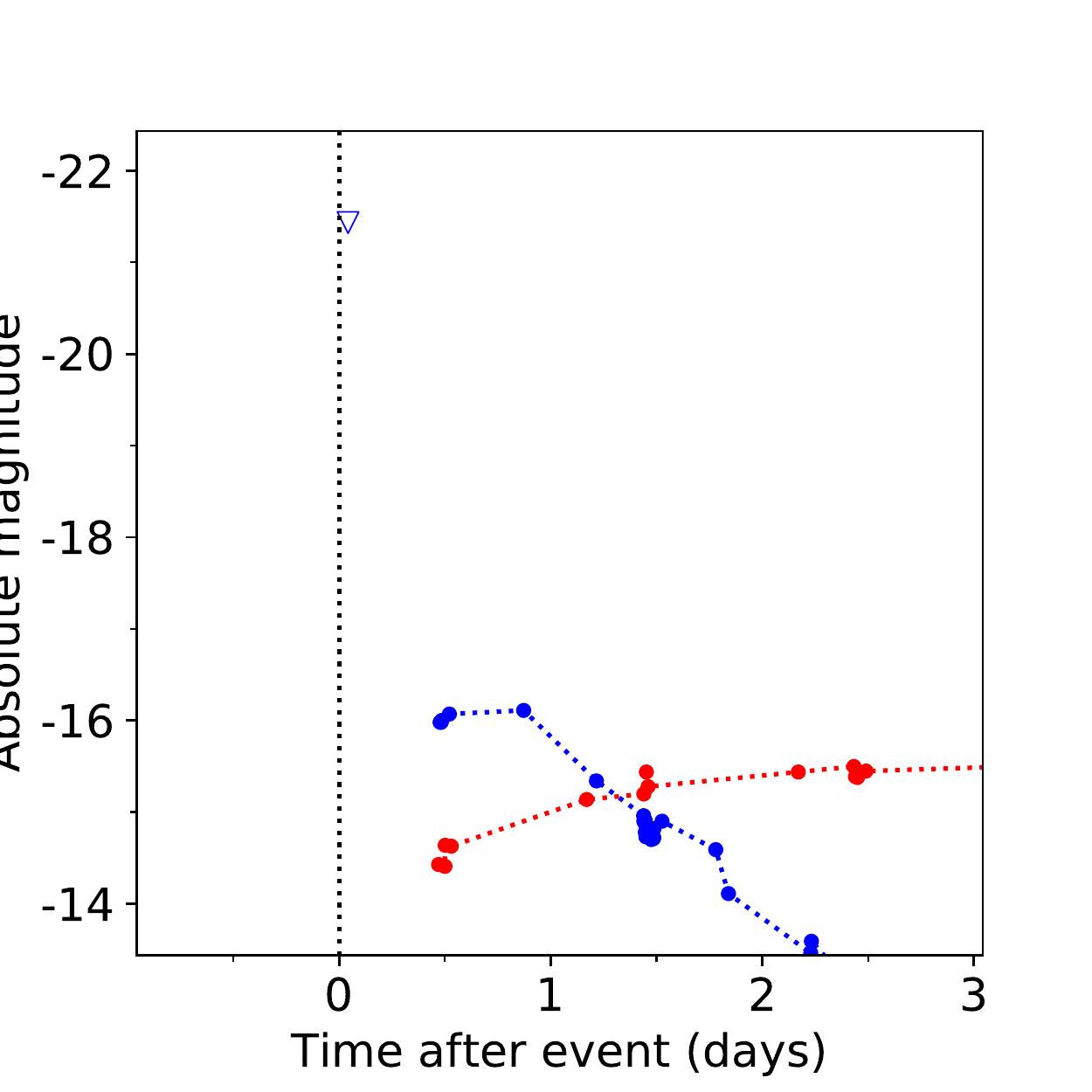}
    \subcaption{S190408an light curve}
  \end{minipage}
  \begin{minipage}[b]{0.24\linewidth}
    \centering
    \includegraphics[keepaspectratio, scale=0.24]{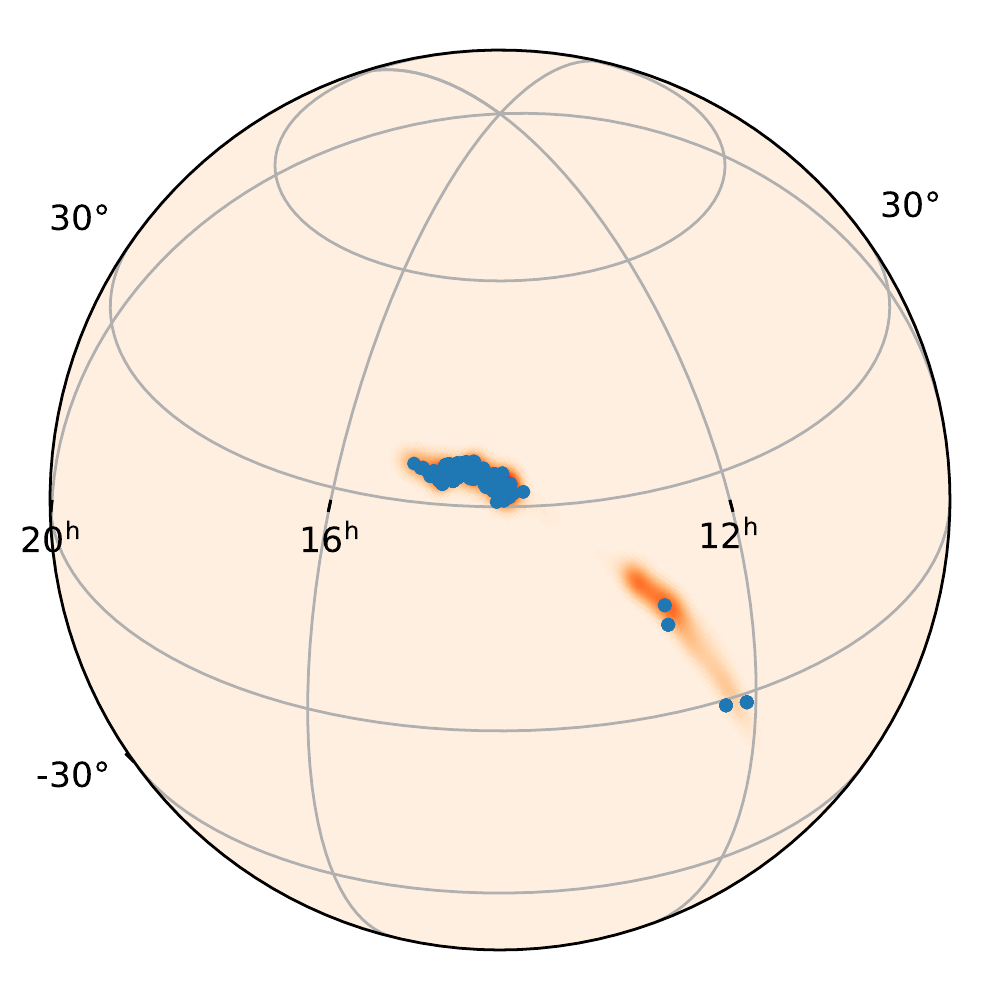}
    \subcaption{GW190412 Skymap}
  \end{minipage}
  \begin{minipage}[b]{0.24\linewidth}
    \centering
    \includegraphics[keepaspectratio, scale=0.24]{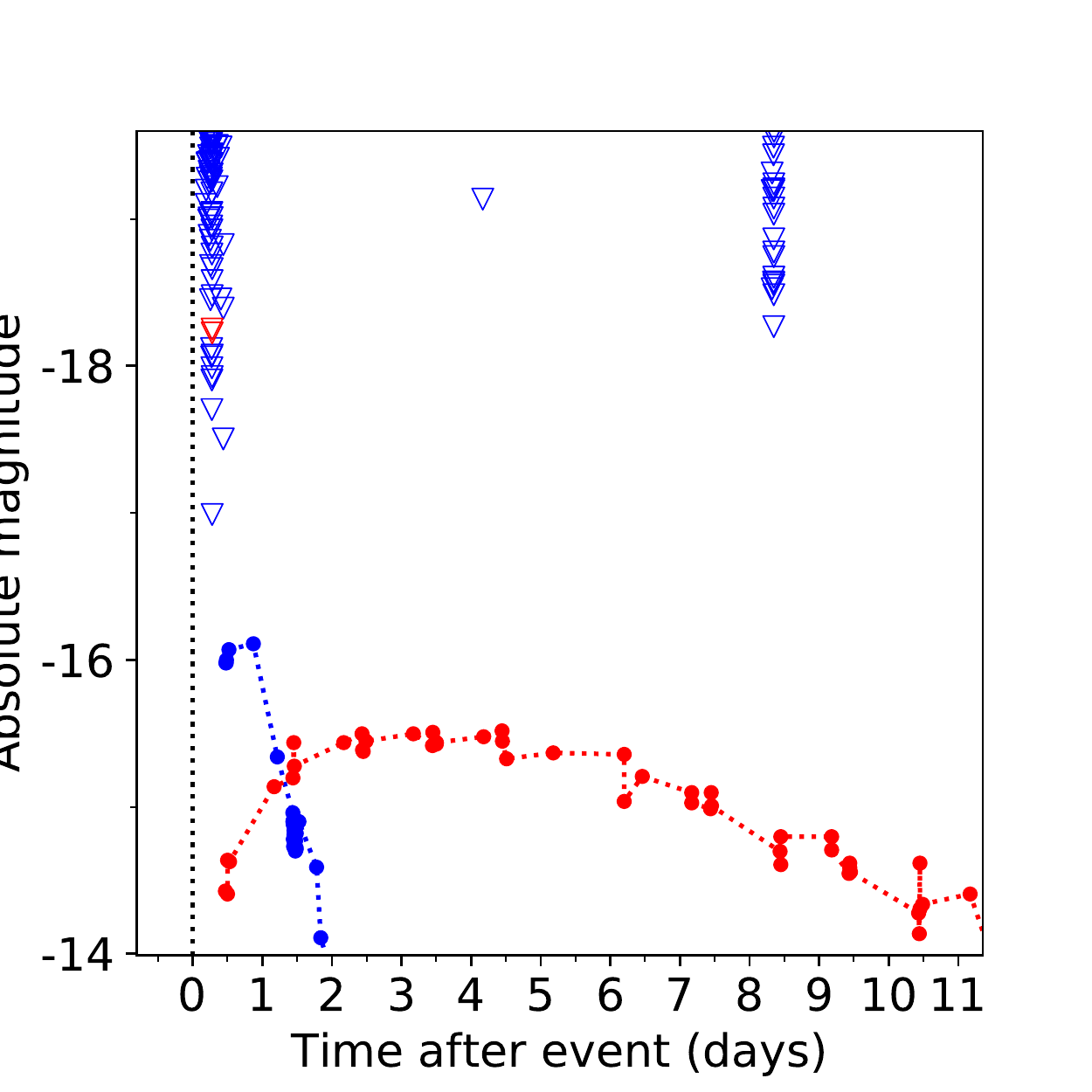}
    \subcaption{GW190412 light curve}
  \end{minipage}\\
  \begin{minipage}[b]{0.24\linewidth}
    \centering
    \includegraphics[keepaspectratio, scale=0.24]{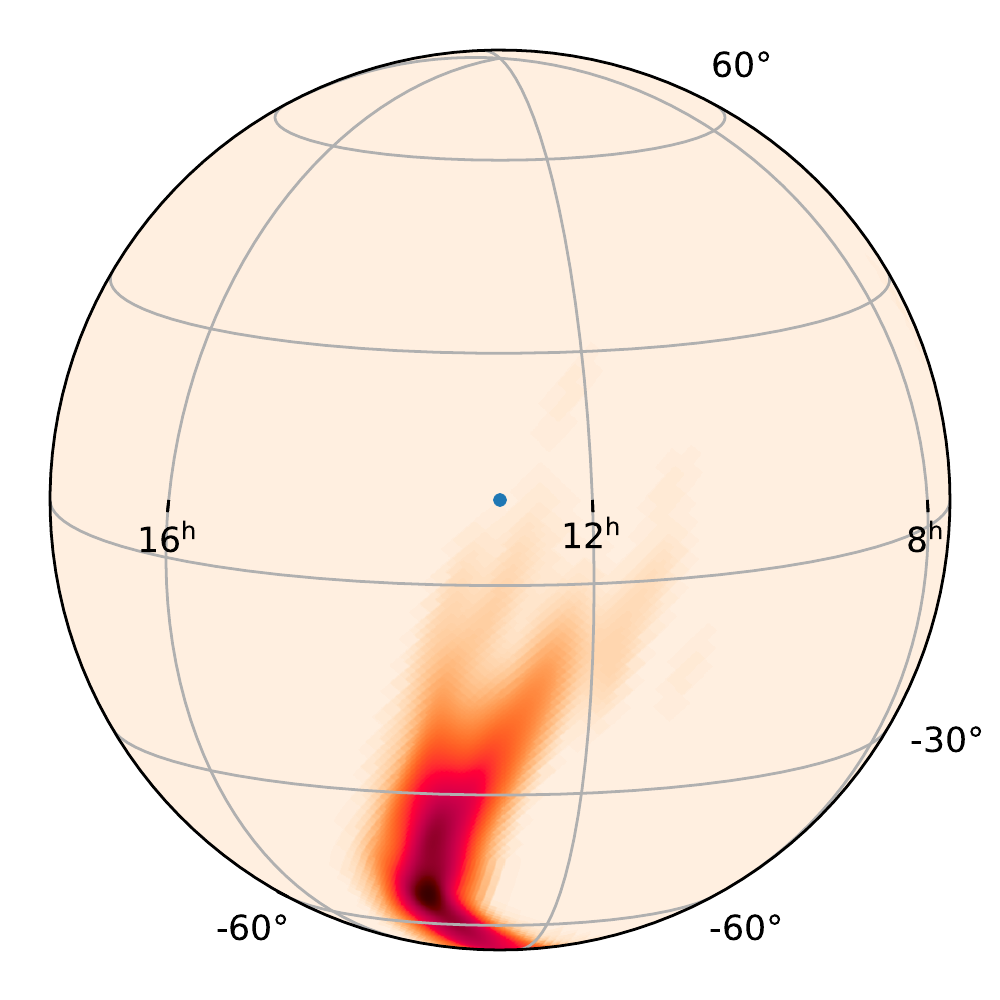}
    \subcaption{S190421ar Skymap}
  \end{minipage}
  \begin{minipage}[b]{0.24\linewidth}
    \centering
    \includegraphics[keepaspectratio, scale=0.24]{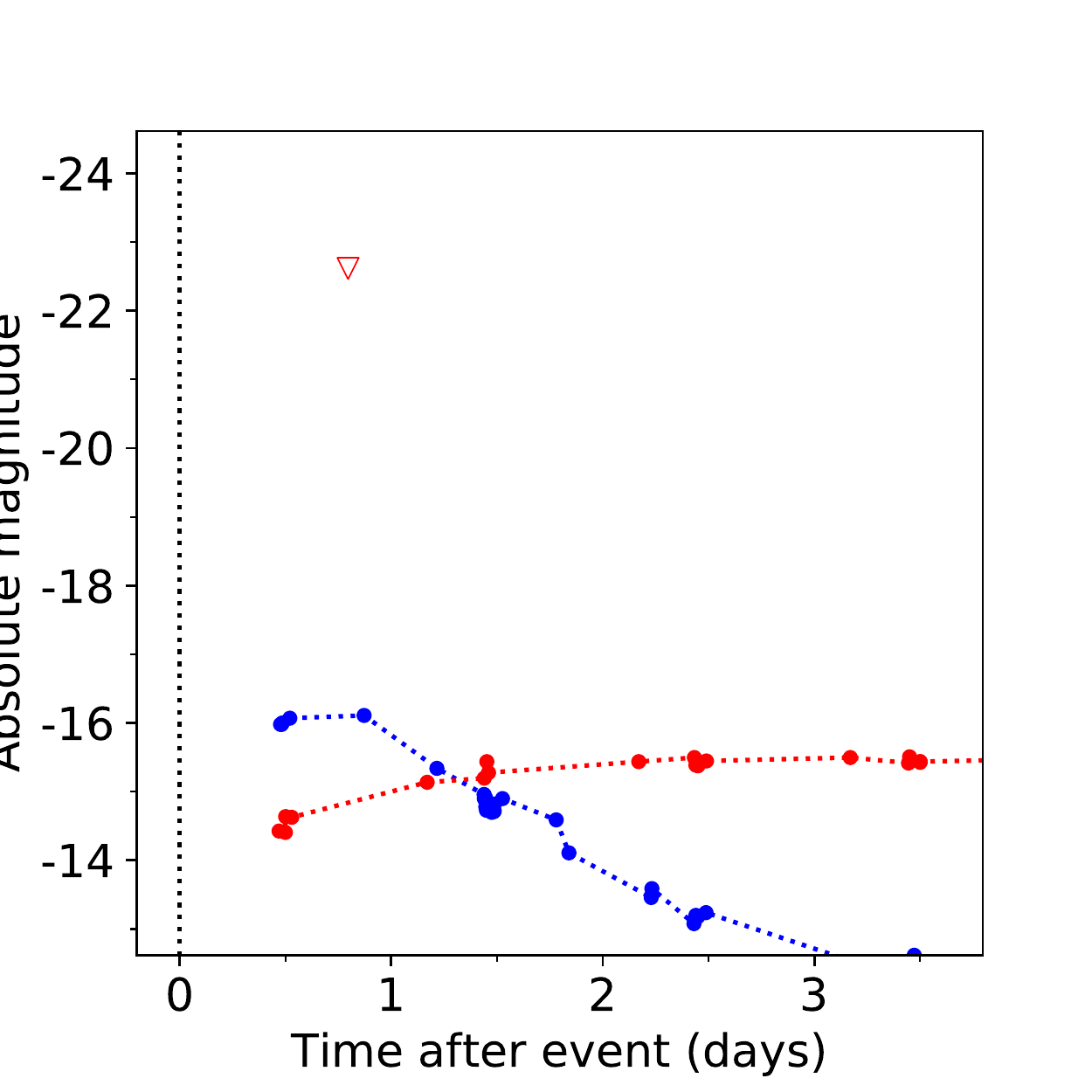}
    \subcaption{S190421ar light curve}
  \end{minipage}
  \begin{minipage}[b]{0.24\linewidth}
    \centering
    \includegraphics[keepaspectratio, scale=0.24]{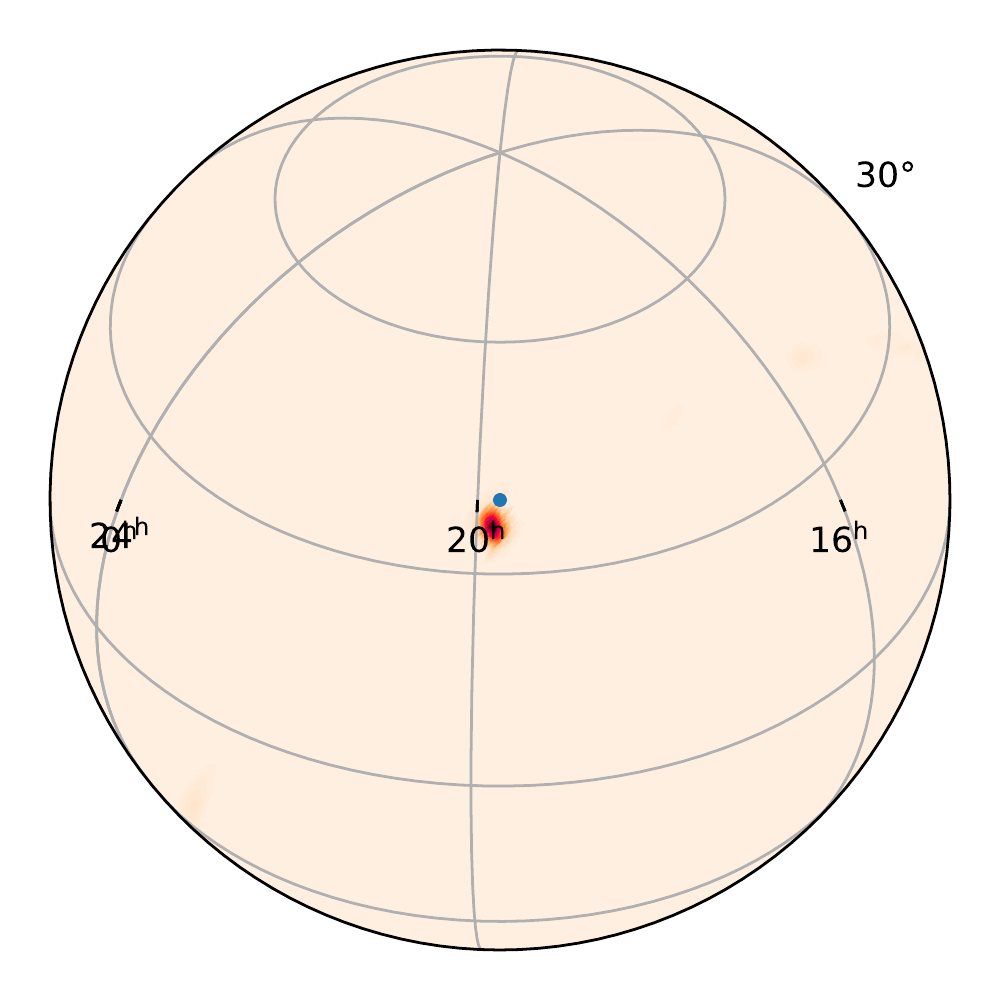}
    \subcaption{S190720a Skymap}
  \end{minipage}
  \begin{minipage}[b]{0.24\linewidth}
    \centering
    \includegraphics[keepaspectratio, scale=0.24]{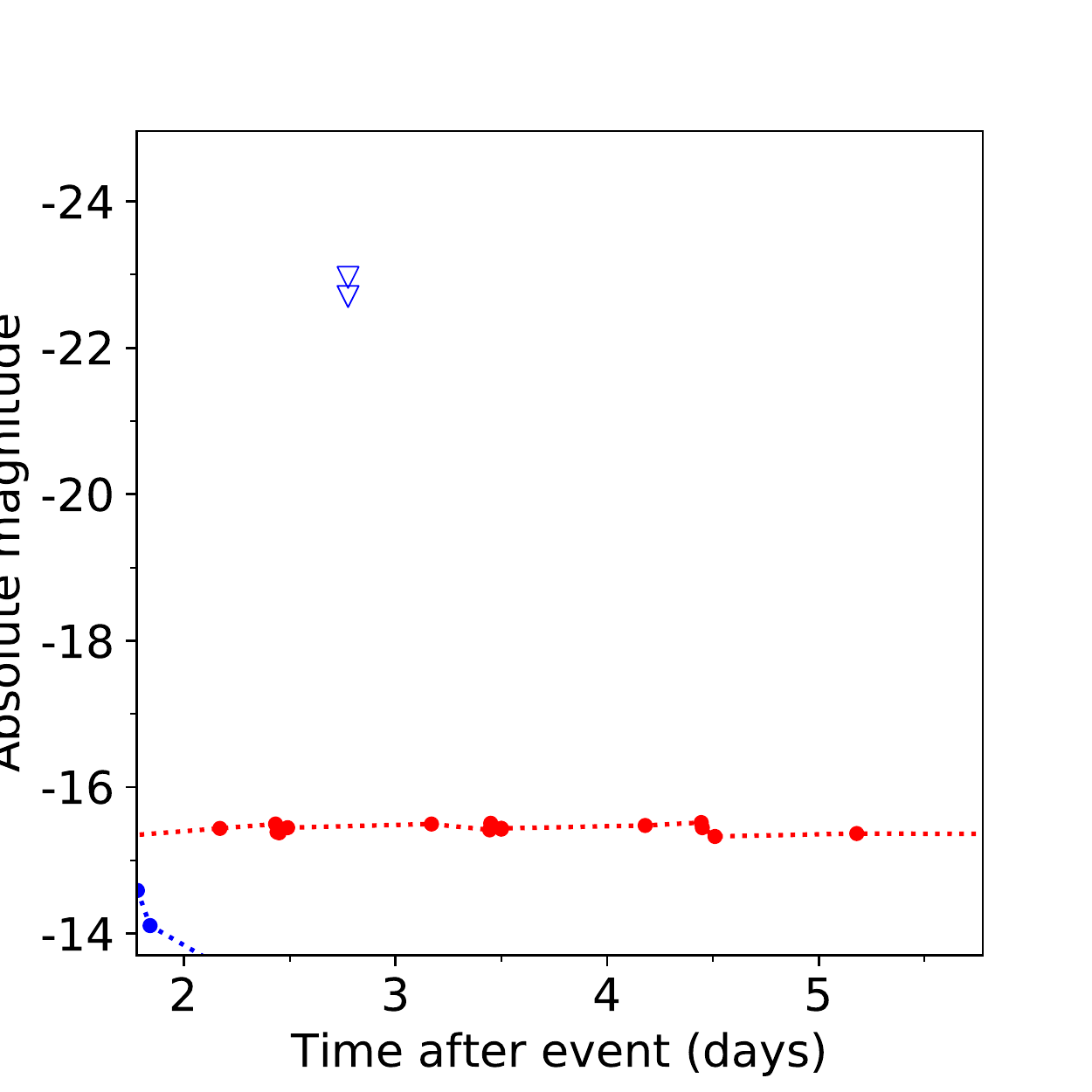}
    \subcaption{S190720a light curve}
  \end{minipage}\\
  \begin{minipage}[b]{0.24\linewidth}
    \centering
    \includegraphics[keepaspectratio, scale=0.24]{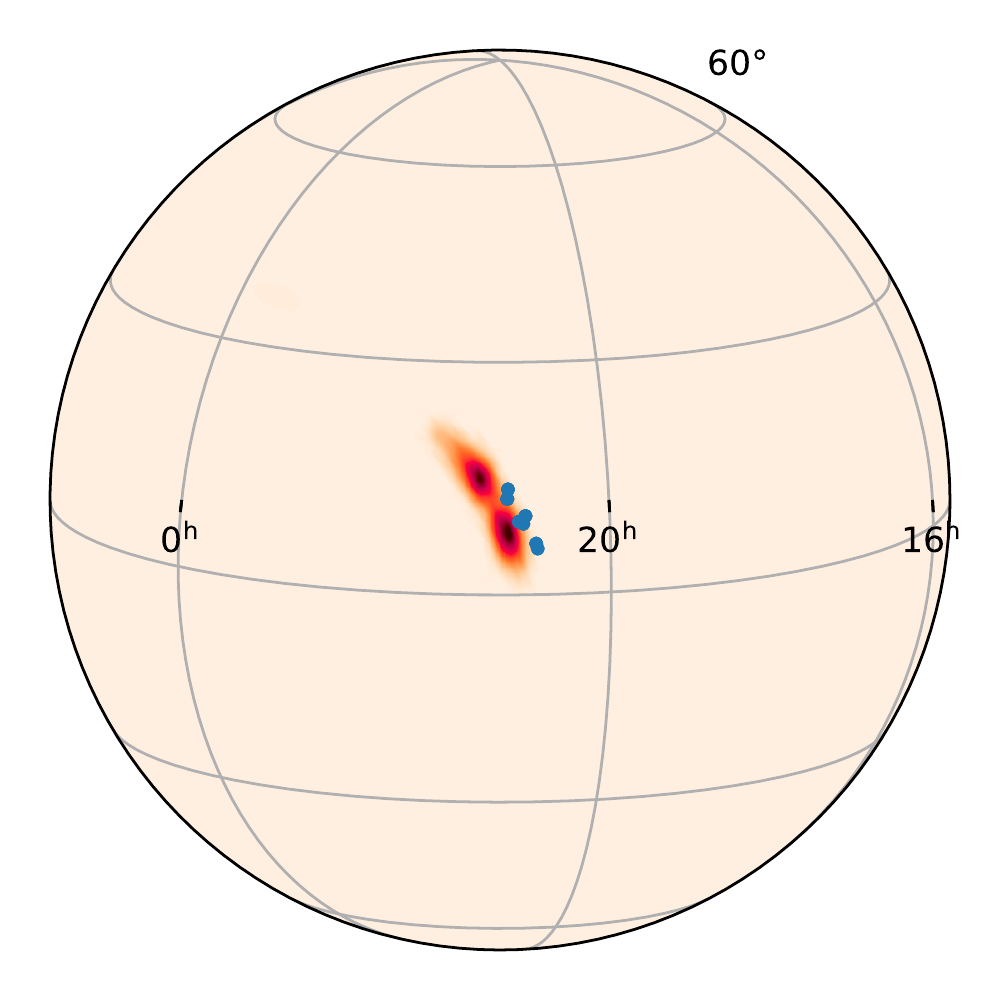}
    \subcaption{S190728q Skymap}
  \end{minipage}
  \begin{minipage}[b]{0.24\linewidth}
    \centering
    \includegraphics[keepaspectratio, scale=0.24]{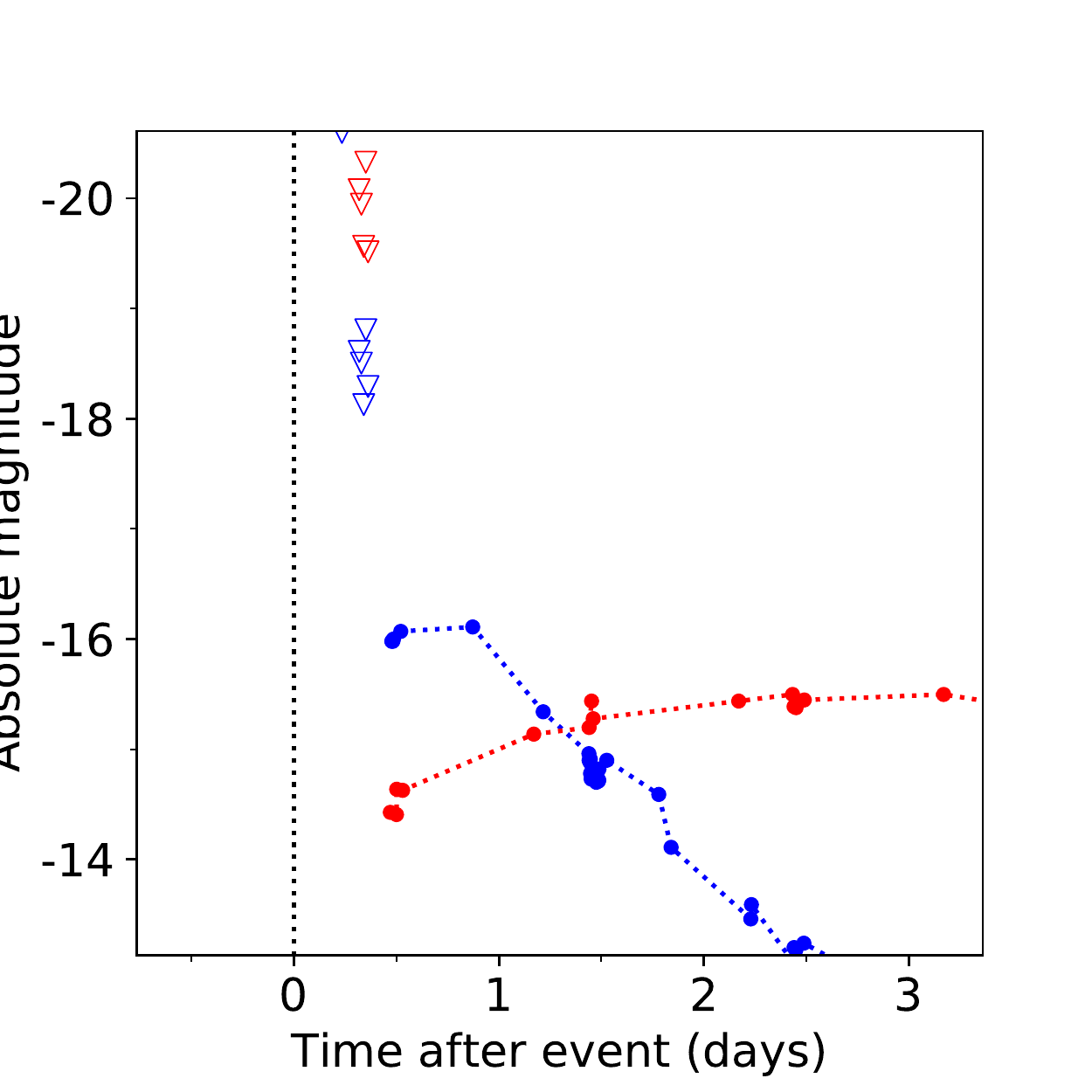}
    \subcaption{S190728q light curve}
  \end{minipage}
  \begin{minipage}[b]{0.24\linewidth}
    \centering
    \includegraphics[keepaspectratio, scale=0.24]{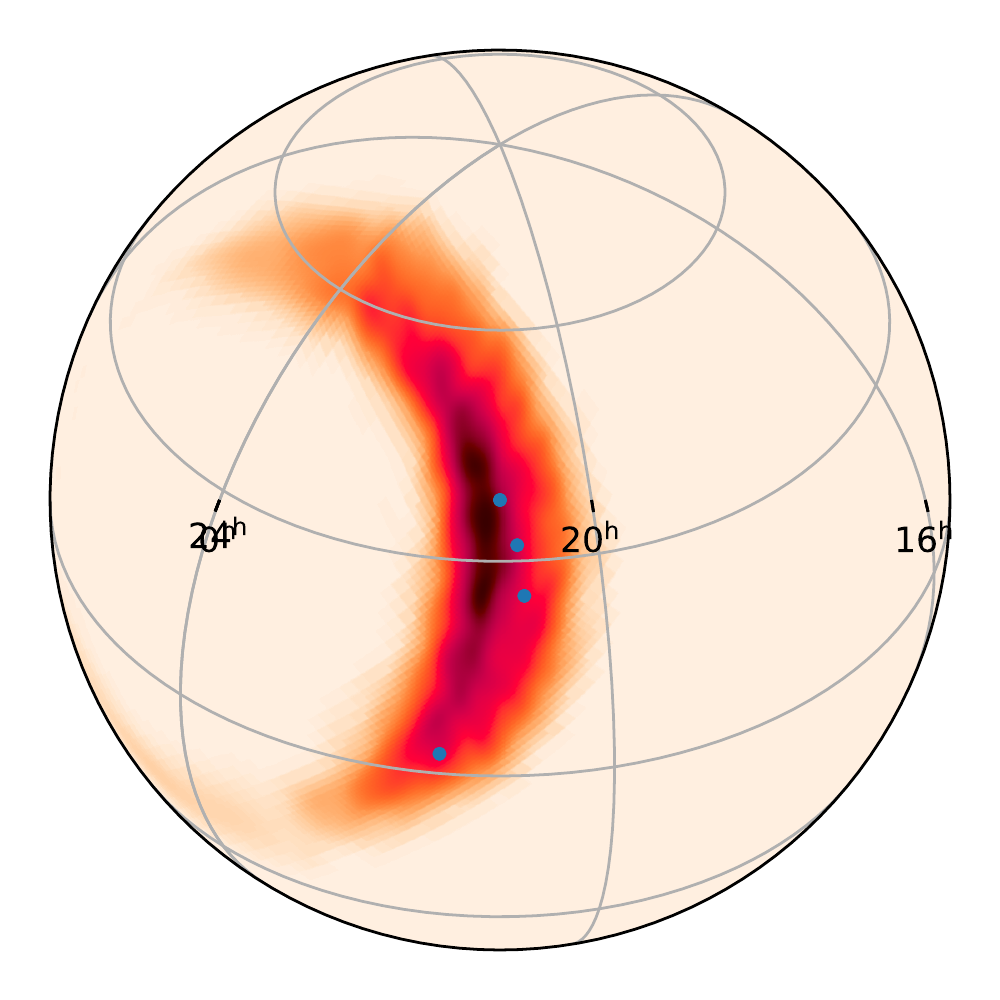}
    \subcaption{S190930s Skymap}
  \end{minipage}
  \begin{minipage}[b]{0.24\linewidth}
    \centering
    \includegraphics[keepaspectratio, scale=0.24]{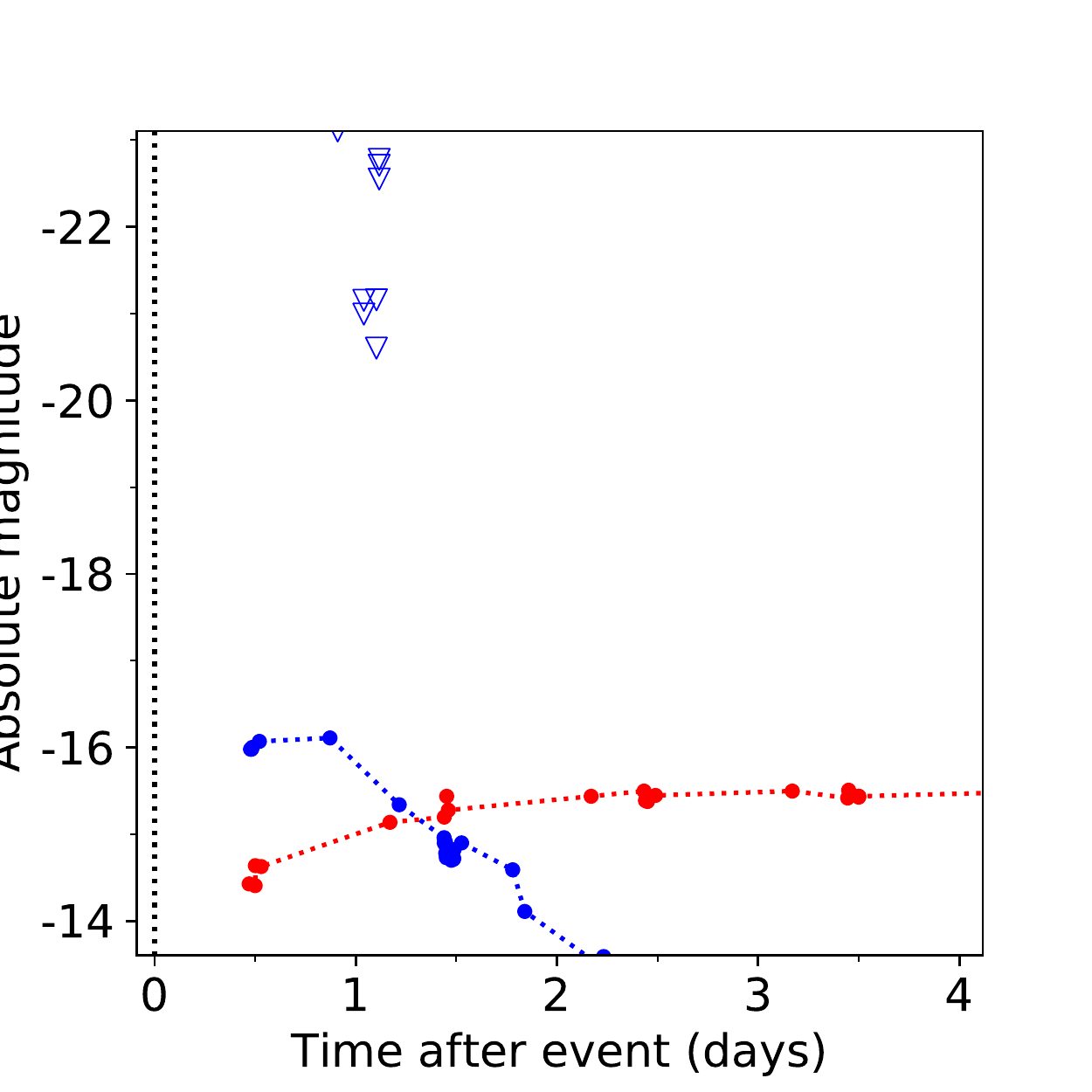}
    \subcaption{S190930s light curve}
  \end{minipage}\\
  \begin{minipage}[b]{0.24\linewidth}
    \centering
    \includegraphics[keepaspectratio, scale=0.24]{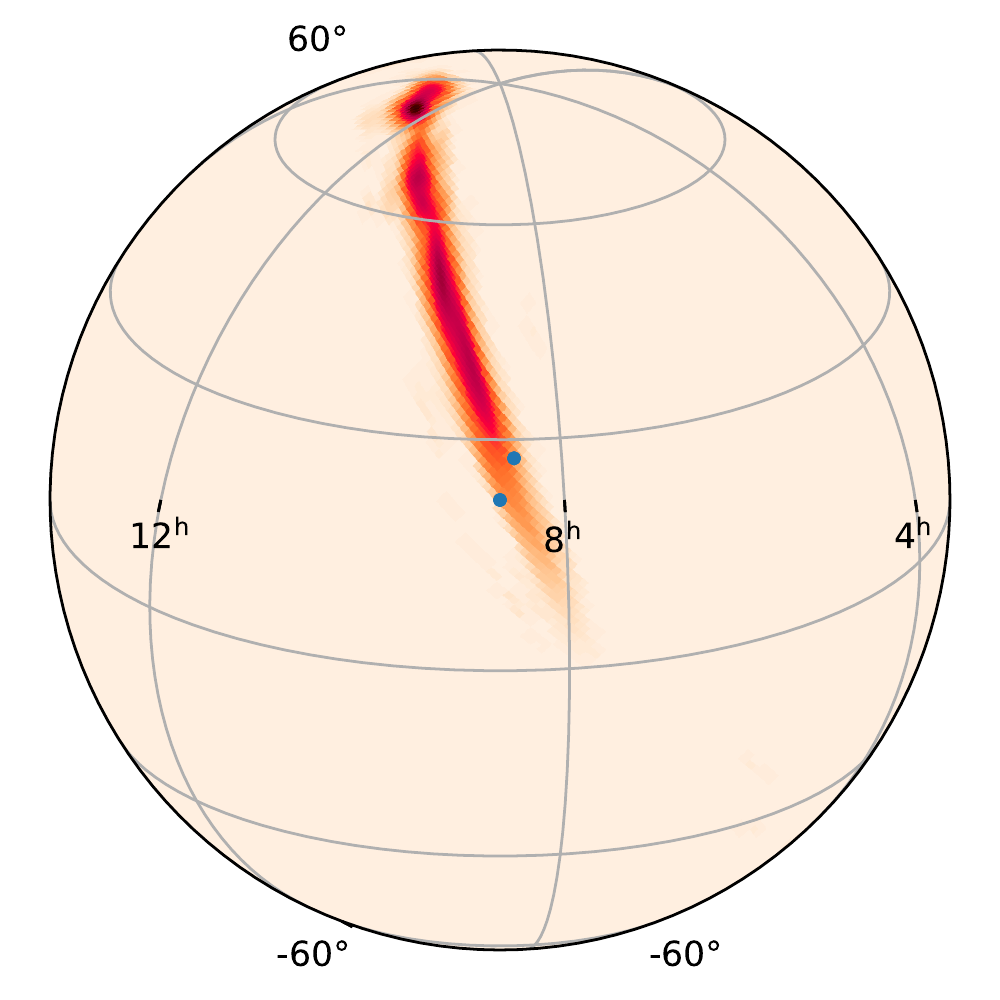}
    \subcaption{S191105e Skymap}
  \end{minipage}
  \begin{minipage}[b]{0.24\linewidth}
    \centering
    \includegraphics[keepaspectratio, scale=0.24]{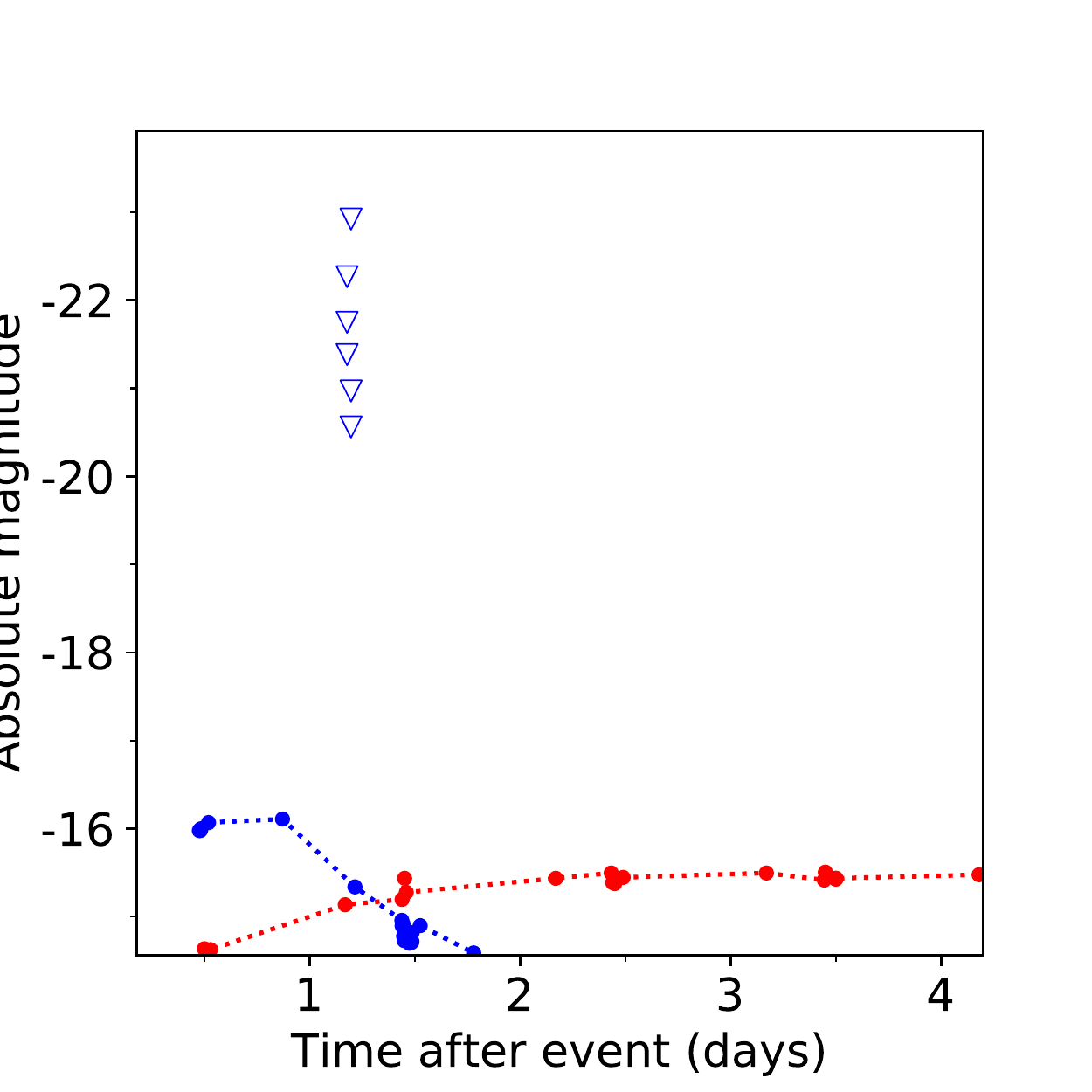}
    \subcaption{S191105e light curve}
  \end{minipage}
  \begin{minipage}[b]{0.24\linewidth}
    \centering
    \includegraphics[keepaspectratio, scale=0.24]{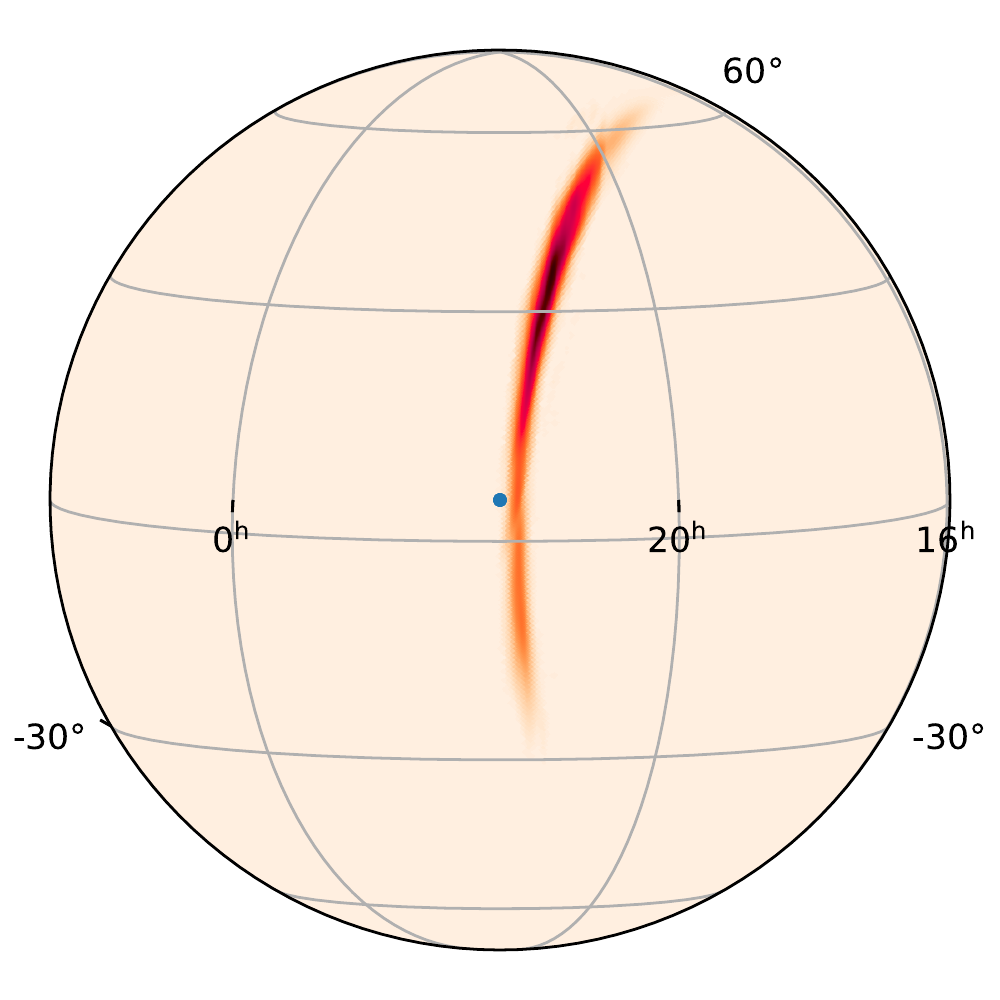}
    \subcaption{S191216ap Skymap}
  \end{minipage}
  \begin{minipage}[b]{0.24\linewidth}
    \centering
    \includegraphics[keepaspectratio, scale=0.24]{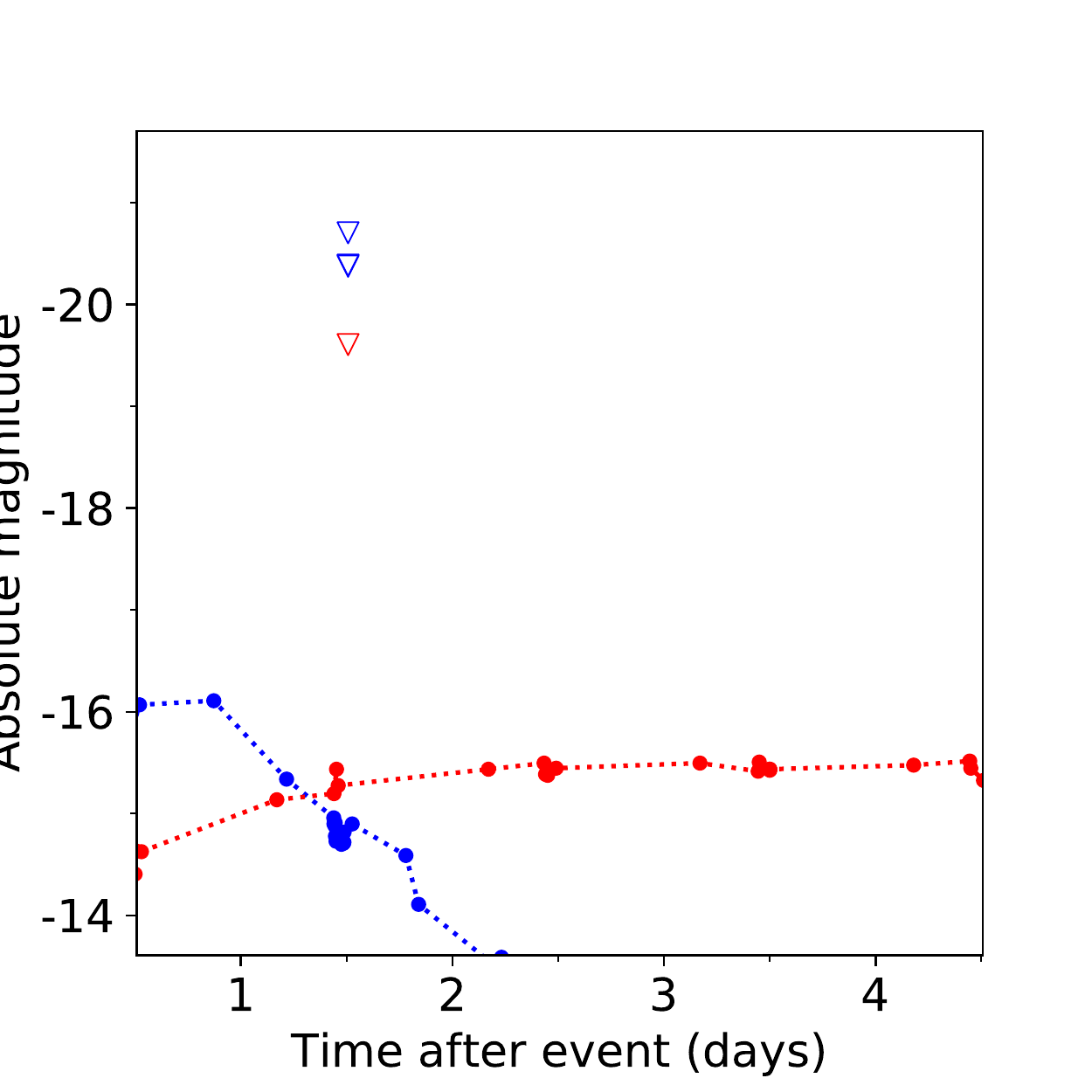}
    \subcaption{S191216ap light curve}
  \end{minipage}\\
  \begin{minipage}[b]{0.24\linewidth}
    \centering
    \includegraphics[keepaspectratio, scale=0.24]{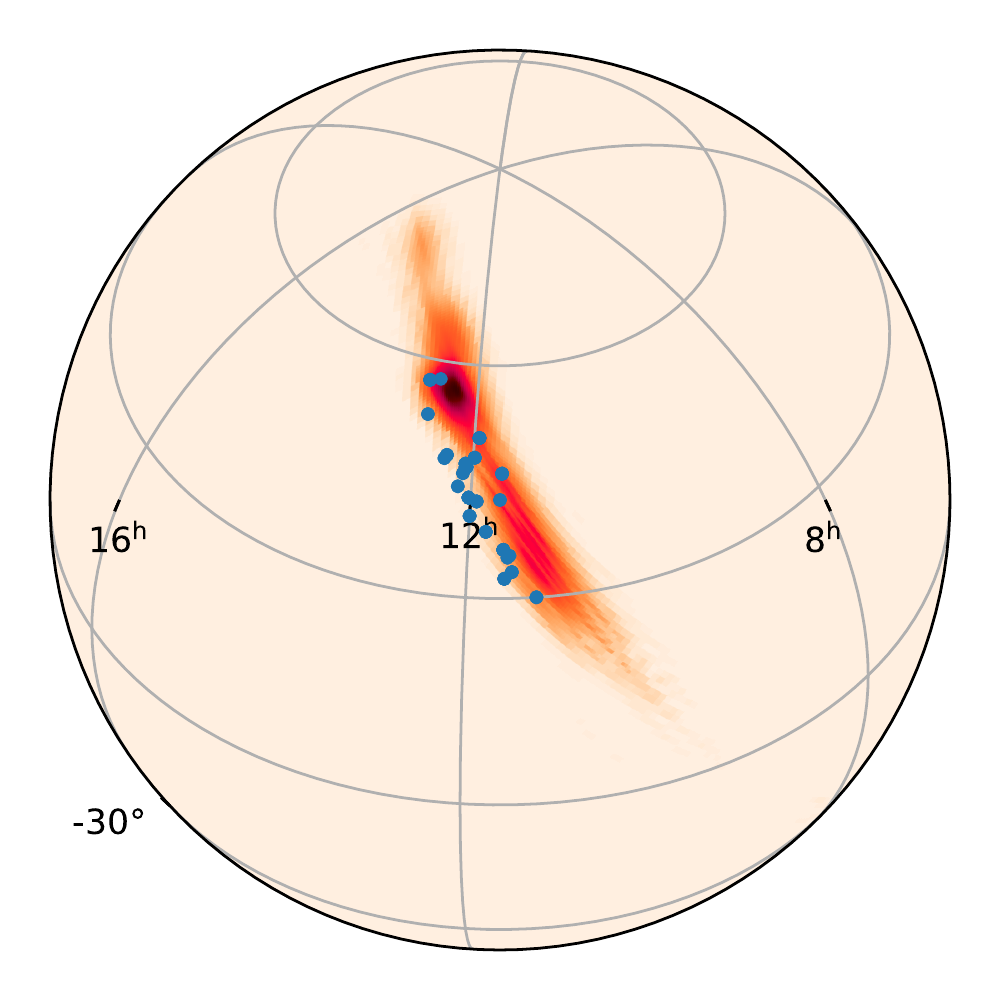}
    \subcaption{S200219ac Skymap}
  \end{minipage}
  \begin{minipage}[b]{0.24\linewidth}
    \centering
    \includegraphics[keepaspectratio, scale=0.24]{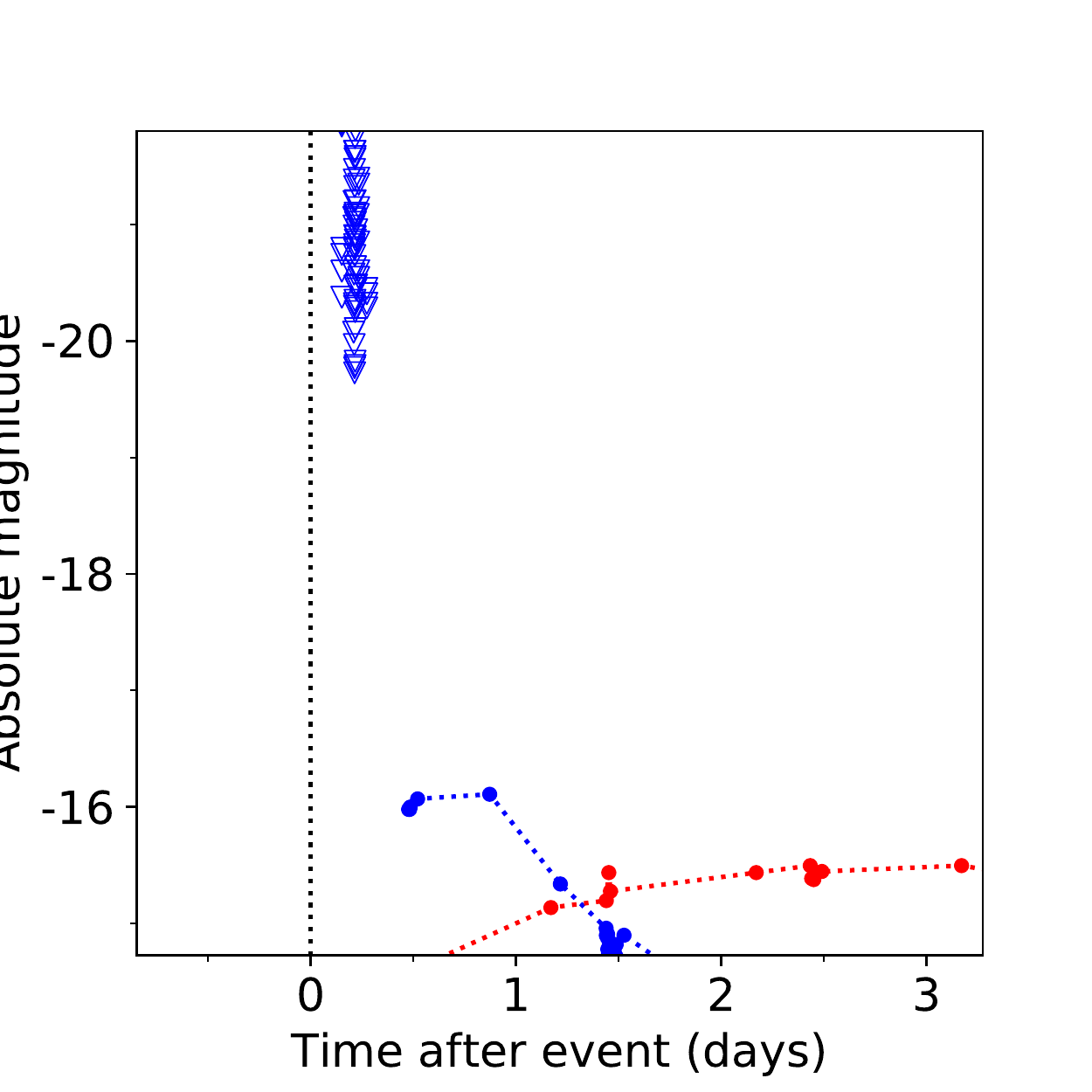}
    \subcaption{S200219ac light curve}
  \end{minipage}
  \begin{minipage}[b]{0.24\linewidth}
    \centering
    \includegraphics[keepaspectratio, scale=0.24]{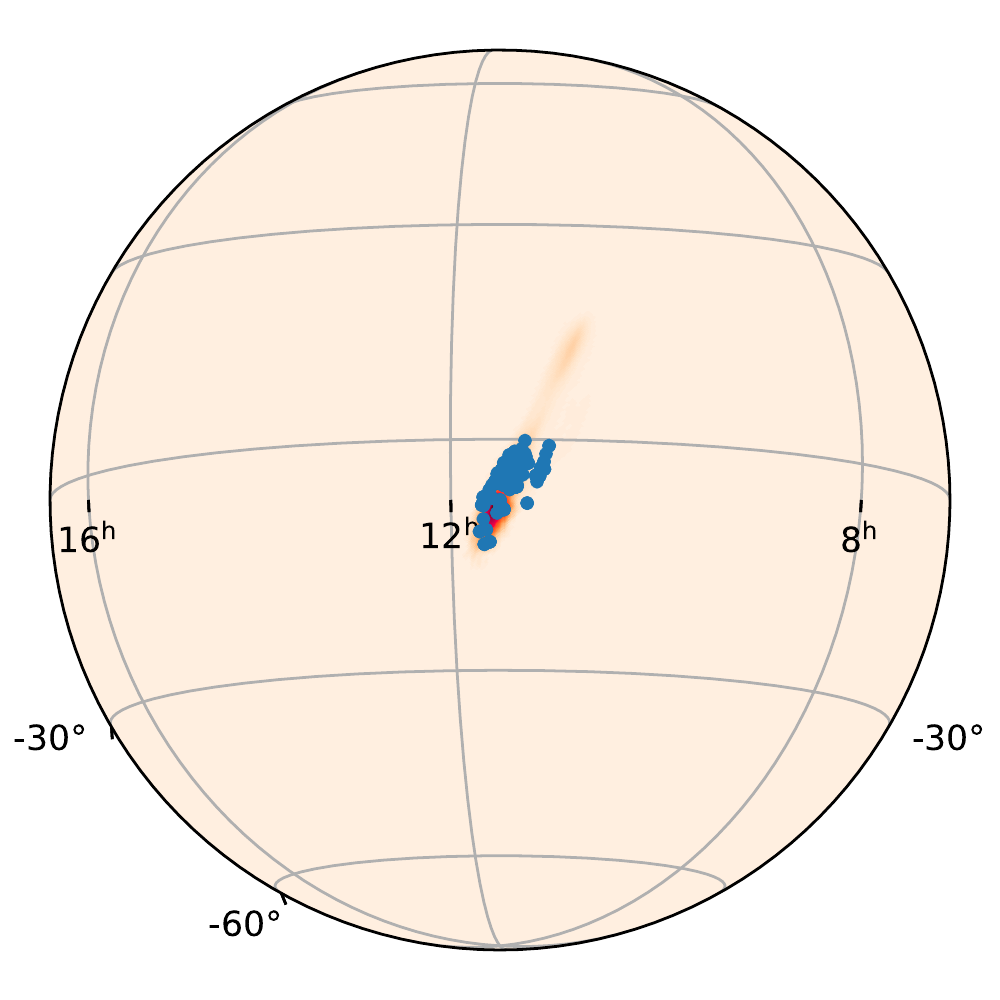}
    \subcaption{S200224ca Skymap}
  \end{minipage}
  \begin{minipage}[b]{0.24\linewidth}
    \centering
    \includegraphics[keepaspectratio, scale=0.24]{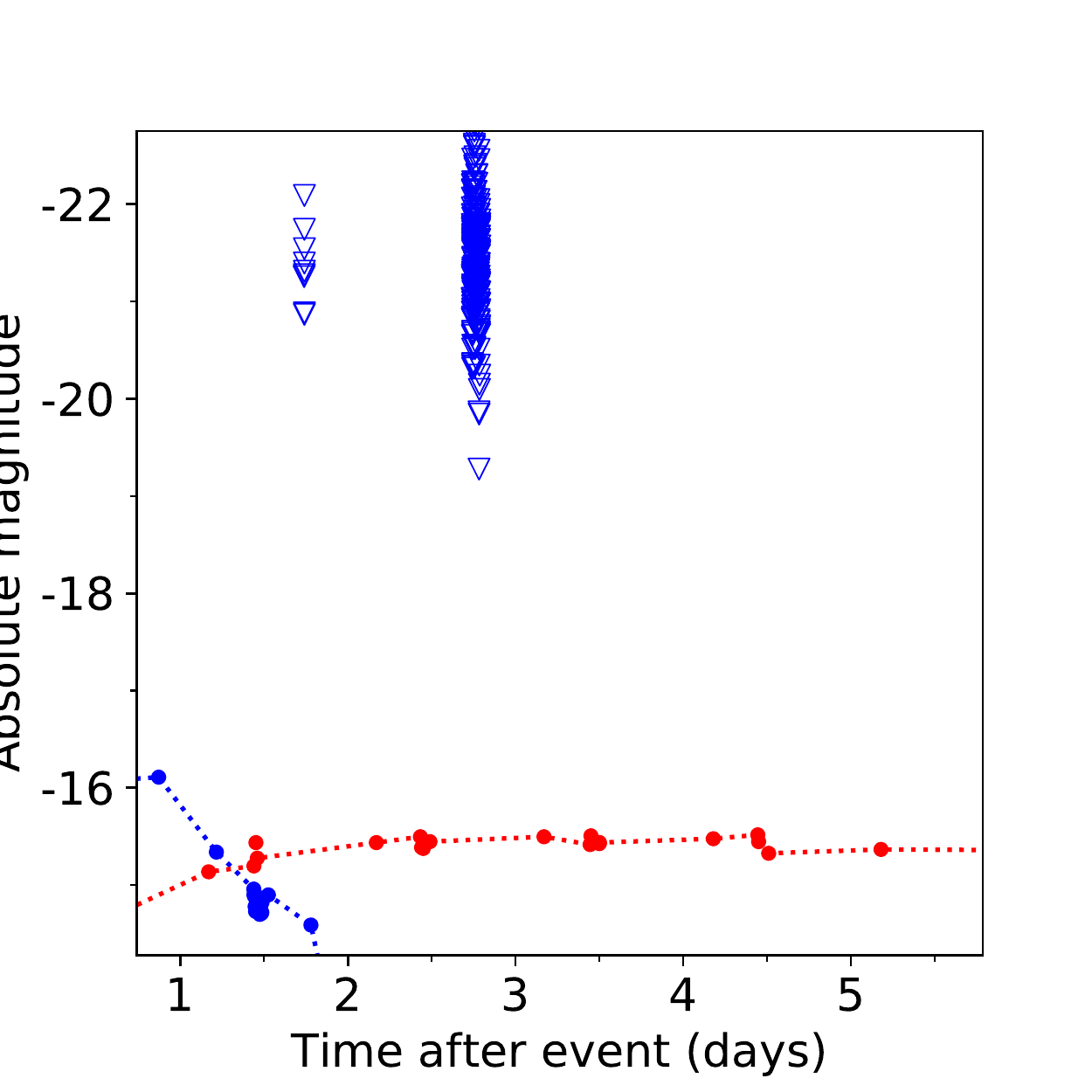}
    \subcaption{S200224ca light curve}
  \end{minipage}\\
  \begin{minipage}[b]{0.24\linewidth}
    \centering
    \includegraphics[keepaspectratio, scale=0.24]{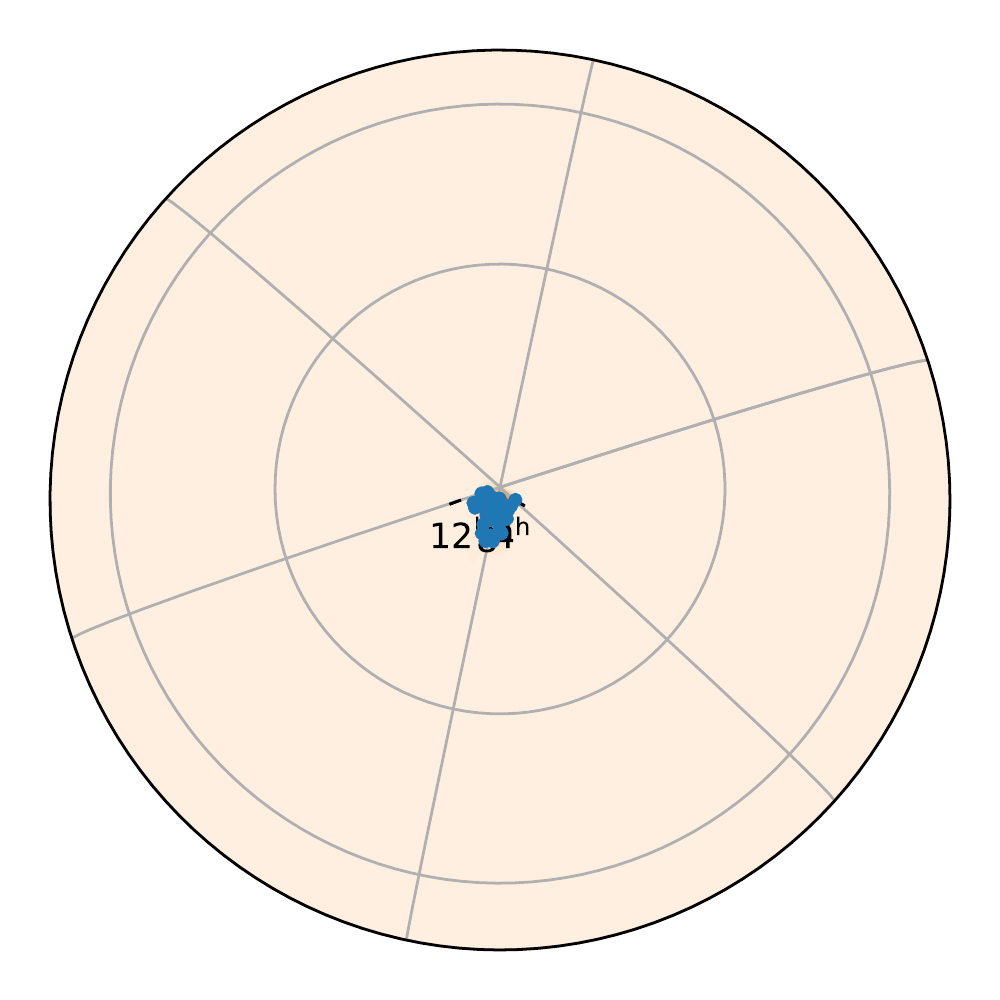}
    \subcaption{S200225q Skymap}
  \end{minipage}
  \begin{minipage}[b]{0.24\linewidth}
    \centering
    \includegraphics[keepaspectratio, scale=0.24]{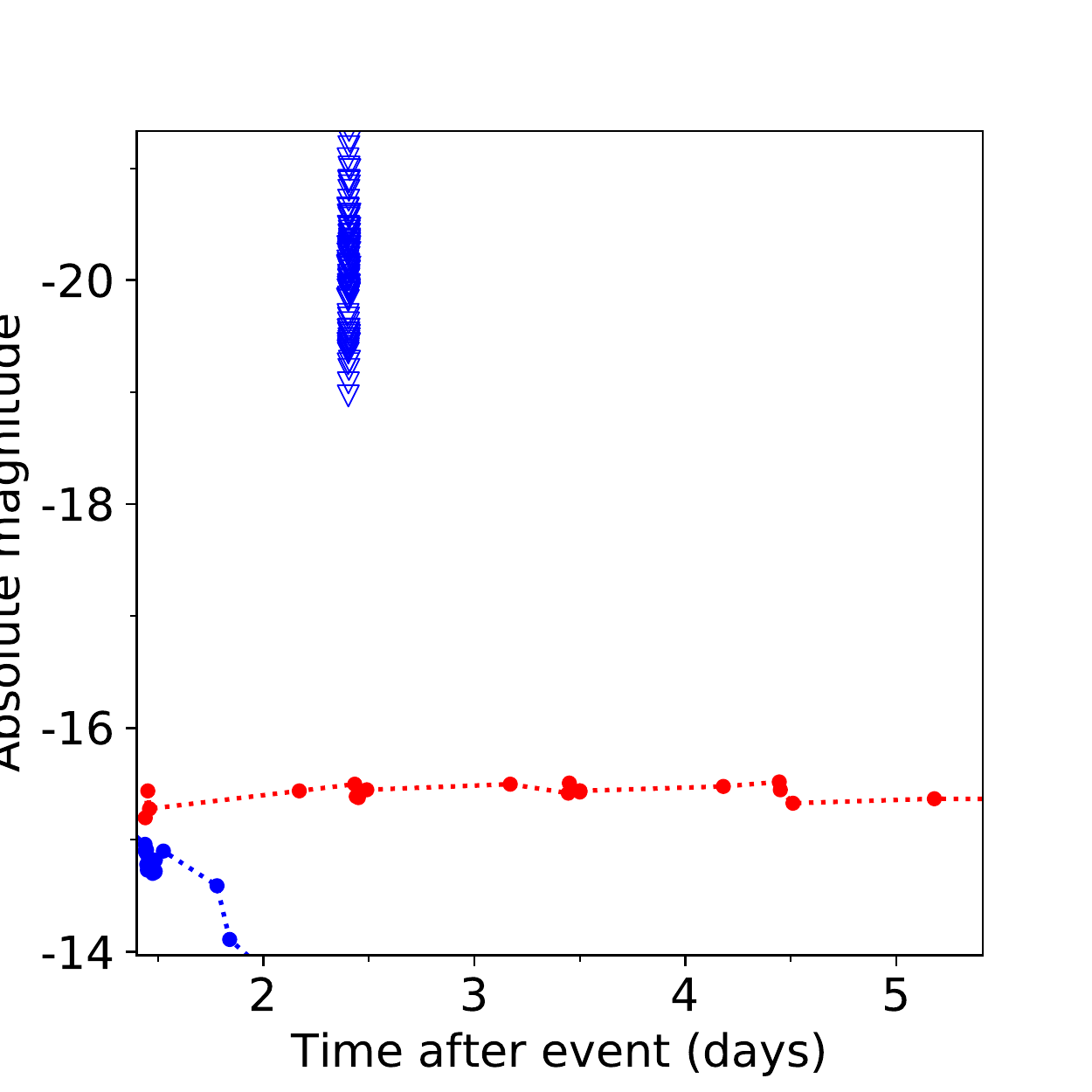}
    \subcaption{S200225q light curve}
  \end{minipage}
  \caption{Skymap and time series of our obtained limiting magnitudes for BBH and MassGap events.}\label{fig:skymap-lc1}
\end{figure}

\begin{figure}[htbp]
  \begin{minipage}[b]{0.24\linewidth}
    \centering
    \includegraphics[keepaspectratio, scale=0.24]{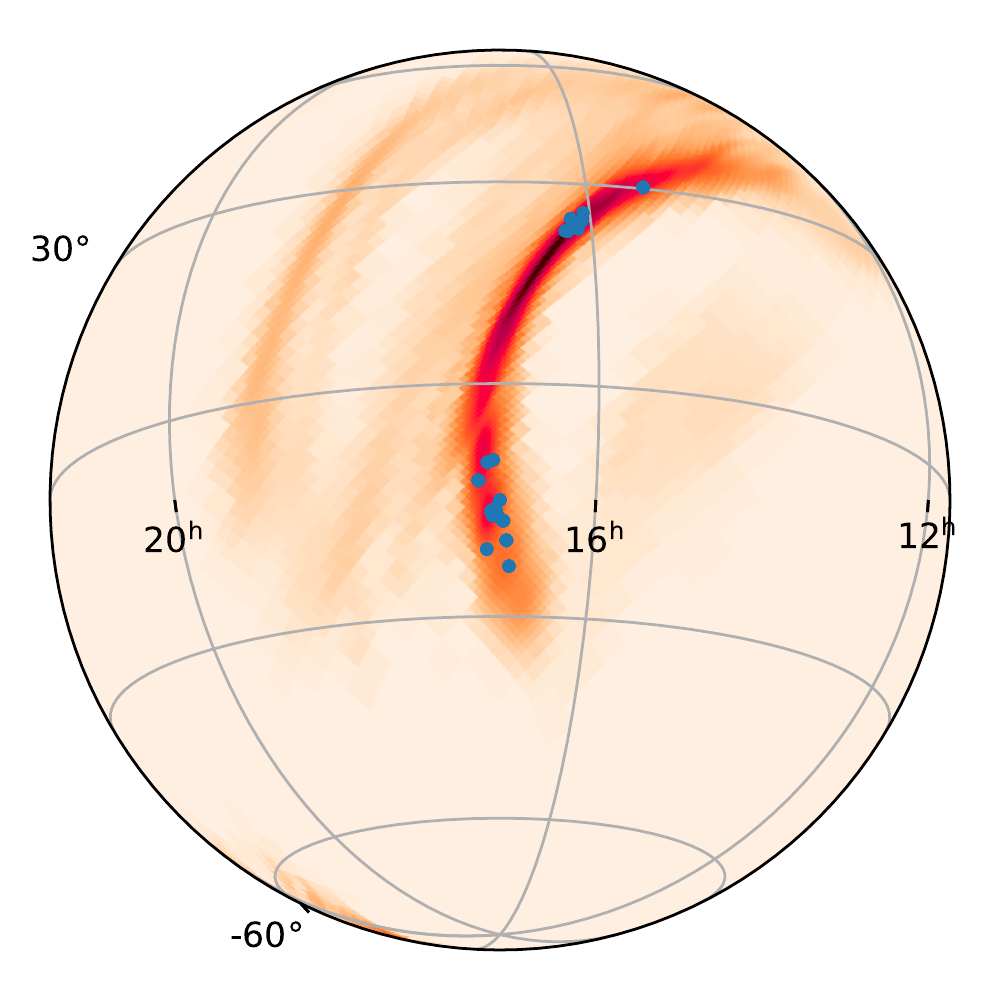}
    \subcaption{GW190425 Skymap}
  \end{minipage}
  \begin{minipage}[b]{0.24\linewidth}
    \centering
    \includegraphics[keepaspectratio, scale=0.24]{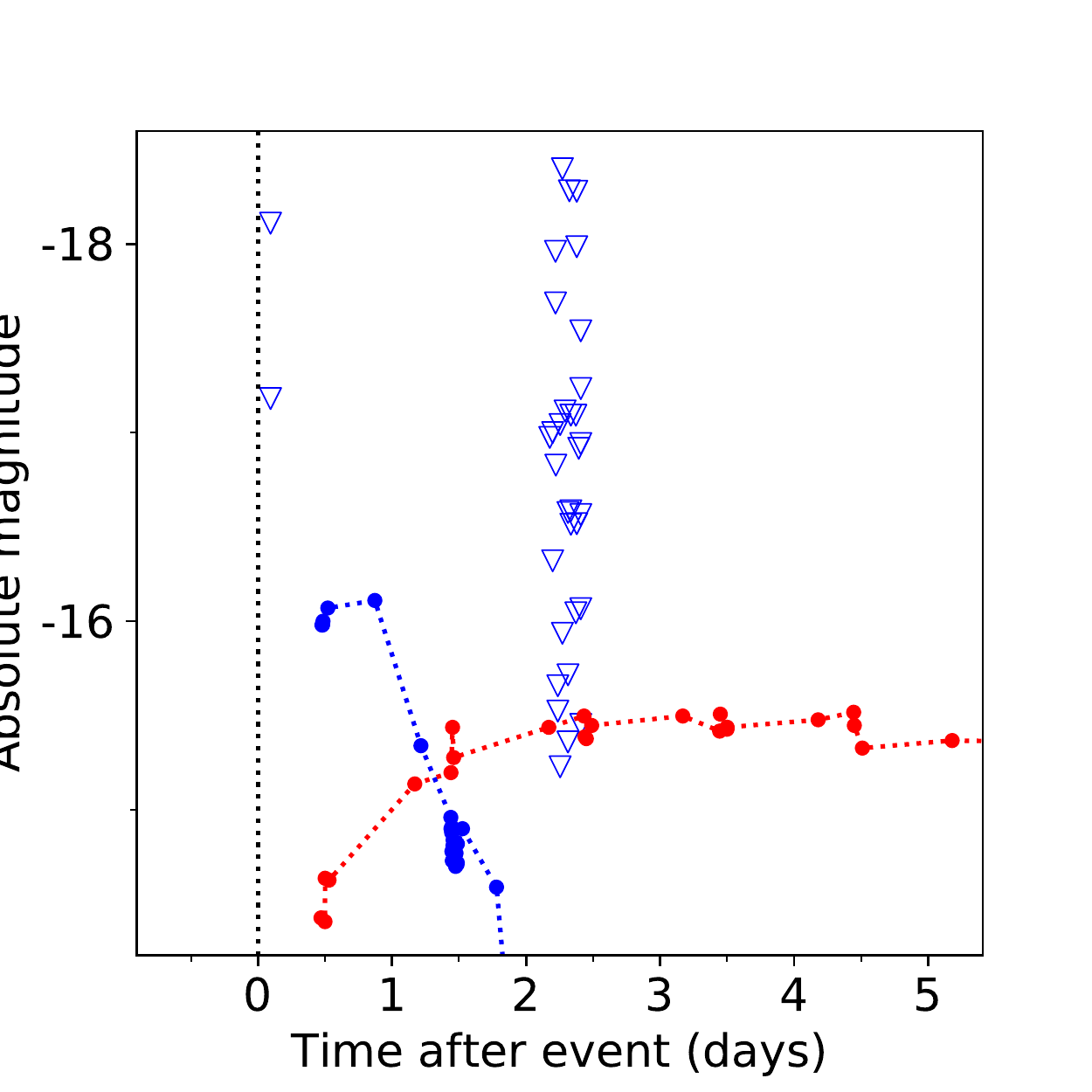}
    \subcaption{GW190425 light curve}
  \end{minipage}
  \begin{minipage}[b]{0.24\linewidth}
    \centering
    \includegraphics[keepaspectratio, scale=0.24]{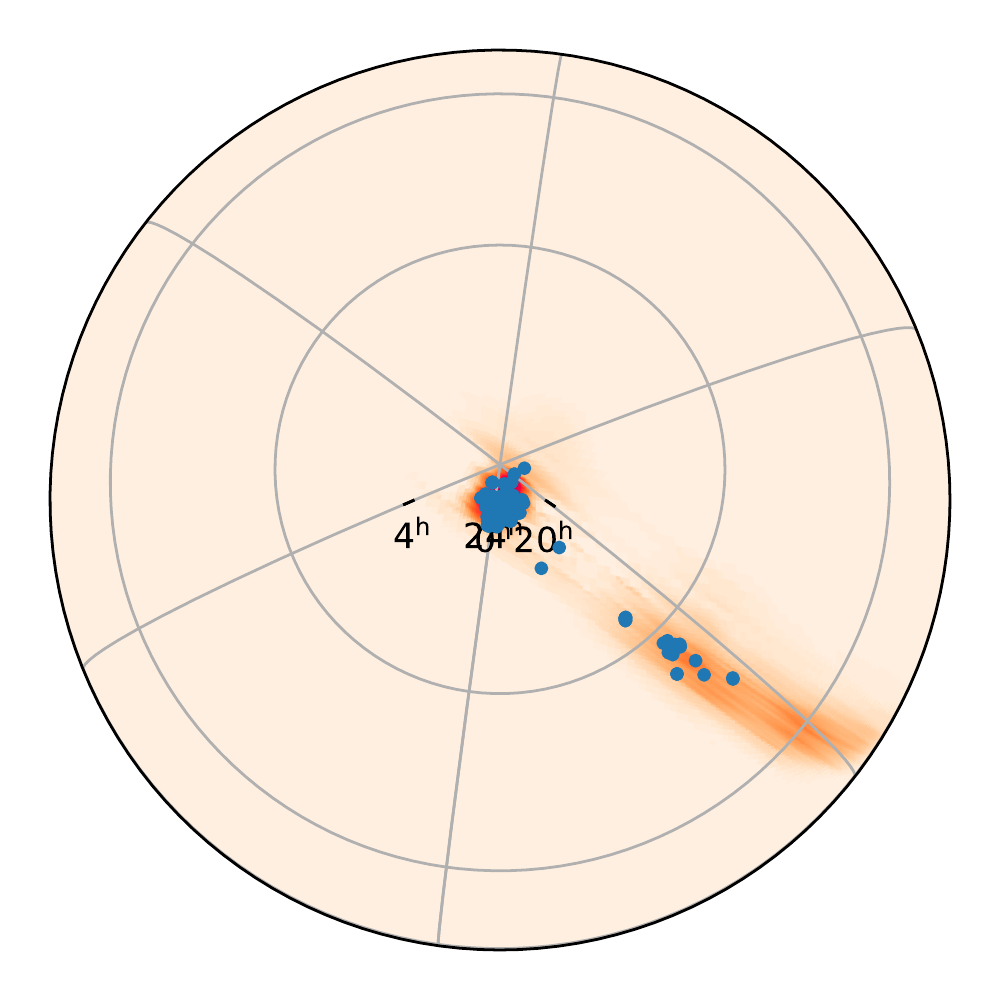}
    \subcaption{S190426c Skymap}
  \end{minipage}
  \begin{minipage}[b]{0.24\linewidth}
    \centering
    \includegraphics[keepaspectratio, scale=0.24]{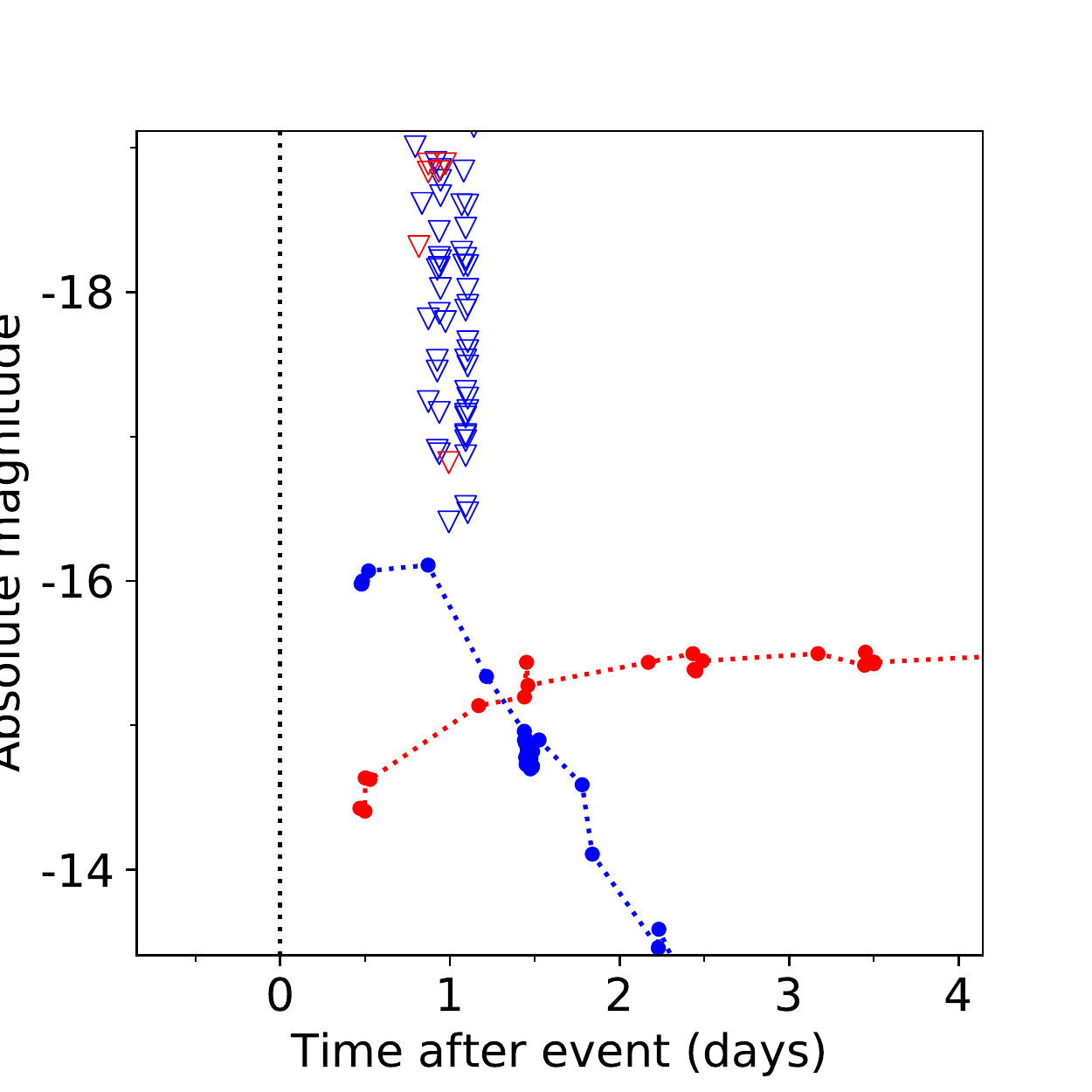}
    \subcaption{S190426c light curve}
  \end{minipage}\\
  \begin{minipage}[b]{0.24\linewidth}
    \centering
    \includegraphics[keepaspectratio, scale=0.24]{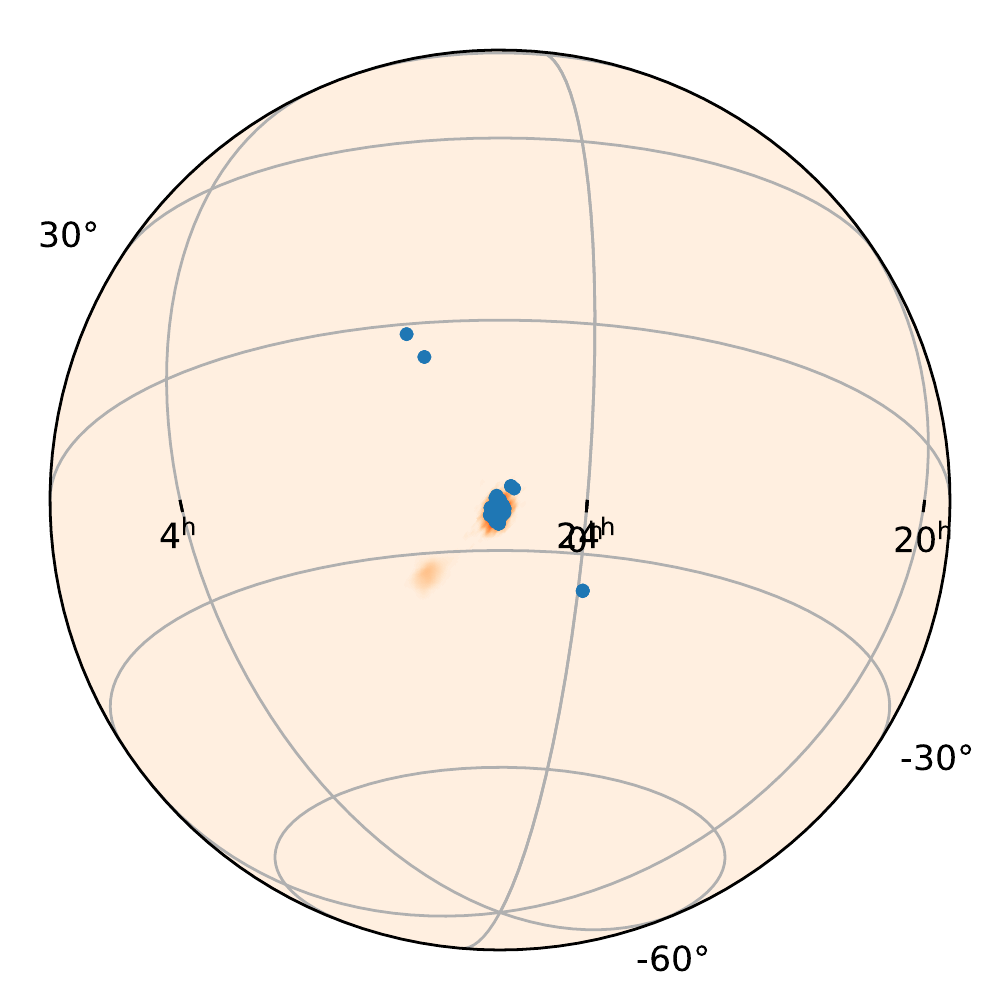}
    \subcaption{S190814bv Skymap}
  \end{minipage}
  \begin{minipage}[b]{0.24\linewidth}
    \centering
    \includegraphics[keepaspectratio, scale=0.24]{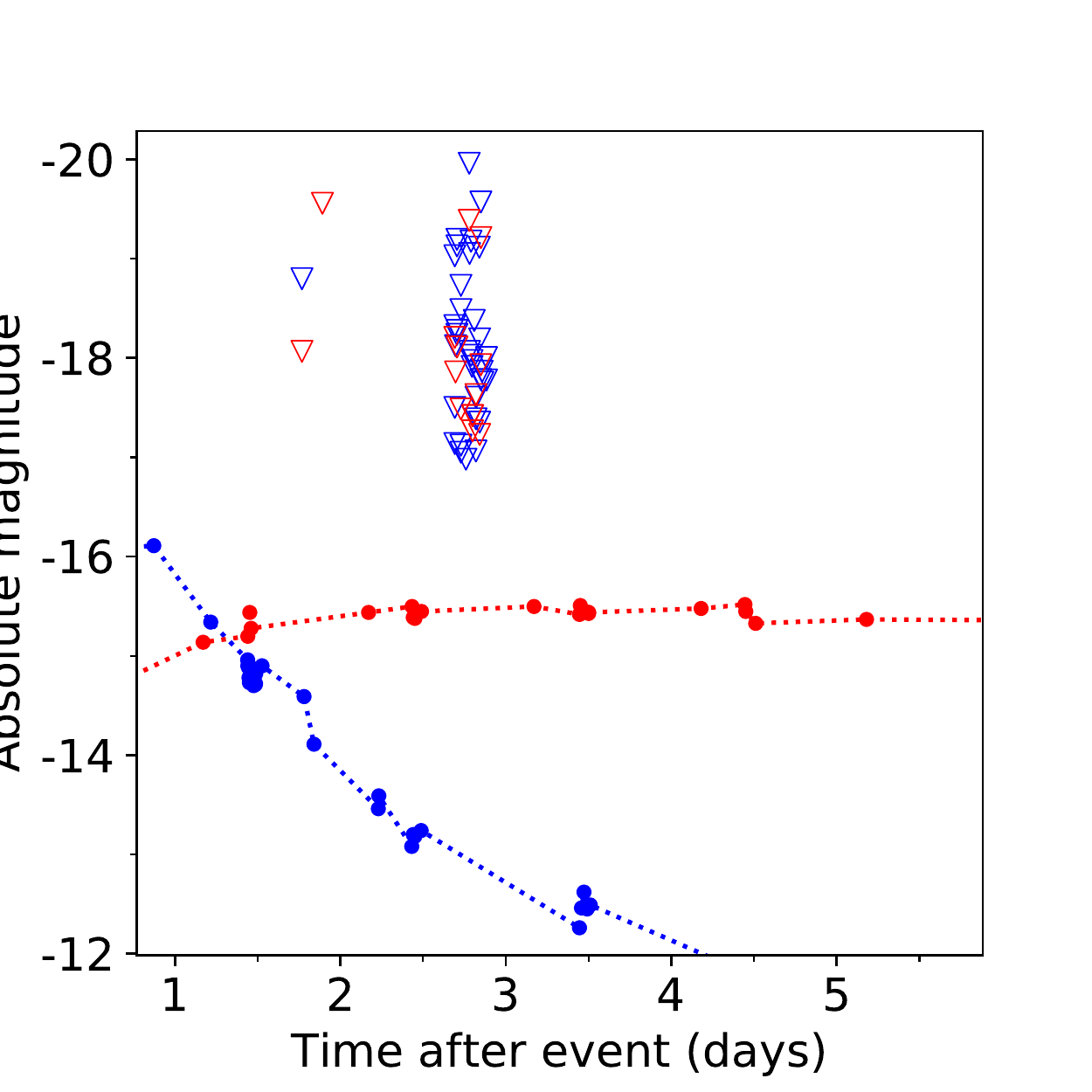}
    \subcaption{S190814bv light curve}
  \end{minipage}
  \begin{minipage}[b]{0.24\linewidth}
    \centering
    \includegraphics[keepaspectratio, scale=0.24]{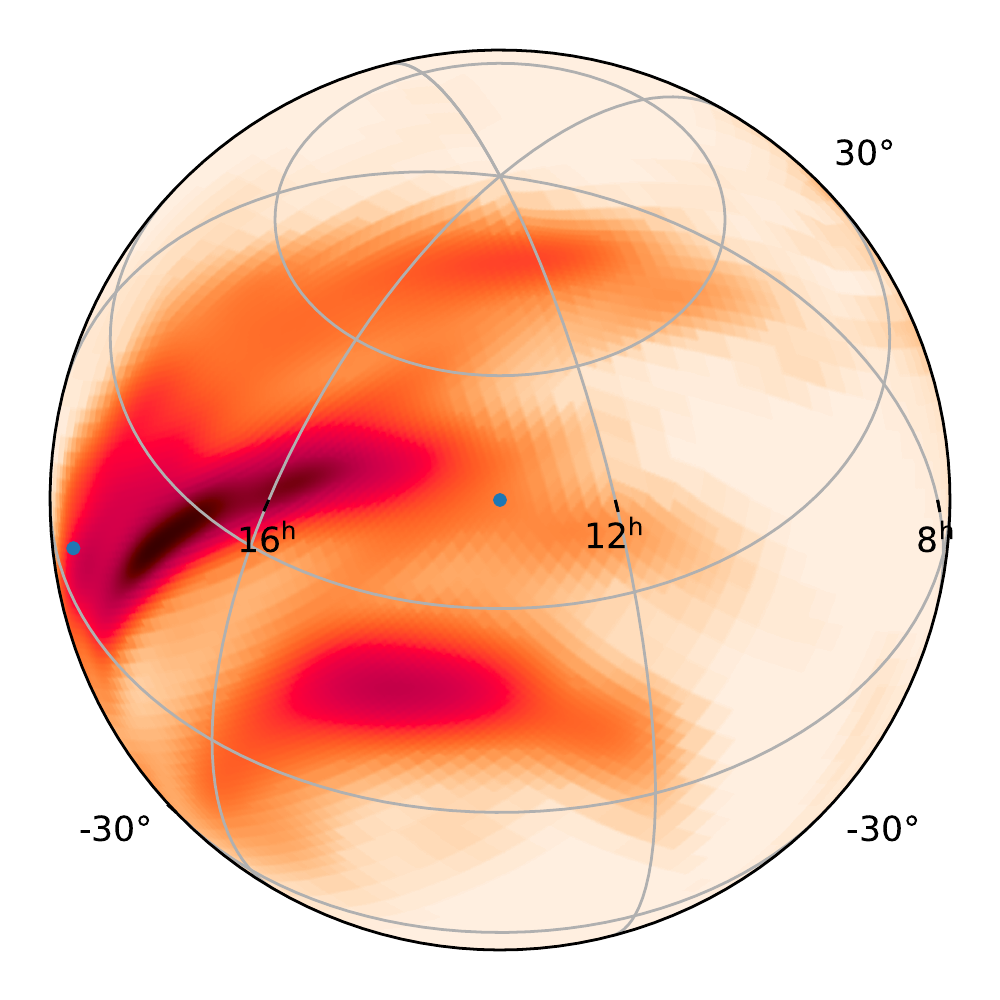}
    \subcaption{S190901ap Skymap}
  \end{minipage}
  \begin{minipage}[b]{0.24\linewidth}
    \centering
    \includegraphics[keepaspectratio, scale=0.24]{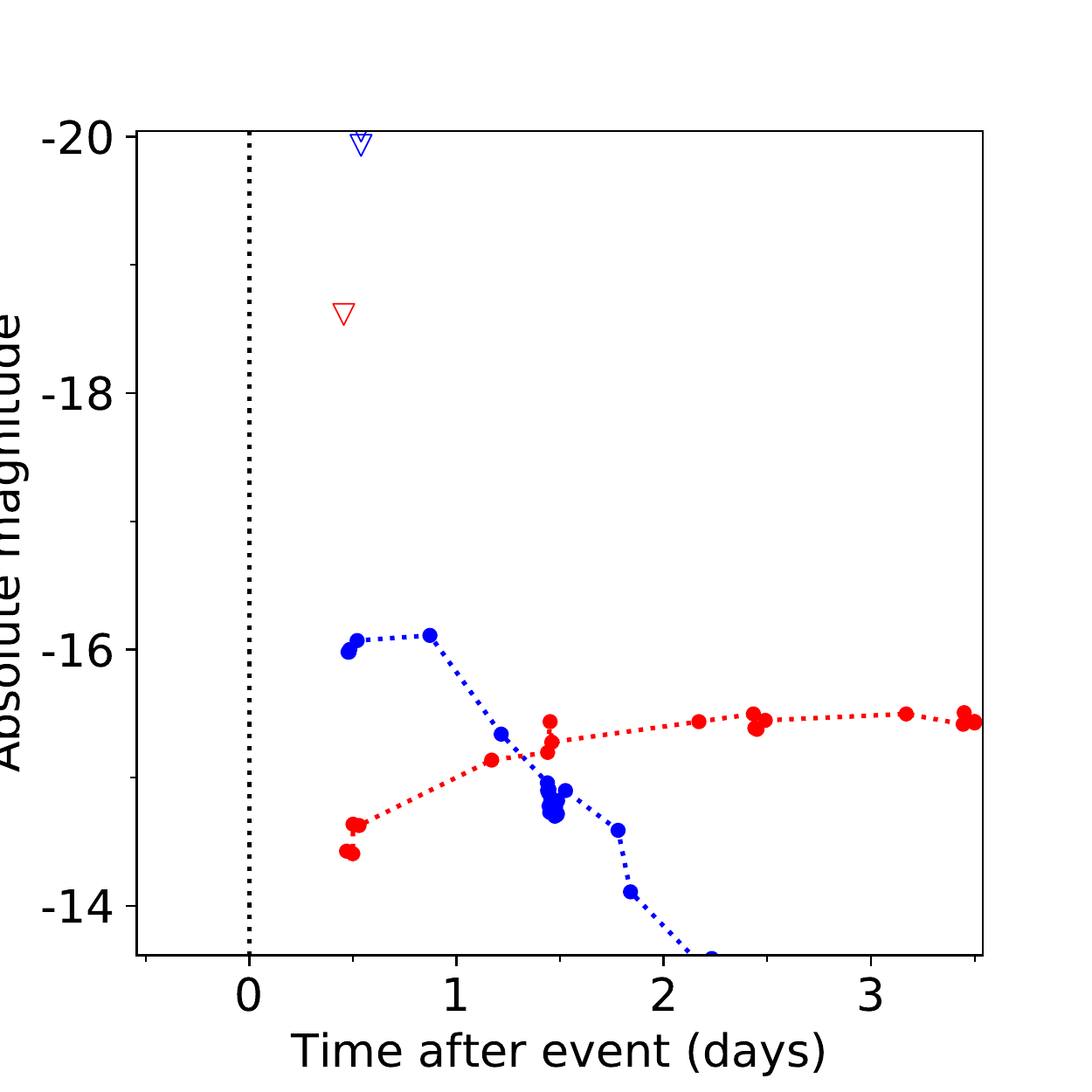}
    \subcaption{S190901ap light curve}
  \end{minipage}\\
  \begin{minipage}[b]{0.24\linewidth}
    \centering
    \includegraphics[keepaspectratio, scale=0.24]{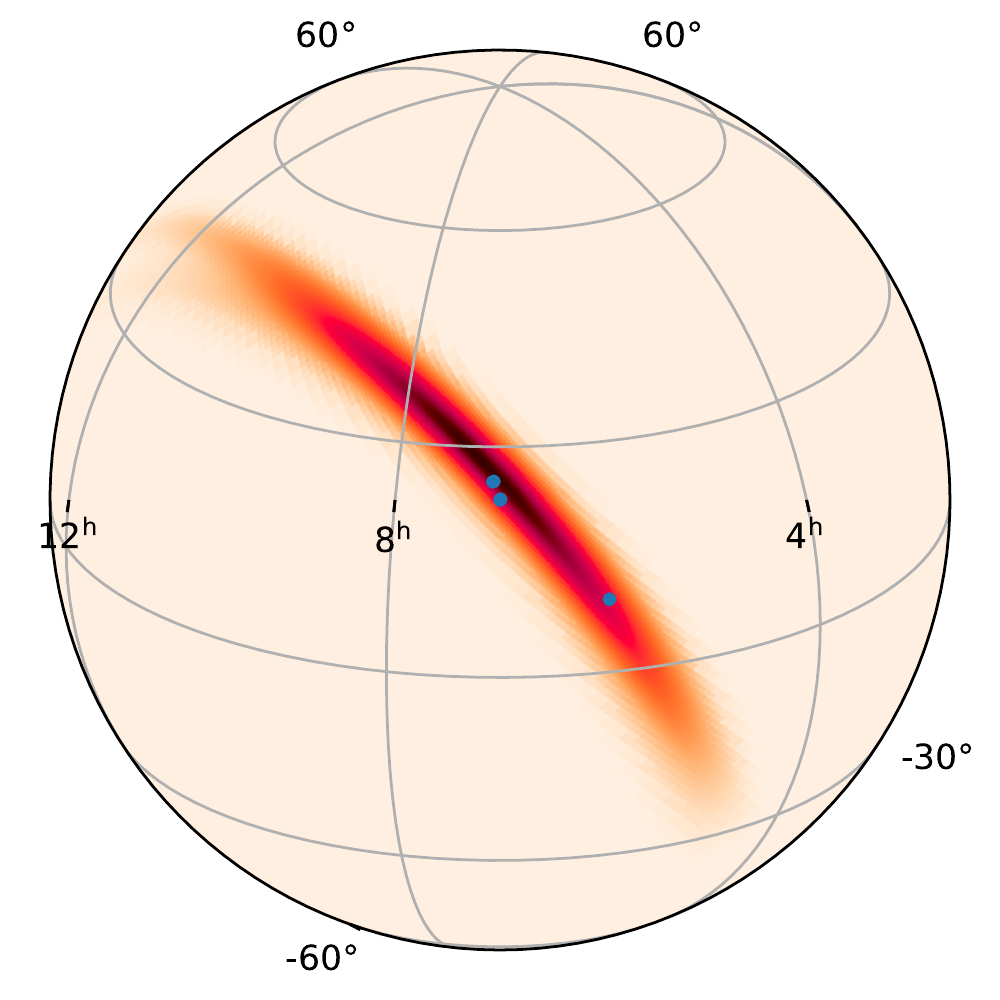}
    \subcaption{S190923y Skymap}
  \end{minipage}
  \begin{minipage}[b]{0.24\linewidth}
    \centering
    \includegraphics[keepaspectratio, scale=0.24]{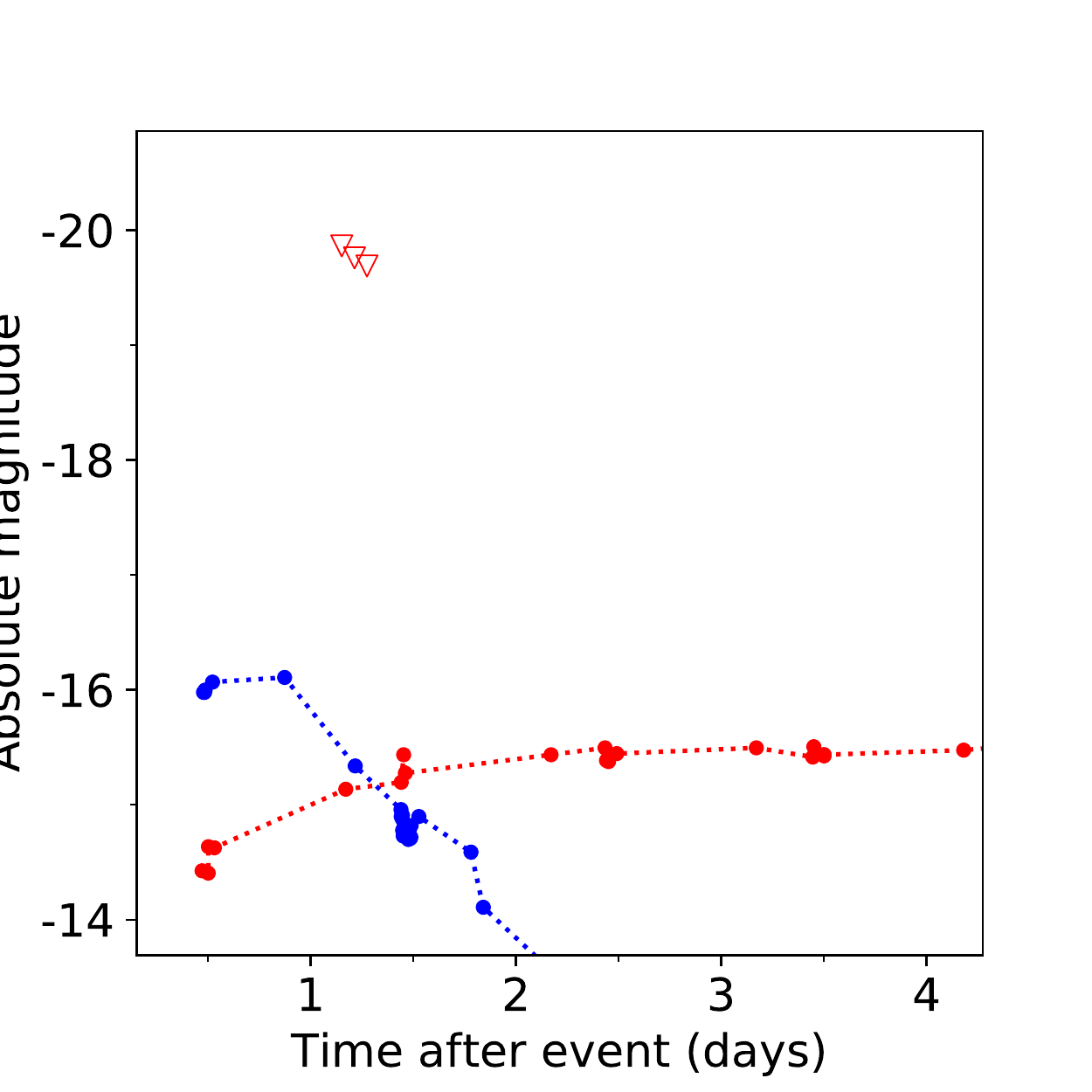}
    \subcaption{S190923y light curve}
  \end{minipage}
  \begin{minipage}[b]{0.24\linewidth}
    \centering
    \includegraphics[keepaspectratio, scale=0.24]{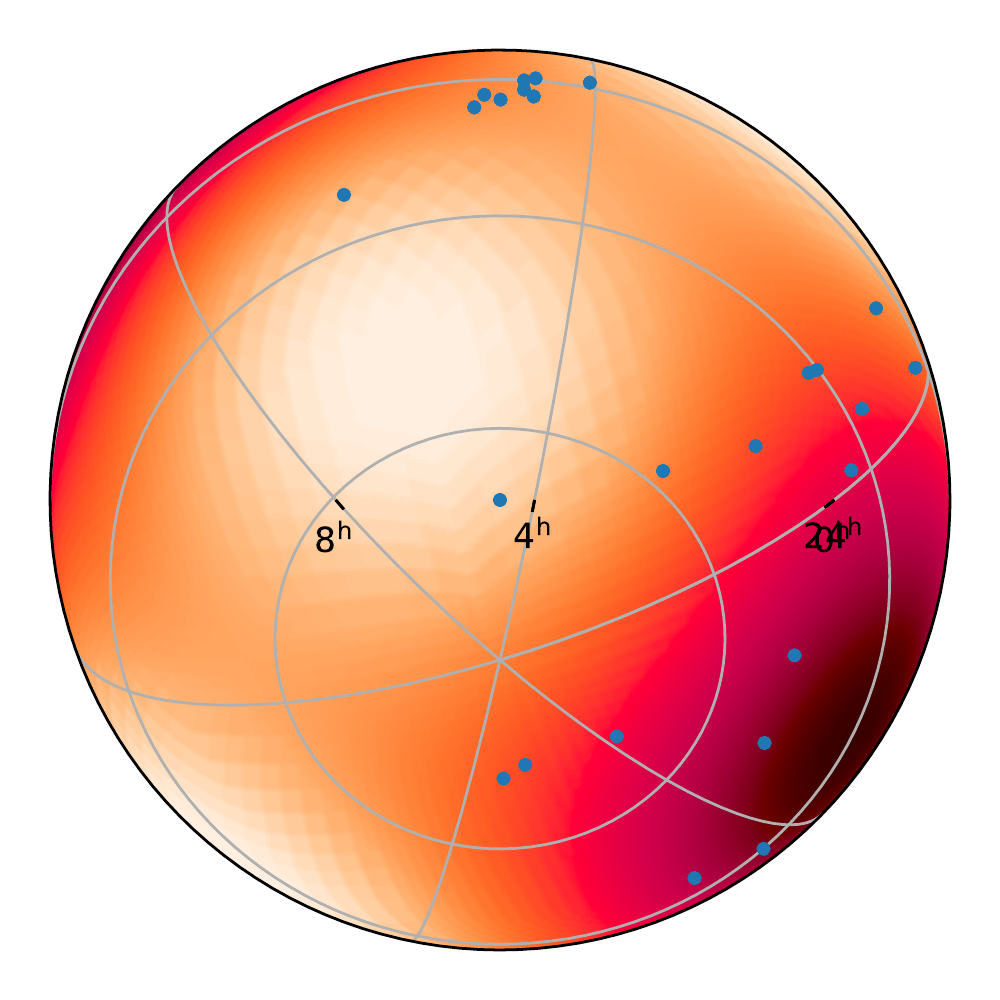}
    \subcaption{S190930t Skymap}
  \end{minipage}
  \begin{minipage}[b]{0.24\linewidth}
    \centering
    \includegraphics[keepaspectratio, scale=0.24]{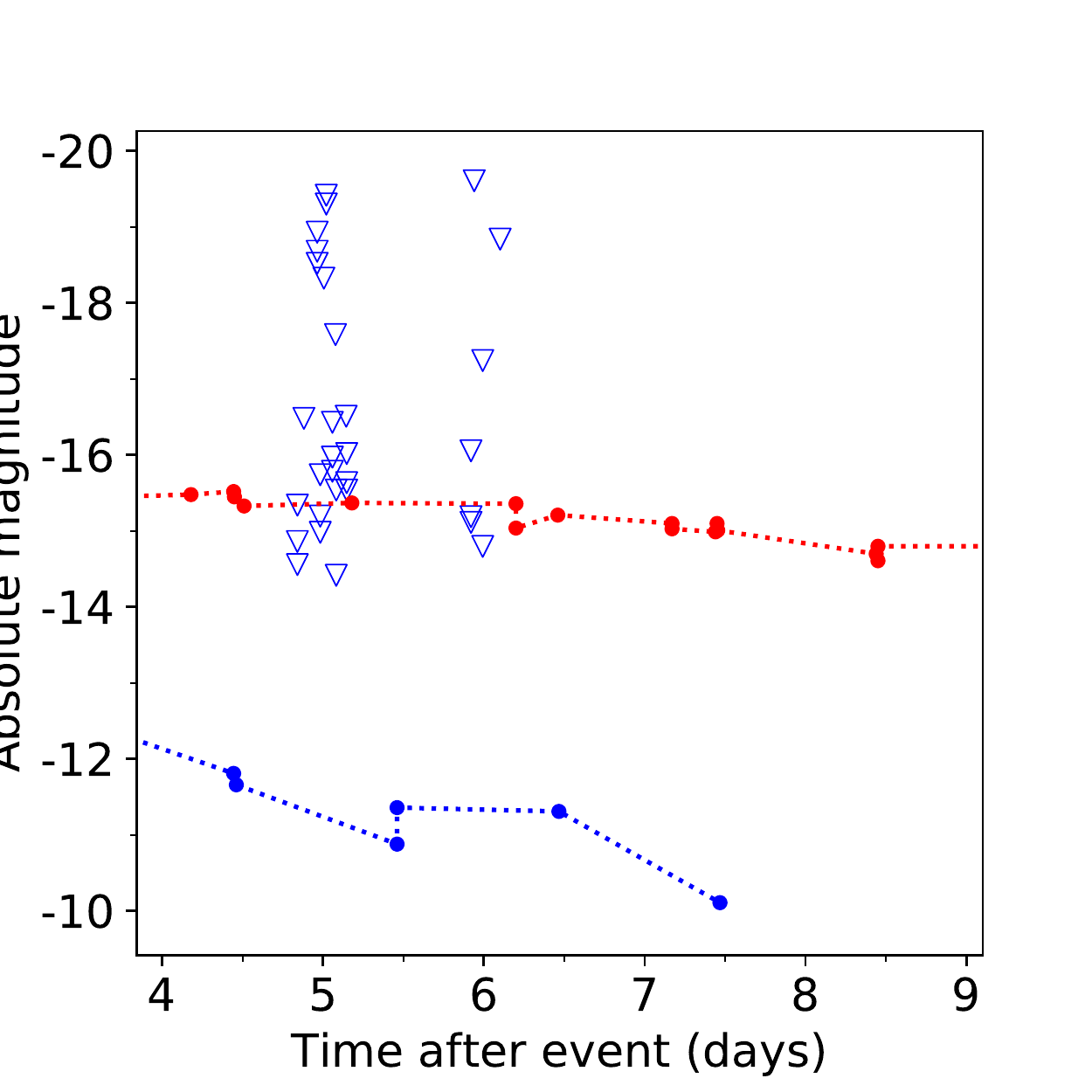}
    \subcaption{S190930t light curve}
  \end{minipage}\\
  \begin{minipage}[b]{0.24\linewidth}
    \centering
    \includegraphics[keepaspectratio, scale=0.24]{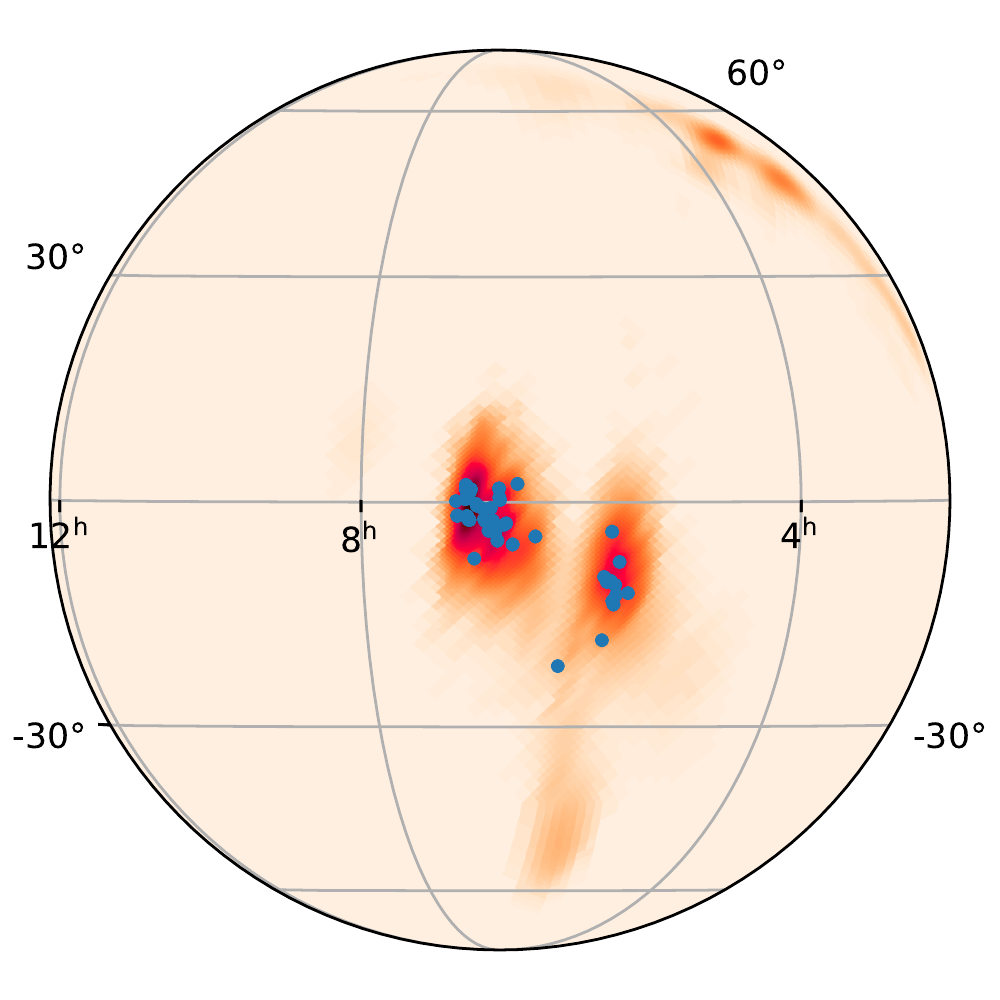}
    \subcaption{S191213g Skymap}
  \end{minipage}
  \begin{minipage}[b]{0.24\linewidth}
    \centering
    \includegraphics[keepaspectratio, scale=0.24]{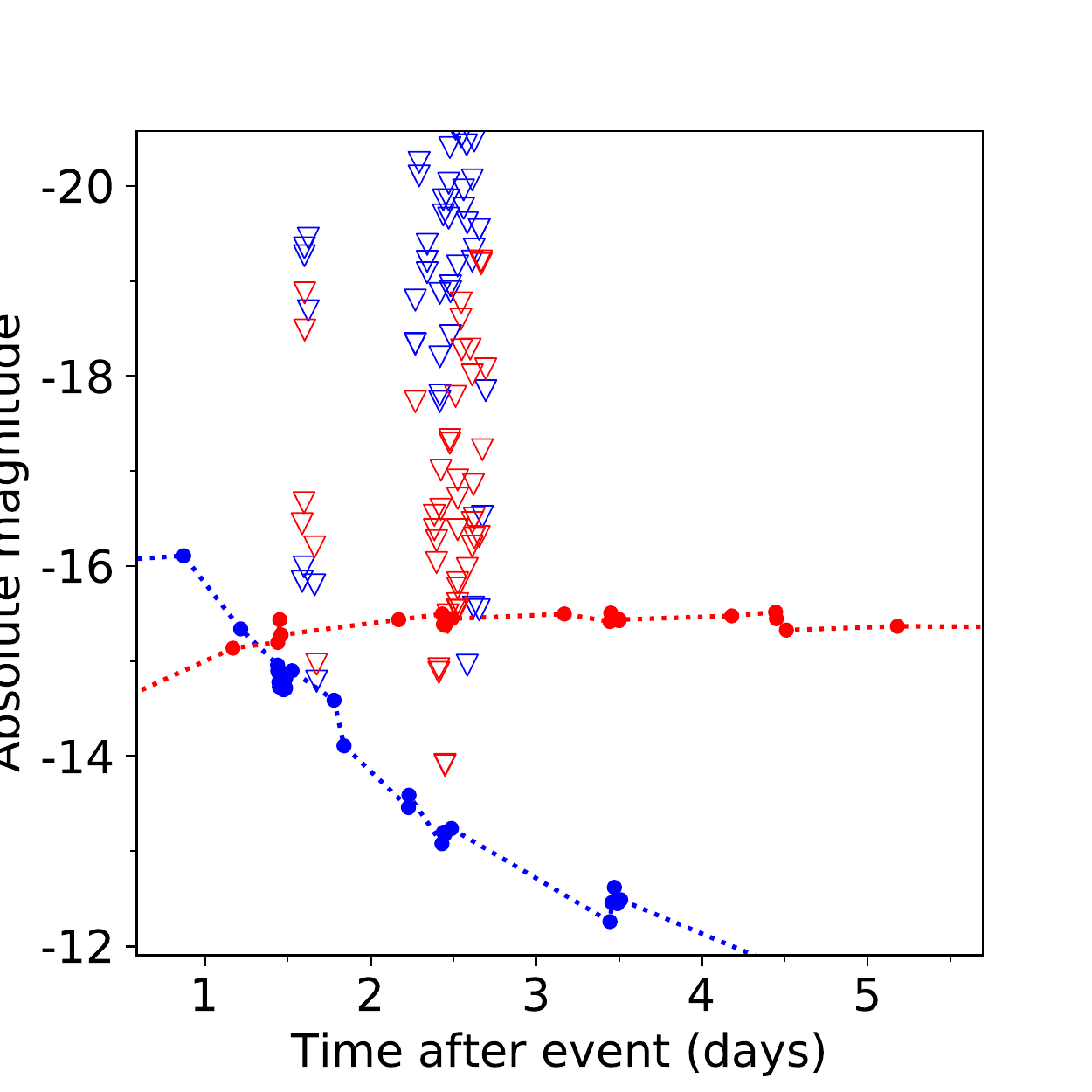}
    \subcaption{S191213g light curve}
  \end{minipage}
  \begin{minipage}[b]{0.24\linewidth}
    \centering
    \includegraphics[keepaspectratio, scale=0.24]{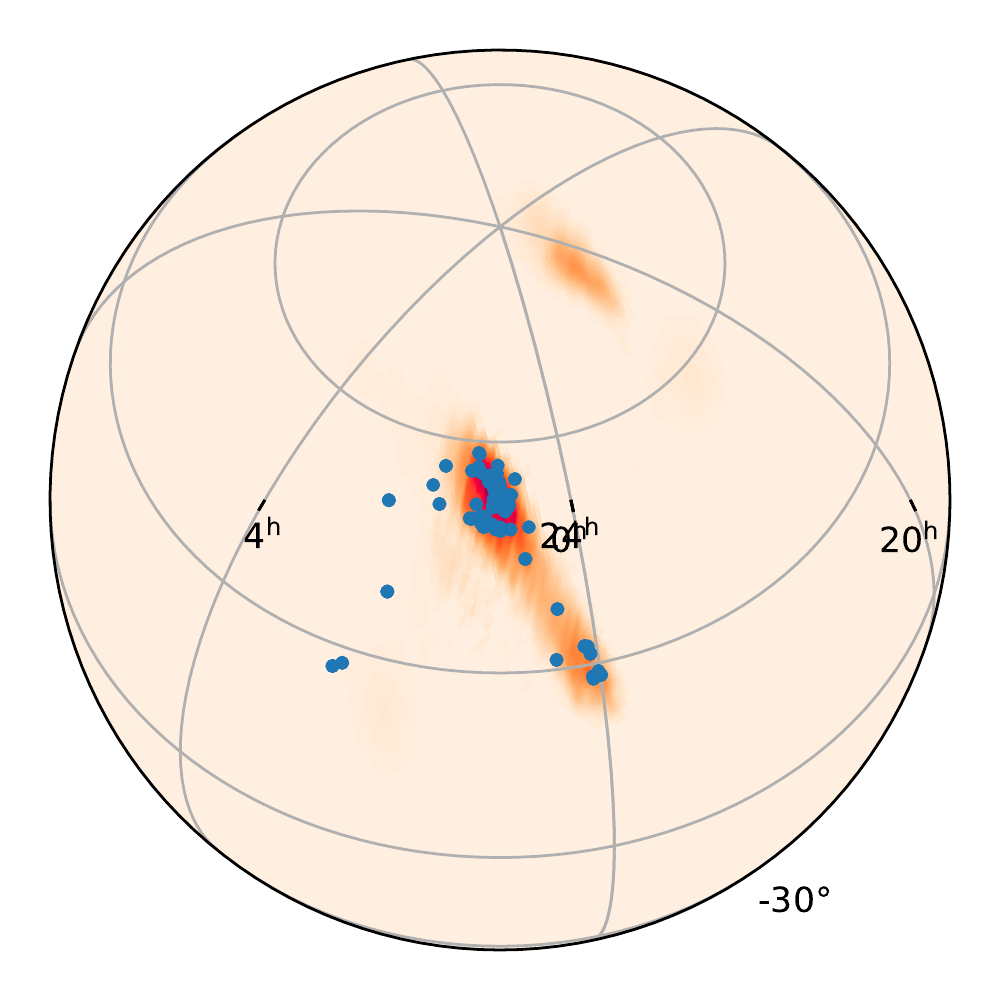}
    \subcaption{S200213t Skymap}
  \end{minipage}
  \begin{minipage}[b]{0.24\linewidth}
    \centering
    \includegraphics[keepaspectratio, scale=0.24]{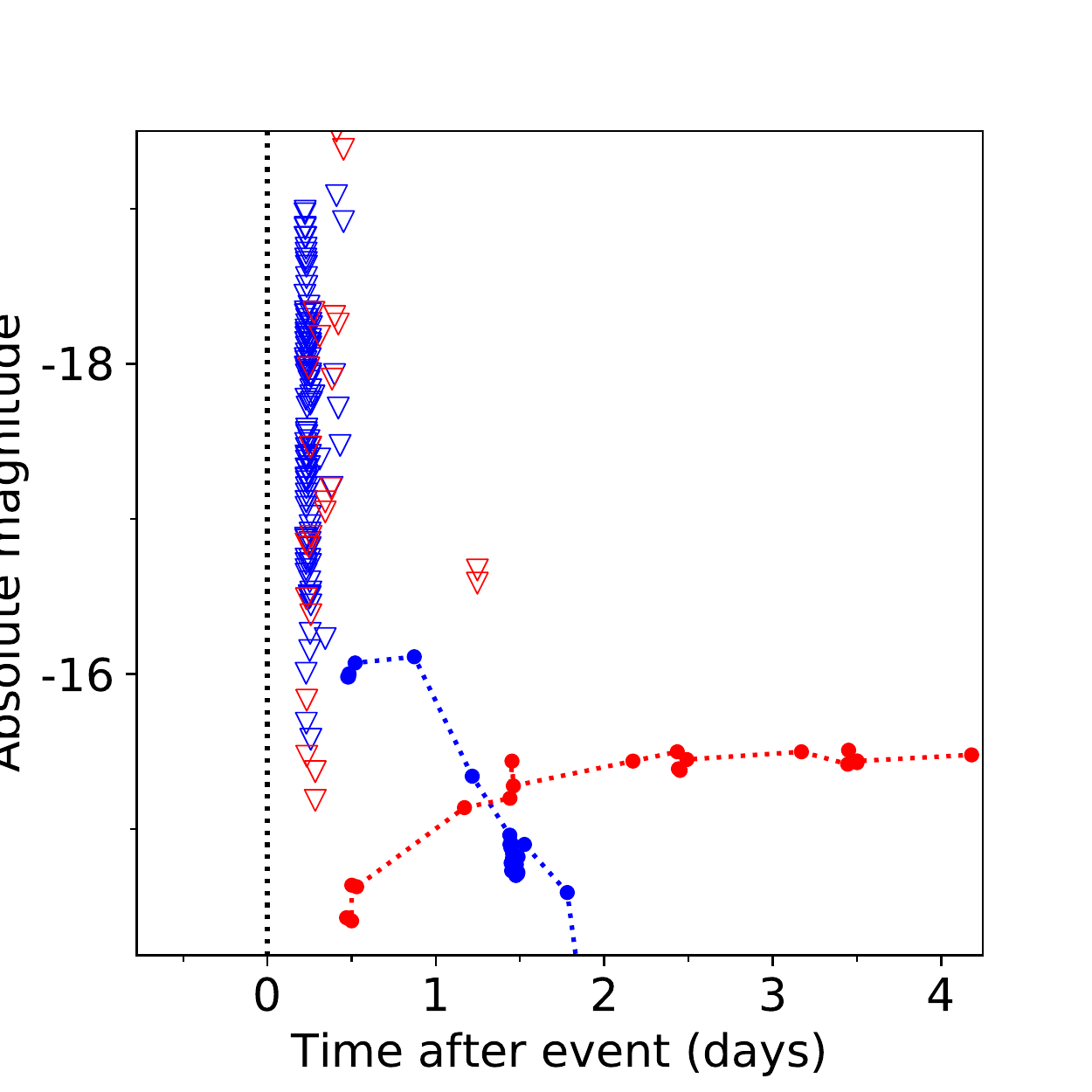}
    \subcaption{S200213t light curve}
  \end{minipage}
  \caption{Skymap and time series of our obtained limiting magnitudes for BNS and NSBH events.}
    \label{fig:skymap-lc2}
\end{figure}

\begin{figure}[htbp]
  \begin{minipage}[b]{0.24\linewidth}
    \centering
    \includegraphics[keepaspectratio, scale=0.24]{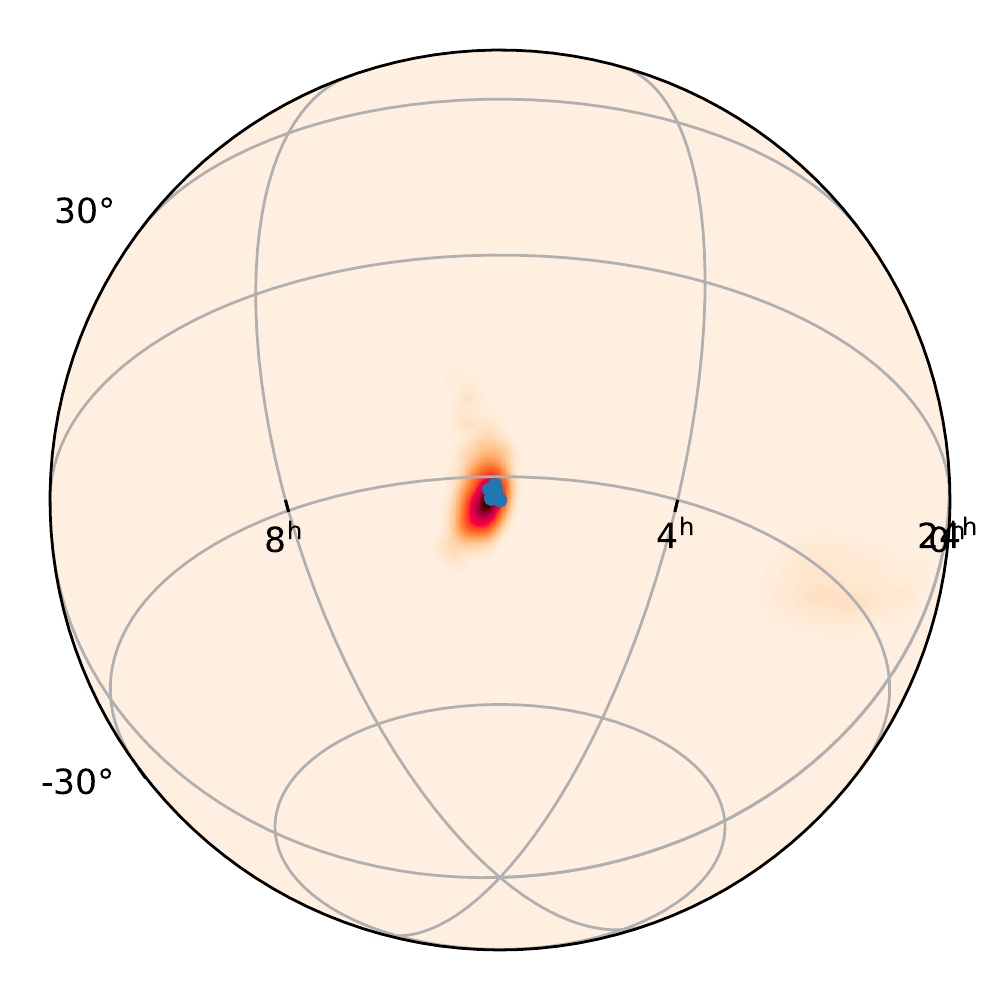}
    \subcaption{S190510g Skymap}
  \end{minipage}
  \begin{minipage}[b]{0.24\linewidth}
    \centering
    \includegraphics[keepaspectratio, scale=0.24]{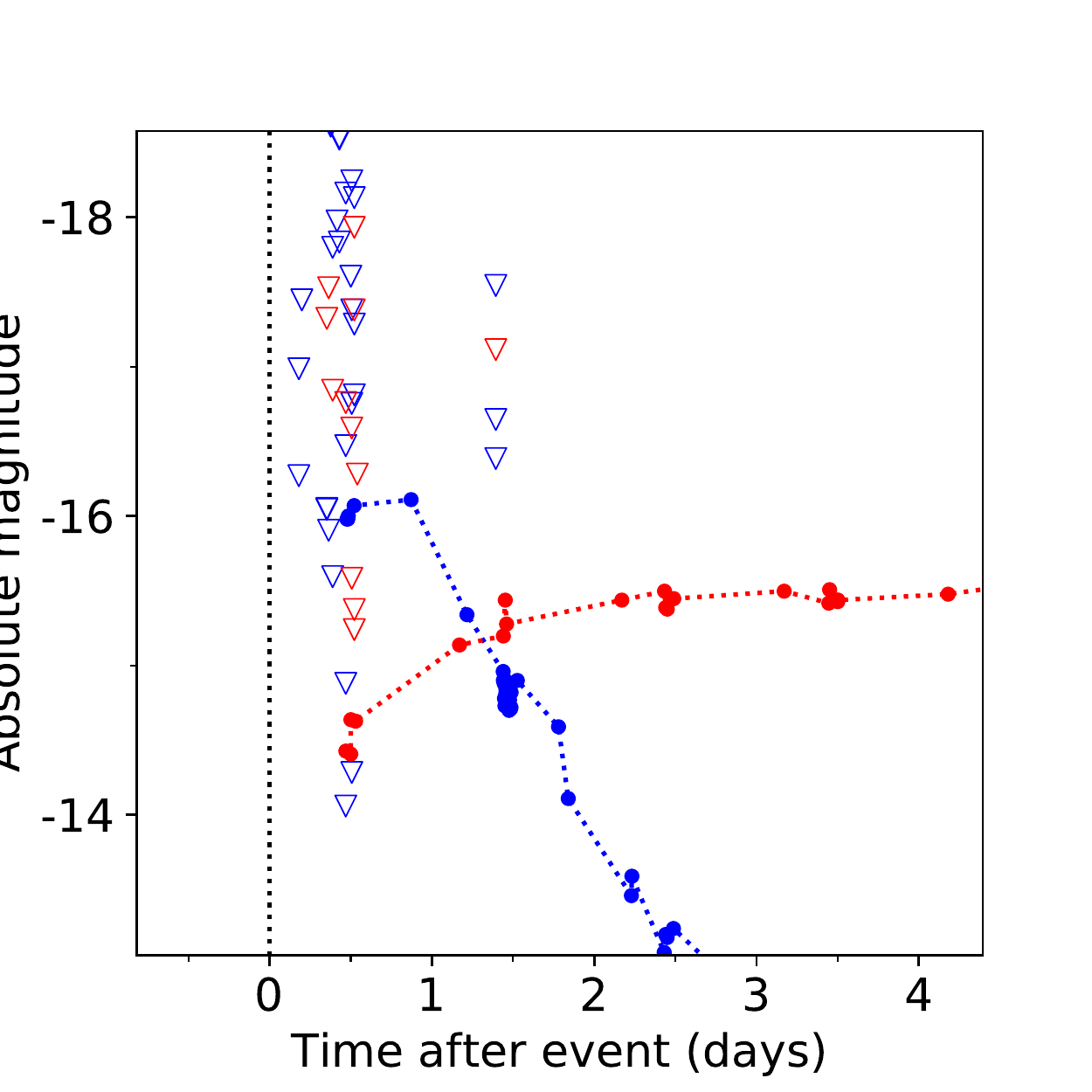}
    \subcaption{S190510g light curve}
  \end{minipage}
  \begin{minipage}[b]{0.24\linewidth}
    \centering
    \includegraphics[keepaspectratio, scale=0.24]{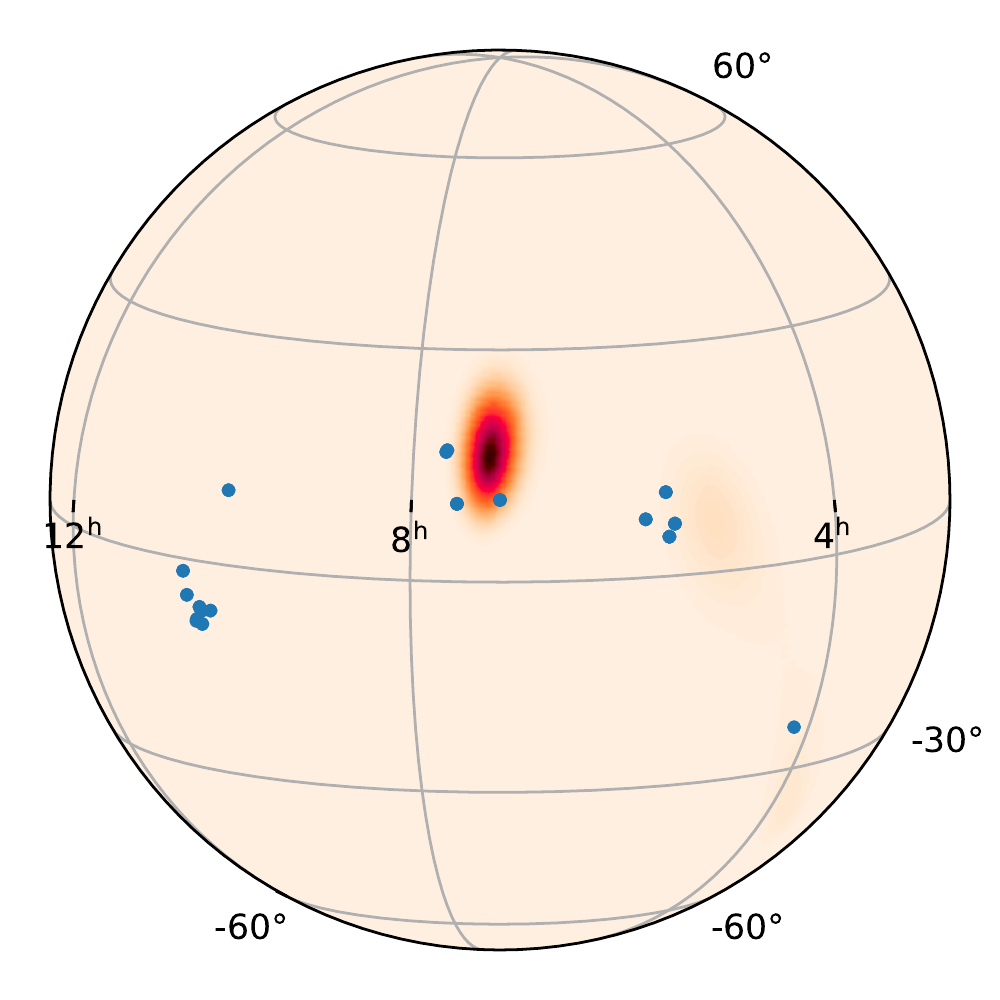}
    \subcaption{S200114f Skymap}
  \end{minipage}
  \begin{minipage}[b]{0.24\linewidth}
    \centering
    \includegraphics[keepaspectratio, scale=0.24]{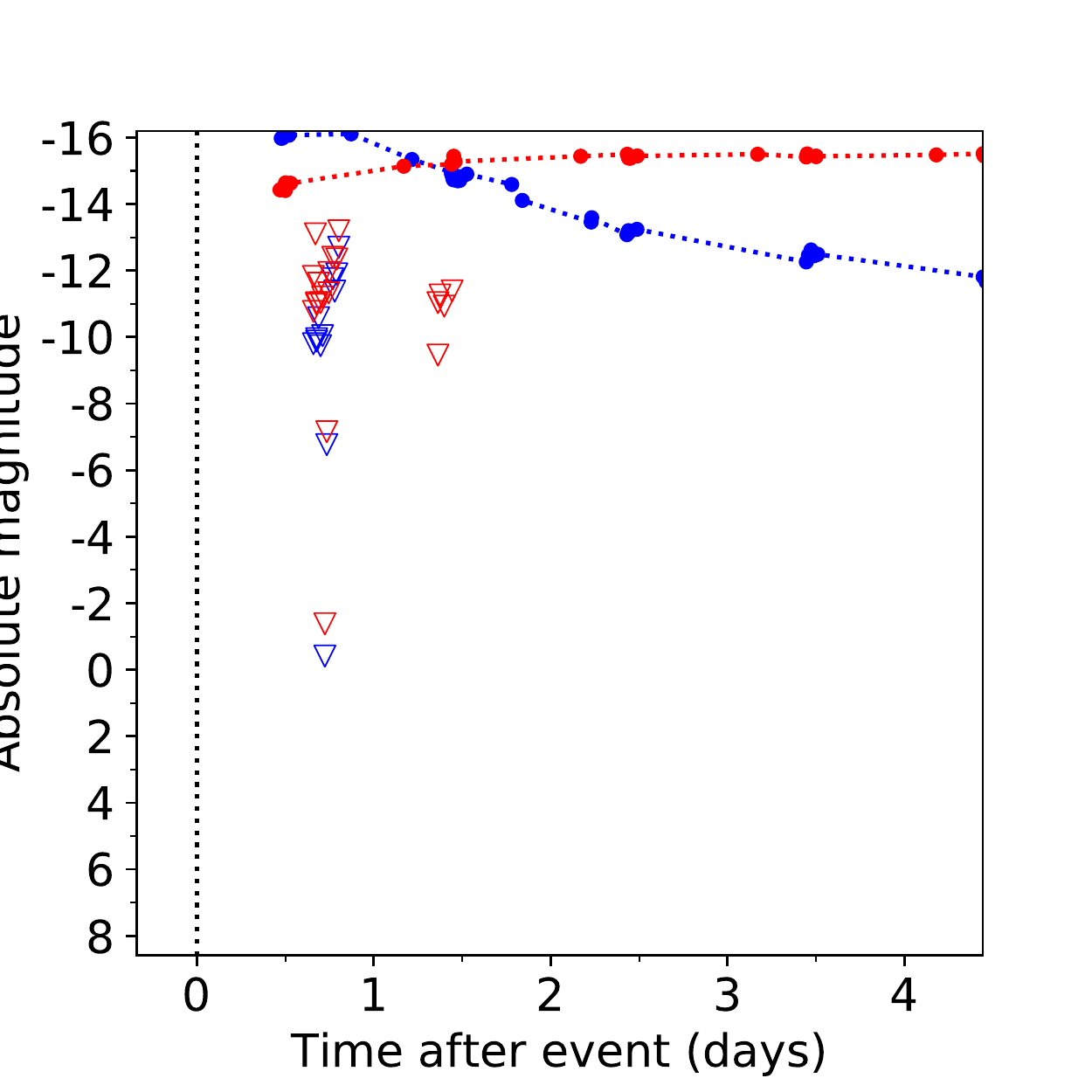}
    \subcaption{S200114f light curve}
  \end{minipage}
  \caption{Skymap and time series of our obtained limiting magnitudes for Terrestrial and Burst events.}
    \label{fig:skymap-lc3}
\end{figure}

We displayed skymaps and time series of the obtained limiting magnitudes obtained in our followed-up events during O3 in Figs.~\ref{fig:skymap-lc1}, \ref{fig:skymap-lc2}, \ref{fig:skymap-lc3}. Filled circles in the skymap show the positions of observed galaxies, and the colored regions display the GW localization maps. Open triangles and filled circles in the time series figure show the limiting magnitudes and light curve of AT~2017gfo in absolute magnitude, respectively.

\end{document}